\documentclass[12pt,amssymb]{article} %Latex2e

\usepackage{mathrsfs}
\usepackage{amssymb}
\usepackage[dvips]{graphicx}
\newcommand{\newsection}[1]{
\addtocounter{section}{1} \setcounter{equation}{0}
\setcounter{subsection}{0} \addcontentsline{toc}{section}{\protect
\numberline{\arabic{section}}{{\rm #1}}} \vglue .6cm \pagebreak[3]
\noindent{ \bf  \thesection. #1}\nopagebreak[4]\par\vskip .3cm}
\newcommand{\newsubsection}[1]{
\addtocounter{subsection}{1}\setcounter{subsubsection}{0}
\addcontentsline{toc}{subsection}{\protect
\numberline{\arabic{section}.\arabic{subsection}}{#1}} \vglue .4cm
\pagebreak[3] \noindent{\it \thesubsection.
#1}\nopagebreak[4]\par\vskip .3cm}

\newcommand{\newsubsubsection}[1]{
\addtocounter{subsubsection}{1}
\addcontentsline{toc}{subsubsection}{\protect
\numberline{\arabic{section}.\arabic{subsection}.\arabic{subsubsection}}{ #1}} \vglue .4cm
\pagebreak[3] \noindent{\it \thesubsubsection.
#1}\nopagebreak[4]\par\vskip .3cm}

\makeatletter
\newcommand{\seclabel}[1]{%
  \@bsphack
  \protected@write\@auxout{}%
     {\string\newlabel{#1}{{\thesection}{\thepage}}}
  \@esphack
  }
\newcommand{\subseclabel}[1]{%
  \@bsphack
  \protected@write\@auxout{}%
     {\string\newlabel{#1}{{\thesubsection}{\thepage}}}
  \@esphack
  }
\newcommand{\subsubseclabel}[1]{%
  \@bsphack
  \protected@write\@auxout{}%
     {\string\newlabel{#1}{{\thesubsubsection}{\thepage}}}
  \@esphack
  }
\newcommand{\tablabel}[1]{%
  \@bsphack
  \protected@write\@auxout{}%
     {\string\newlabel{#1}{{\arabic{tabnum}}{\thepage}}}
  \@esphack
  }
\makeatother

\renewcommand{\theequation}{\thesection.\arabic{equation}}

\newlength{\extraspace}
\newlength{\extraspaces}
\newcounter{dummy}
\newcommand{\bc}{\begin{center}}
\newcommand{\ec}{\end{center}}
\newcommand{\be}{\begin{equation}
\addtolength{\abovedisplayskip}{\extraspaces}
\addtolength{\belowdisplayskip}{\extraspaces}
\addtolength{\abovedisplayshortskip}{\extraspace}
\addtolength{\belowdisplayshortskip}{\extraspace}}
\newcommand{\ee}{\end{equation}}

\newcommand{\ba}{\begin{eqnarray}
\addtolength{\abovedisplayskip}{\extraspaces}
\addtolength{\belowdisplayskip}{\extraspaces}
\addtolength{\abovedisplayshortskip}{\extraspace}
\addtolength{\belowdisplayshortskip}{\extraspace}}
\newcommand{\ea}{\end{eqnarray}}

\newcommand{\ban}{\begin{eqnarray*}
\addtolength{\abovedisplayskip}{\extraspaces}
\addtolength{\belowdisplayskip}{\extraspaces}
\addtolength{\abovedisplayshortskip}{\extraspace}
\addtolength{\belowdisplayshortskip}{\extraspace}}
\newcommand{\ean}{\end{eqnarray*}}
\newcommand{\baa}{
\addtocounter{equation}{1} \setcounter{dummy}{\value{equation}}
\setcounter{equation}{0}
\renewcommand{\theequation}{\thesection.\arabic{dummy}\alph{equation}}
\begin{eqnarray}
\addtolength{\abovedisplayskip}{\extraspaces}
\addtolength{\belowdisplayskip}{\extraspaces}
\addtolength{\abovedisplayshortskip}{\extraspace}
\addtolength{\belowdisplayshortskip}{\extraspace}}
\newcommand{\eaa}{
\end{eqnarray}
\setcounter{equation}{\value{dummy}}
\renewcommand{\theequation}{\thesection.\arabic{equation}}}

\input epsf

\newcounter{fignum}
\newcounter{tabel}

\newcounter{tabnum}
\newcommand{\vev}[1]{\left\langle #1\right\rangle}

\newcommand{\half}{\frac{1}{2}}
\newcommand{\del}{\partial}
\newcommand{\delb}{\bar{\del}}
\newcommand{\eol}{\nonumber \\}

\newcommand{\cO}{{\cal O}}
\newcommand{\Ext}{{\rm Ext}}
\newcommand{\Hom}{{\rm Hom}}
\newcommand{\ch}[1]{{\bf ch}( #1 )}

\newcommand{\bt}{{\bf 10}}
\newcommand{\bfv}{{\bf 5}}
\newcommand{\bfb}{{\overline{\bf 5 \!}\,}}
\newcommand{\btb}{{\overline{\bf 10 \!}\,}}

\begin{document}

%%%%%%%%%%%%%%%%%%%%%%%%%%%%
%
% Date and preprint number
%
%%%%%%%%%%%%%%%%%%%%%%%%%%%%%

\begin{flushright}
February 2008\\
%{\tt hep-th/yymmnnn}\\
AEI-2007-174
\end{flushright}
\vspace{2cm}

\thispagestyle{empty}

%%%%%%%%%%%%%%%%%%%%%%
%
% Title & abstract
%
%%%%%%%%%%%%%%%%%%%%%%%

\begin{center}
{\Large\bf  Model Building with F-Theory
 \\[13mm] }

{\sc Ron Donagi}\\[2.5mm]
{\it Department of Mathematics, University of Pennsylvania \\
Philadelphia, PA 19104-6395, USA}\\[9mm]

{\sc Martijn Wijnholt}%
\footnote{Current address: Ludwig Maximilians Universit\"at, Theresienstrasse 37, D-80333 M\"unchen, Germany}\\[2.5mm]
{\it Max Planck Institute (Albert Einstein Institute)\\
Am M\"uhlenberg 1 \\
D-14476 Potsdam-Golm, Germany }\\
[30mm]

 {\sc Abstract}

\end{center}

Despite much recent progress in model building with $D$-branes, it
has been problematic to find a completely convincing explanation of
gauge coupling unification. We extend the class of models by
considering $F$-theory compactifications, which may incorporate
unification more naturally. We explain how to derive the charged
chiral spectrum and Yukawa couplings in $N=1$ compactifications of
$F$-theory with ${ G}$-flux. In a class of models which admit
perturbative heterotic duals, we show that the $F$-theory and
heterotic computations match.

  \vfill

\newpage

\renewcommand{\Large}{\normalsize}

\tableofcontents

\newpage

\newsection{Introduction}

String theory is an extension of quantum field theory which
incorporates quantum gravity. In the process it reformulates many
questions about field theory into questions about the geometry of
extra dimensions. Recently, ten-dimensional string backgrounds were
found that reproduce the MSSM at low energies
\cite{Bouchard:2005ag,Wijnholt:2007vn}.

However finding realizations of the MSSM is merely an intermediate
step, because we would like to answer questions that the MSSM does
not explain. Indeed there are quite a number of hard issues to
tackle. Many of these have to do with the intrinsic difficulties of
a theory of quantum gravity. Thus some of these issues may probably
be resolved by a better conceptual understanding of quantum gravity.
However as in \cite{Verlinde:2005jr} we would like to take a more
practical perspective with regards to the phenomenological
requirements which have a direct impact on particle physics. We will
assume that they can be understood in a framework where
four-dimensional gravity is effectively turned off. That is, we do
not yet want to be pushed into having to specify a complete model of
physics at the Planck scale, while there are still many issues in
particle physics that presumably can be explained without referring
to a full UV completion. Interestingly, string theory allows us to
think in such a framework and in the process provides an intuitive
geometric picture through the brane world scenario.

One of the first coincidences that one would like to address is the
issue of gauge coupling unification. The most natural scenario is
still some type of Grand Unified Theory. In particular, one would
like to have realizations of such models in type IIb string theory,
where most of the recent progress in moduli stabilization, mediation
of SUSY breaking and other issues has recently taken place. There
have in fact been attempts to construct D-brane GUT models, but
these suffer from a number of inherent difficulties, such as the
lack of a spinor representation for $SO(10)$ or the perturbative
vanishing of top quark Yukawa couplings for $SU(5)$ models. These
difficulties arise because past constructions have relied on
mutually local 7-branes. There is however a natural way to evade
these obstacles, which is by incorporating mutually non-local
7-branes. This enlarged class of models goes under the name of
$F$-theory \cite{Vafa:1996xn}. In fact, in certain limits $F$-theory
is dual to the heterotic string, which ``explains'' why $F$-theory
should be able to circumvent the no-go theorems.

It is then surprising that, despite the potentially promising
phenomenology of the $F$-theory set-up, some important issues in
$F$-theory compactifications have not been addressed. Foremost among
these, it is not currently known how to derive the spectrum of
quarks and leptons. It is the purpose of this paper to explain the
origin of charged chiral matter and to provide tools for computing
the spectrum and the couplings. Our approach is to deduce everything
from the eight-dimensional Yang-Mills-Higgs theory living on the
7-branes, and our results are therefore quite general. As expected
from type IIb string theory, we can get chiral matter spread in the
bulk of a 7-brane or localized on the intersection of 7-branes by
turning on suitable fluxes. Also as expected, the Yukawa couplings
are computed simply from the overlap of the chiral zero modes on the
7-brane. These results should be helpful in putting many
extra-dimensional phenomenological models, in which localization of
wave functions was used to explain differences in couplings, on a
firmer footing. In order to make sure that our results are correct,
we test our formulae for $F$-theory compactifications which are dual
to the heterotic string. We will see that the computations on both
sides of the duality match. Along the way we clarify several issues
in heterotic/$F$-theory duality.

In this paper we emphasize mostly conceptual issues. In section
\ref{ModelF} we discuss the main model building ingredients of
$F$-theory. In particular we explain how charged chiral matter
arises and how we can compute the spectrum and the supersymmetric
Yukawa couplings. We also discuss constructive techniques for
$F$-theory vacua with Grand Unified gauge groups. Section
\ref{Duality} is mostly devoted to $F$-theory/heterotic duality.
After reviewing the spectral cover approach to constructing
heterotic vacua, we show that the heterotic computation of massless
matter matches exactly with our $F$-theory prescription. We also
discuss the matching of superpotentials under the duality. In
section \ref{Examples} we briefly discuss some simple explicit
examples of GUT models. Finally in section \ref{GUTbreaking} we
discuss how to break the GUT group to the Standard Model gauge
group. In the appendices we collect some properties of spinors and
the Dirac operator that we will use in the text.

{\it Note for revision:} The present version
of this paper includes significant clarifications
and improvements over the original version of arXiv:0802.2969. Most of
these were designed to bring the paper more in line with the
subsequent papers \cite{Donagi:2009ra,Donagi:2011jy}.
We are also grateful to T. Watari for suggested
improvements to Appendix C. Shortly after the posting of
arXiv:0802.2969, similar results were obtained in the work \cite{BHV} by the
Harvard group.

\newpage

\newsection{Model building with $F$-theory}\seclabel{ModelF}

The purpose of this section is to discuss how to engineer gauge
groups and charged chiral matter from $F$-theory.

\newsubsection{Gauge fields}

Let us consider an $F$-theory compactification to four dimensions with
$N=1$ supersymmetry.
This consists of a Calabi-Yau fourfold $Y_4$, which is elliptically
fibered $\pi:Y_4 \to B_3$ with a section $\sigma: B_3 \to Y_4$. The
base $B_3$, or more precisely the section, is the space-time visible
to type IIB, and the complex structure of the $T^2$ fibre encodes
the dilaton and axion at each point on $B_3$:
\be
\tau = {e^{-\phi}\,i} + C_{(0)}
\ee
It is convenient to represent
the four-fold in Weierstrass form:
\be
y^2  = x^3 + f x  + g
\ee
Requiring the four-fold to be Calabi-Yau
implies that $f$ and $g$ are sections
of $K_{B_3}^{-4}$ and $K_{B_3}^{-6}$ respectively.
The complex structure of the fibre is given by
\be j(\tau) = {4 (24\,f)^3 \over \Delta}, \qquad \Delta = 4 f^3 + 27
g^2 \ee
At the discriminant locus $\{\Delta = 0\} \subset B_3$ the $T^2$ degenerates by pinching
one of its cycles. Let us label the one-cycles by $(p,q) = p\, \alpha + q \, \beta$, and suppose
we pick a local coordinate $z$ on $B_3$ such that the $(1,0)$-cycle pinches as $z \to 0$.
Then $\tau$ has a monodromy around this locus:
\be \tau \sim {1\over 2 \pi i} \log(z) \ee
This is a shift in the axion. It means that the brane at $z=0$ is a source for
one unit of $RR$-flux, and so we identify it with an ordinary D7-brane, a brane on which a
$(1,0)$ string (i.e. a fundamental string)
can end. For more general $(p,q)$ we can do an $Sl(2, {\bf Z})$ transform, and we find
that the brane is a locus where a $(p,q)$-string can end. This is called
a $(p,q)$ 7-brane.

In perturbative string theory, the worldvolume of an isolated
7-brane contains a $U(1)$ gauge field $A_\mu$, from quantising an
open string with both ends on the brane. Non-perturbatively this
gauge field is encoded in the so-called $G$-flux
\cite{Dasgupta:1999ss}. To see this, let us compactify on an extra
$S^1$ with radius $R$. This is dual to $M$-theory on $Y_4$, where
the area of the elliptic fiber is now proportional to $R^{-1}$. In
$M$-theory gauge fields arise from expanding the three-form ${\sf
C}_3$ along harmonic two-forms $\omega$, and the four-form flux of
this field is called the $G$-flux. So the same must be true in the
$F$-theory limit. Therefore on the $F$-theory side we formally
introduce a three-form field ${\sf C}_3$ and expand it along
harmonic two-forms to get the gauge fields.

However we should only expand around a subset of the harmonic
two-forms on $Y_4$, because some of the modes of the $M$-theory
compactification do not survive in the $F$-theory limit. The easiest
way to see this is by following various BPS states through the
duality. If ${\sf C}_3$ has both spatial indices on $B_3$ then it
couples to an $M2$-brane wrapped on a cycle $\alpha_2$ in $B_3$.
This gets mapped to a $D3$-brane wrapping $\alpha_2 \times S^1_R$,
which becomes a string in four dimensions as $R\to \infty$,
therefore couples to a pseudo-scalar (more precisely its dual
two-form field) but not a four-dimensional vector. Similarly if
${\sf C}_3$ has two spatial indices on the elliptic fiber, it
couples to an $M2$-brane wrapping this fiber which gets mapped to a
fundamental string with momentum along $S^1_R$. Therefore it couples
to the KK gauge field associated to $S^1_R$, and in the limit $R\to
\infty$ we just recover a component of the four-dimensional metric,
not a four dimensional vector field. Finally, membranes wrapping the
remaining cycles of $Y_4$ get mapped to $(p,q)$-strings. If they map
to open strings, the ends of such a string are electric charges on
the worldvolume of 7-branes, therefore they couple to the gauge
fields on the 7-branes. If they map to closed strings, then they
couple to some linear combination of the NS and RR two-forms with
one index on $B_3$ and thus we get a gauge field in four-dimensions
also. Thus the relevant harmonic forms constitute the lattice
\be\label{Lambdadef} \Lambda = \{\, \omega \in H^2(Y_4)\,|\, \omega
\cdot \alpha = 0 \ {\rm when } \  \alpha \in  H_2(B_3)\ {\rm or}\
\alpha = [T^2]\, \} \ee
This lattice is the coroot lattice of the four-dimensional gauge
group originating from the 7-branes.

From the above argument, we see that the three-form field actually
has a simple physical interpretation: it gives the non-perturbative
description of the NS and RR two-forms of type IIb. Recall that
these two-forms transform as a doublet under the $Sl(2, {\bf Z})$
duality group. Due to the branch cuts of the axio-dilaton on $B_3$,
they are hard to describe on the IIb space-time directly. However
when we add the elliptic fibration, we can get a simple global
description as follows. We define
\be {\sf H} = H_{RR} - \tau H_{NS} \ee
Then the three-form fluxes in type IIb can be encoded in $F$-theory
as \cite{Gukov:1999ya}
\be {\sf G} = {\sf H} \wedge dz + {\sf \bar{H}} \wedge d\bar{z} \ee
where $dz$ and $d\bar{z}$ are the normalized harmonic $(1,0)$ and
$(0,1)$ forms on an elliptic curve.

Furthermore in $F$-theory we must identify the abelian gauge field
on the 7-brane as a bound state of the closed string fields of the
IIb theory; it can not be added by hand. To see this, we note that
in $F$-theory a supersymmetric 7-brane is described as a cosmic
string solution, which is then further lifted to an elliptically
fibered Calabi-Yau metric in two dimensions higher
\cite{Greene:1989ya}. Since the brane is described as a soliton of
IIb supergravity, all the degrees of freedom we associate with it
must be made from $g_{ij}, \tau, H_{RR},H_{NS}$ and the IIb spinors.
And indeed this works in the standard way. At the center of the
cosmic string, where an $S^1 \subset T^2$ shrinks to zero size, the
Calabi-Yau geometry is similar to a Taub-NUT space and supports a
harmonic two-form $\omega$ of type $(1,1)$ which peaks when the
$S^1$ shrinks to zero. Then the collective coordinates of the
7-brane may be interpreted from the worldvolume perspective by
expanding in this harmonic form. Thus the $U(1)$ gauge field on the
7-brane is obtained by expanding ${\sf C}_3$ as
\be\label{locallift}
 {{\sf C}_3}\ =\ {\sf A} \wedge \omega .
\ee
The $G$-flux ${\sf G}_4 = d{\sf C}_3$ then describes the magnetic
flux along the 7-brane. Note that this explicitly identifies the
gauge field on a 7-brane as a collective coordinate constructed from
the RR and NS two-form fields of type IIb supergravity.

We may similarly recover the remaining fields of the $8d$
supersymmetric Yang-Mills multiplet as collective coordinates of the
7-brane. The adjoint field $\Phi$ describing deformations of the
7-brane comes from deformations of the discriminant locus, that is
from complex structure deformations $\delta g_{ij}$ of the
four-fold. Using the holomorphic $(4,0)$ form, these can also be
written as harmonic forms $\alpha^{3,1}$ of Hodge type $(3,1)$.
Expanding
\be\label{complexmodexp} \alpha^{3,1}\ =\ \Phi^{2,0} \wedge \omega
\ee
we see that the adjoint field corresponds to a section of the
canonical bundle of the surface which the 7-brane wraps. Note this
means that the $U(1)_R$ symmetry of the eight-dimensional gauge
theory is identified with the structure group of the canonical
bundle, not with the structure group of the normal bundle, as is the
case for many lower dimensional branes. Finally the spinors will be
sections of the spinor bundle of the wrapped surface tensored with
the spinor bundle associated to the canonical bundle (to account for
their $R$-charges). That is, they are sections of the gauge bundle
tensored with
\be\label{S2spinors} \Omega^{0,p}(K_{S_2}^{1/2})\otimes
(K_{S_2}^{-1/2}\oplus K_{S_2}^{1/2}) = \Omega^{0,p}(S_2) \oplus
\Omega^{2,p}({S_2}) \ee
for $p=0,1,2$, and $S_2$ is the surface that the 7-branes wrap.
Sections related by Serre duality correspond to CPT conjugates
rather than independent fields.  Clearly the unique generator of
$h^{0,0}(S_2)$ (together with its CPT conjugate in $h^{2,2}(S_2)$)
corresponds to the four-dimensional gaugino, and the generators of
$h^{2,0}(S_2)$ and $h^{0,1}(S_2)$ yield adjoint valued
four-dimensional chiral superfields from deformations of the
7-branes and continuous moduli of the line bundles on the 7-branes
respectively.

In a cosmic string background the two-form $\omega$ is not
normalizable. This implies that the number of massless $U(1)$ gauge
fields is always less than the number of singular fibers, counted
with appropriate multiplicity. In fact since a configuration of
7-branes is labelled asymtotically by its total dyonic $(p,q)$
charge, we expect that at least two non-normalizable modes will get
lifted.

The allowed $G$-fluxes are constrained by the equations of motion
\cite{Becker:1996gj,Gukov:1999ya}. There is a superpotential
coupling
\be\label{fluxW} W\ =\ {1\over 2\pi}\int \Omega^{4,0} \wedge {\sf G}
\ee
Varying with respect to the complex structure moduli, we see that
the flux should be of type $(2,2) + (4,0) + (0,4)$. If $Y_4$ is
compact, supersymmetry and the requirement of a Minkowski vacuum
also imply that $W=0$, leading to the vanishing of the $(4,0)$ and
$(0,4)$ parts. We also have to satisfy the $D$-terms. When $Y_4$ is
smooth and there are no parametrically light states coupling to
${\sf C}_3$, these are given by
\be\label{Dterm} J \wedge {\sf G}\ =\ 0 \ee
i.e. ${\sf G}$ is required to be primitive with respect to $J$,
where $J$ is the K\"ahler form on $B_3$. Equivalently we can require
that the contraction $\imath_J {\sf G} = 0$. This is a packaging of
the $D$-terms for many $U(1)$ gauge fields into one equation. Note
though that the conditions we imposed for validity are strictly
necessary. We will find that when $Y_4$ is singular, there are
corrections to this equation. Indeed we will typically be interested
in models with $4d$ non-abelian gauge fields, in which case $Y_4$ is
not smooth. If $Y_4$ is compact, there is a tadpole cancellation
condition
\be N_{D3}\ =\ {\chi(Y_4)\over 24} - {1\over 8\pi^2}\int_{Y_4} {\sf
G} \wedge {\sf G} \ee
where $N_{D3}$ is the number of $D3$ branes filling ${\bf R}^4$, not including possible
instantons which are already described by ${\sf G}$. Finally,
the $G$-flux must be properly quantized \cite{Witten:1996md}
\be\label{Gquantized}
\left[{{\sf G}\over 2\pi}\right] - {p_1(Y_4) \over 4} \ \in \ H^4(Y_4,{\bf Z})
\ee

Because ${\sf G}/2\pi$ is generally half-integer quantized
(\ref{Gquantized}), and the flux on the 7-branes is deduced by
expanding ${\sf G} = {\sf F} \wedge \omega$, we anticipate that
${\sf F}/2\pi$ is also half-integer quantized, and so does not
generally correspond to a good line bundle on the 7-brane. Indeed it
has been argued in \cite{Gukov:2001hf} that the induced flux is
half-integer quantized precisely when the tangent bundle to the
brane  does not admit a spin structure. See also \cite{Curio:1998bv}
where the shift in the quantization law of the `spectral line
bundle' (an object that we will discuss later) is related to the
shift in the quantization law of the $G$-flux by $c_2/2=p_1/4$ on
the Calabi-Yau four-fold. Thus it is useful to split up the induced
gauge field on the 7-brane into two pieces:
\be\label{FWshift}
  {\sf A}\  =\ {\sf A}_E -  \half {\sf A}_{K_S}
\ee
where ${\sf A}_{K_S}$ is the connection on the canonical bundle of
the cycle $S$ that the 7-brane is wrapped on, and ${\sf A}_E$ is the
connection for a well-defined bundle $E$ on $S$. The shift in the
quantization law of the gauge field on a brane for zero $B$-field is
known as the Freed-Witten anomaly \cite{Freed:1999vc}.

Similarly we may consider configurations with multiple branes. The
$U(1)$ gauge field associated to each brane can be decomposed into a
well-defined piece and a correction given by half the connection of
the canonical bundle of the four-cycle that the brane is wrapped on.
The Cartan generators are linear combinations of these $U(1)$'s, so
as long as all the branes are wrapped on the same cycle the shifts
cancel out when we compare $G$-fluxes with line bundles, and may be
ignored. However in an intersecting brane configuration the branes
are wrapped on different four-cycles, and the shifts do not cancel.

\begin{figure}[t]\label{vectors}
\addtocounter{tabnum}{1} \tablabel{vectors}
\begin{center}
\renewcommand{\arraystretch}{1.5}
\begin{tabular}{|c|c|}
\hline {\it number of $U(1)$'s} & {\it origin} \\ \hline\hline
\strut $h^{1,1}(Y_4) - h^{1,1}(B_3) -1$ &   7-branes and two-forms\\
\hline $h^{2,1}(B_3) $   & four-form RR-potential \\ \hline
\end{tabular}\\[5mm]
\parbox{10cm}
{\bf Table \arabic{tabnum}: \it Abelian vector multiplets in
$F$-theory compactifications.}
\renewcommand{\arraystretch}{1.0}
\end{center}
%\end{figure}
%
%
%
%\begin{figure}[t]
\addtocounter{tabnum}{1} \tablabel{moduli}
\begin{center}
\renewcommand{\arraystretch}{1.5}
\begin{tabular}{|c|c|}
\hline {\it number of moduli} & {\it origin} \\ \hline\hline \strut
$h^{2,1}(Y_4) - h^{2,1}(B_3) -1  $ &  Wilson lines  on 7-branes
\\[-2mm]
                            & and two-form periods \\
\hline $ h^{1,1}(B_3)$   & K\"ahler moduli of $B_3$ \\ \hline
$h^{3,1}(Y_4)  $    & complex structure of $B_3$          \\[-2mm]
                    & and 7-brane deformations \\ \hline
\end{tabular}\\[5mm]
\parbox{10cm}
{ \bf Table \arabic{tabnum}: \it Moduli of $F$-theory. The
axio-dilaton is usually stabilized and so not included here.}
\renewcommand{\arraystretch}{1.0}
\end{center}
\end{figure}

Besides the $U(1)$ gauge fields from the 7-branes, we get additional
$U(1)$ factors from expanding the RR four-form along harmonic
three-forms. This is summarized in table \ref{vectors}. In addition
we will get neutral chiral fields from the moduli of the
compactification. This is summarized in table \ref{moduli} (see eg.
\cite{Andreas:1997pd,Curio:1998bv,Andreas:1998zf}).

%%%%
 \begin{figure}[t]
\begin{center}
        %\resizebox{\textwidth}{!}{
            \scalebox{.2}{
               \includegraphics[width=\textwidth]{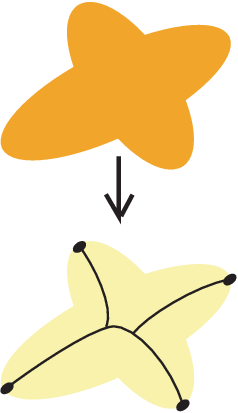}
               }
\end{center}
\vspace{-.5cm}
\begin{center}
\parbox{14cm}{\caption{ \it Multi-pronged strings in $B_3$ lift to
curves in $Y_4$, allowing for matter and gauge groups which cannot
be obtained from ordinary open strings.}\label{multipronged}}
\end{center}
 \end{figure}

Next we would like to discuss how non-abelian gauge symmetries are
encoded in $F$-theory. We expect to find non-abelian gauge bosons
from open strings stretching between 7-branes. It is well-known that
a perturbative open string has two ends and so cannot give rise to a
spinor representation or an adjoint of an exceptional group. This
gets evaded in $F$-theory because the branes are generically not
mutually local, so the dilaton can not be taken small and there is
no perturbative description. Then the missing open string states
which are needed to get a spinor or an exceptional adjoint can be
realized as {\sc BPS} junctions, i.e. open strings with multiple
ends \cite{Johansen:1996am,Gaberdiel:1997ud,DeWolfe:1998zf}. They
correspond to minimal area two-cycles $C$ in $Y_4$ which are
projected to a multi-pronged string in $B_3$. When 7-branes approach
each other, some of these minimal area cycles shrink to zero size
and create an enhanced singularity. Given a set of generators
$\vec{\omega}$ of the lattice $\Lambda$, the charges of these BPS
states associated to vanishing curves under the Cartan generators in
$\Lambda$ are given by
\be\label{vanishingweights}
\vec{w} = \int_C \vec{\omega}.
\ee
As the notation suggests, these vanishing curves will be in
one-to-one correspondence with weights of some non-abelian Lie
algebra (and also, as we will see in the next section, with weights
of matter representations).

The dictionary between singularities of the elliptic fibration and
enhanced gauge symmetries has been worked out in some detail. The
basic starting point is the Kodaira classification of singular
fibers which we have reproduced in table \ref{Kodaira}. To first
approximation, we would associate an ADE gauge group to an ADE
singularity. However if the dimension of the base is larger than one
then there can be monodromies which act as automorphisms on the
algebra and reduce the group to a non-simply laced version. We will
not review this in detail (see \cite{Bershadsky:1996nh}) but we will
quote some results on the form of the singularities in a moment.

\begin{figure}[t]
\addtocounter{tabnum}{1} \tablabel{Kodaira}
\begin{center}
\renewcommand{\arraystretch}{1.5}
\begin{tabular}{|c|c|c|c|c|}
\hline

 {\it ord(f)} & {\it ord (g)} & {\it ord($\Delta$)}   &
                               {\it fiber type} & {\it singularity type} \\
                               \hline \hline
$\geq 0$ & $\geq 0$ & $0$ & smooth & $-$\\
$0$ & $0$ & $n$ &  $I_n$  & $A_{n-1}$ \\
$\geq 1$ & $1$ & $2$& $II$ & $-$\\
$1$ & $\geq 2$ & $3$ &  $III$ & $A_1$ \\
$\geq 2$ & $2$ & $4$ &   $IV$  & $A_2$\\
$2$ & $\geq 3$ & $n+6$ &  $I_n ^*$ & $D_{n+4}$ \\
$\geq 2$ & $3$ & $n+6$ &  $I_n ^*$ & $D_{n+4}$ \\
$\geq 3$ & $4$ & $8$ &  $IV^*$ & $E_6$ \\
$3$ & $\geq 5$ & $9$ &  $III^*$  & $E_7$\\
$\geq 4$ & $5$ & $10$ &  $II^*$  & $E_8$ \\ \hline
\end{tabular}\\[5mm]
\parbox{10cm}
{ \bf Table \arabic{tabnum}: \it Kodaira classification.}
\renewcommand{\arraystretch}{1.0}
\end{center}
\end{figure}
Later we will be interested in comparison with the heterotic string.
Such a comparison can be made using heterotic/$F$-theory duality in
eight dimensions, which states that the heterotic string on $T^2$ is
equivalent to $F$-theory on an elliptically fibered $K3$ surface
with a choice of section. By fibering this duality over a complex
surface $B_2$, we get a four dimensional duality between the
heterotic string on a Calabi-Yau three-fold $Z = (T^2 \to B_2)$ and
$F$-theory on a Calabi-Yau four-fold $Y_4 = (K3 \to B_2)$ where $K3$
itself is elliptically fibered. One may match the analytic data on
both sides of the duality in a certain limit on the boundary of
moduli space, where the $K3$ surface undergoes a stable degeneration
to two $dP_9$-surfaces glued along a common elliptic curve $E$
\cite{Morrison:1996pp,Friedman:1997yq,Donagi:1997,Bershadsky:1997zs,
Curio:1998bv,Donagi:1998aa,Donagi:1998bb}.\footnote{The duality map
is expected to receive various corrections away from this limit.
Indeed, on the heterotic side $T$-dualities mix the bundle and
geometric data for finite size $T^2$, so one can not unambiguously
reconstruct a geometry.} On the heterotic side this corresponds to
compactifying on an elliptic curve with the same complex structure
as $E$ and taking the limit where the volume of the $T^2$ goes to
infinity. More details of this duality will be discussed in section
\ref{Duality} after we review the construction of bundles on the
heterotic side.

%%%%
 \begin{figure}[th]
\begin{center}
        %\resizebox{\textwidth}{!}{
            \scalebox{.3}{
               \includegraphics[width=\textwidth]{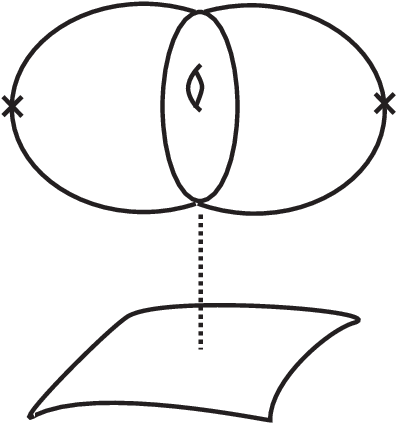}
               }
\end{center}
\vspace{-.5cm}
\begin{center}
\parbox{14cm}{\caption{ \it One can degenerate the $K3$ surface into two $DP_9$ surfaces
glued along an elliptic curve, with non-abelian gauge symmetries
localized at the crosses. In this limit one may compare with the
$E_8\times E_8$ heterotic string.}\label{Fdegeneration}}
\end{center}
 \end{figure}

In the stable degeneration limit, we may choose good coordinates on
the moduli space by unfolding a $dP_9$ surface with an $E_8$
singularity, keeping fixed a canonical divisor $E$. We consider a
degree six equation in ${\bf WP}^3_{(1,1,2,3)}$:
\ba\label{Eeightunfold}
 0&=& y^2 + x^3 + \alpha_1 xyv + \alpha_2 x^2 v^2 + \alpha_3 y v^3 + \alpha_4 xv^4 + \alpha_6 v^6 \eol
 & & +\, p_i(v,x,y)\, u^i
\ea
This is actually a $dP_8$ surface; one may obtain a $dP_9$ by
blowing up the point $u=v=0$. Intersection with the hyperplane $u=0$
yields the elliptic curve $E$ that we will keep fixed. The $p_i,\
i>0$, are polynomials of degree $6-i$ that describe the unfolding of
the $E_8$ singularity, which lives at $v = x = y = 0$. As discovered
in \cite{Candelas:1996su,Bershadsky:1996nh}, and further elucidated
in \cite{Perevalov:1997vw,Katz:1997eq,Berglund:1998ej}, the
coefficients in the $p_i$ depend on the choice of a group $H$ which
will play a role similar to the holonomy group in the heterotic
string.\footnote{ The description using a weighted projective bundle
in the following discussion does not quite apply to $E_8$
\cite{Friedman:1997yq}. However, for all other cases except $E_8$
there is indeed a description by a weighted projective bundle.}
Namely up to a change of variable they are parametrized by
Looijenga's weighted projective space
\be {\cal M}_{H} = {\bf WP}^r_{s_0,\ldots, s_r} \ee
The $s_i$ are the Dynkin indices and are listed in table
\ref{Dynkinpolys} (the non-simply laced cases will be relevant for
compactifications to less than eight dimensions). This is of course
also precisely the moduli space of flat $H$-bundles on $T^2$, which
is how it will show up on the heterotic side. For instance for $H =
SU(n)$, one has all $p_i = 0$ except
\be p_1 = v^{5-n}\left(a_0\, v^n + a_2\, x v^{n-2} + a_3\, y v^{n-3} + \ldots +
a_n\, x^{n/2}\right) \ee
(the last term being given by $yx^{(n-3)/2}$ when $n$ is odd).
Further dividing by the symmetry $ u \to \lambda^{-1} u$, the
coefficients $a_j$ take values in
\be {\cal M}_{SU(n)} = {\bf WP}^n_{1,\ldots, 1} \ee
This set of deformations preserves a singularity corresponding to an
enhanced gauge group $G$, which is the commutant\footnote{To be more
precise, in eight dimensions we always have the $18+2$ $U(1)$'s from
expansion of ${\sf C}_3$ along harmonic forms. On the heterotic side
this arises because the holonomy group $H$ on $T^2$ reduces to an
abelian group, and so the commutator of $H$ in $E_8$ contains extra
$U(1)$'s. These extra $U(1)$'s are massive for generic
compactifications below eight dimensions.} of $H$ in $E_8$. Again
consider the case $H=SU(n)$. If we turn off all the $a_i$ for $i>0$,
then the geometry is of the form
\be y^2 = x^3 + x v^4 + v^6 + u v^5 \ee
Near $x=y=v=0$, we may drop the $xv^4$ and $v^6$ terms, and we get
to leading order
\be y^2 = x^3 + v^5 \ee
which is an $E_8$ singularity. On the other hand, suppose that we
also turn on $a_5$, so that the geometry is of the form
\be y^2 = x^3 + x v^4 + v^6 + u v^5 + uxy \ee
After redefining $y\to y +\half xu, x \to 2x$ and dropping
subleading terms near $v=x=y=0$, we get
\be y^2 = x^2 + v^5 \ee
which is an $SU(5)$ singularity. Similarly in the intermediate cases
we can get $SO(10),E_6,E_7$ singularities, which are the commutators
of $SU(4),SU(3)$ and $SU(2)$ respectively.

\begin{figure}[t]
\addtocounter{tabnum}{1} \tablabel{Dynkinpolys}
\begin{center}
        %\resizebox{\textwidth}{!}{
            \scalebox{.7}{
               \includegraphics[width=\textwidth]{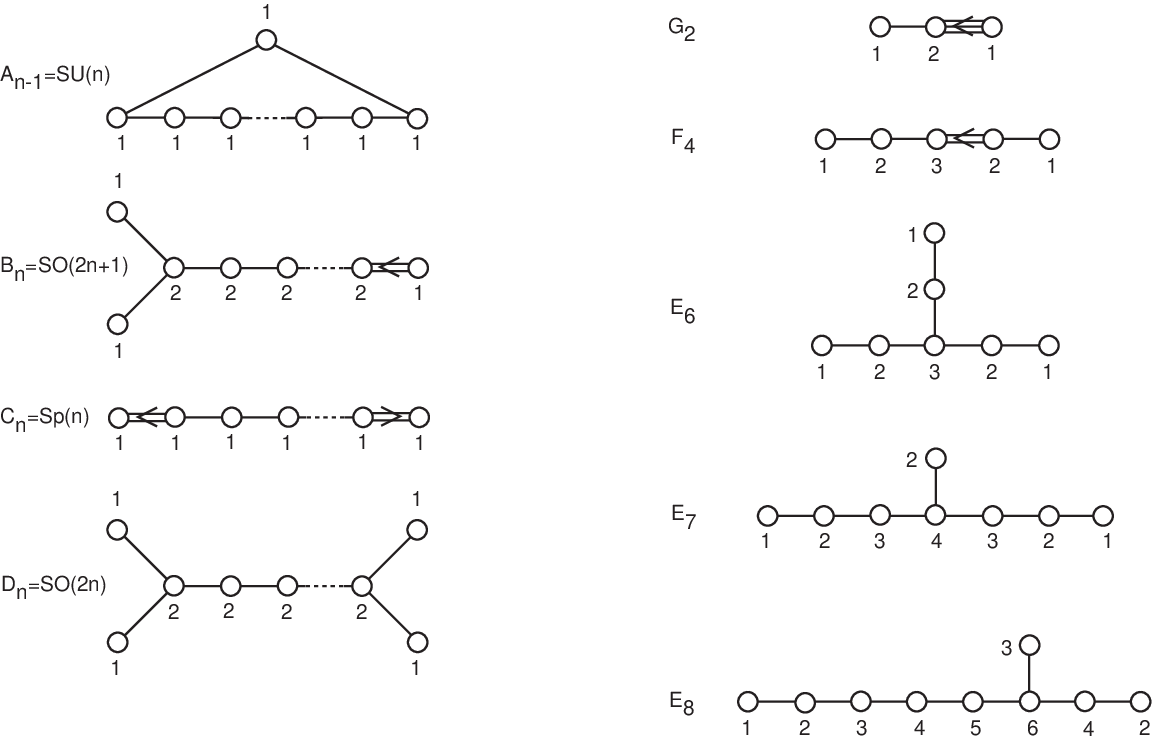}
               }
\end{center}
\vspace{-.5cm}
%$\begin{array}{cc}
%$\insertfig{Dynkinv2}{5}$
% &
%\\
%\end{array} \\[+2mm]
%$
\begin{center}
\parbox{14cm}
{ \bf Table \arabic{tabnum}: \it Dynkin diagrams and Dynkin
indices.}
\end{center}
\end{figure}

We can further fiber this degeneration over $B_2$, arriving at a
stable degeneration of $Y_4$ into two $dP_9$ fibrations $W_1,W_2$
over $B_2$, glued along an elliptically fibered Calabi-Yau
three-fold $Z$. We can write this as $Y_4 = W_1 \cup_Z W_2$, and $Z$
will eventually be identified with the heterotic dual in the limit
of large volume of the elliptic fiber. Then, $\{u,v,x,y \}$ can be
taken as sections of $\{K_{B_2}^{-6},{\cal N},K_{B_2}^{-2}\otimes
{\cal N}^2,K_{B_2}^{-3}\otimes {\cal N}^3\}$ respectively, where
${\cal N}$ is some sufficiently ample line bundle on $B_2$. The
coefficients in equation (\ref{Eeightunfold}) now become sections of
line bundles over $B_2$ as well. However requiring an enhanced gauge
group $G$ over $\sigma_{B_2}$ implies certain restrictions on these
sections. Roughly speaking, just as requiring a singularity of type
$G$ on a $dP_9$ is equivalent to expressing the coefficients of
(\ref{Eeightunfold}) in terms of a reduced number of coefficients
$a_j$, which take values ${\cal M}_H$, so is requiring a singularity
of type $G$ in $W_1$ along $\sigma_{B_2}$ equivalent to expressing
the coefficients of (\ref{Eeightunfold}) in terms of a reduced
number of sections $a_j$, such that $a_j(p)$ take values in ${\cal
M}_H$ for any point $p$\, on the base $B_2$. Now $G$ is allowed to
be non-simply laced, and $H$ is still the commutator of $G$ in
$E_8$. This is not an automatic consequence due to the issue of
monodromies mentioned previously, however it turns out to be true
anyways. The $a_j$ are sections of the line bundles ${\cal N}^{s_j}
\otimes K_{B_2}^{d_j}$. The $d_j$ turn out to be precisely the
degrees of the independent Casimirs of $H$ ($d_0$ is taken to be
zero), so the $a_j$ should be thought of as the Casimirs of the
adjoint field of the eight-dimensional gauge theory on the 7-branes.

So the upshot is that a $dP_9$ fibration $W_1$ with a fixed
hyperplane section $Z$ and a singularity of type $G$ along the zero
section is equivalent to a choice of the $a_j$, that is a choice of
section $s: B_2 \to {\cal W}_H$ of the weighted projective bundle
\be {\cal W}_{H} = {\bf WP}(\cO \oplus \bigoplus_{j>0}
K_{B_2}^{\,d_j}) \ee
where the weights are given by $a_j \to \lambda^{s_j} a_j$. The
fiber of ${\cal W}_H \to B_2$ is given by ${\cal M}_H$. However the
geometry $W_1$ specifies only part of the data of an $F$-theory
compactification, because we are also allowed to turn on Wilson
lines and fluxes along the 7-branes. That is, we can turn on periods
of ${\sf C}_3$ (which are typically trivial in a four-fold
compactification however) and $G$-fluxes. This is called the
`twisting data' for the fibration \cite{Friedman:1997yq} or `Deligne
cohomology'. It was first analyzed in the heterotic context in
\cite{Donagi:1997} and used in \cite{Curio:1998bv}. We will later
return to the issue of which $G$-fluxes one is allowed to switch on
for these geometries, after discussing how matter is engineered.

\newsubsection{Charged chiral matter from intersecting branes}
\subseclabel{IBranes}

There are basically two ways to get charged chiral matter from
7-branes. In this section, we discuss intersecting
7-branes. Some properties of spinors and Dirac operators in complex
geometry that will be heavily used in the following are collected in
the appendices. %\ref{Cspinors}.

Given two 7-branes, with gauge bundles located on them, there will
be massless matter from open string modes living on the
intersection. The idea is very simple: when branes intersect at a
small angle, we can think of them as a non-trivial field
configuration in a gauge theory with an extended gauge group. The
field content of a 7-brane is that of eight-dimensional maximally
SUSY gauge theory. The fields consist of an eight-dimensional vector
field, an adjoint valued complex scalar which is also sometimes
called a Higgs field, and a Weyl spinor with $R$-charge $-1/2$. Let
us first suppose that the eight-dimensional gauge theory has gauge
group $G$. We turn on a constant adjoint VEV for the Higgs field,
breaking $G$ to a subgroup $H_1 \times H_2$. Let's suppose that the
adjoint representation of $G$ decomposes under $H_1 \times H_2$ as
\be\label{adjdec}
R_{\rm adj}(G) = \sum_a R_a(H_1) \otimes R_a'(H_2)
\ee
Then the fermions splits into the massless gauginos of $H_1$ and
$H_2$ and massive fermions in the remaining representations
appearing in (\ref{adjdec}). Geometrically this corresponds to
separating the branes into two parallel stacks.

In order to describe intersecting branes, we need a slightly
different configuration for the Higgs field. To set the notation,
let us suppose we have a stack of branes wrapped on a complex
surface with local coordinates $z_1$ and $z_2$. The holonomy on this
surface will be denoted by $U(2)_S$. Further let us denote the local
coordinate on the canonical bundle of $S$ by $z_3$. The holonomy of
the canonical bundle is contained in $U(1)_R$, and to preserve $N=1$
supersymmetry this may be identified with $\det(U(2)_S)^{-2}$. Using
separation of variables, the gaugino of the $8d$ gauge theory can be
decomposed under $SO(3,1) \times U(2)_S \times U(1)_R$ as
\be\label{8dgaugino} \chi^\pm \psi_1^\pm \psi_2^\pm \psi_3^\pm
\otimes {\rm Ad}({\cal G}) \ee
Here we have introduced the following notations. ${\cal G}$ is the
principle bundle over $S$ with gauge group $G$, and $\psi_i$ is a
positive or negative chirality spinor for the $z_i$-plane. Recall
from (\ref{S2spinors}) that we can identify these spinors with
forms, with $\psi_i^+ \sim 1$, $\psi_i^- \sim d\bar z_i$, and $d\bar
z_3 \sim dz_1 \wedge dz_2$ using contraction with the holomorphic
$(3,0)$-form, which is preserved by the holonomy. The spinors
$\chi^\pm$ are $4d$ chiral/anti-chiral spinors. Furthermore the
above spinor representation is reducible: only the states with an
even number of pluses belong to the $8d$ gaugino.

In the non-degenerate case, we can describe a brane intersection as
a configuration breaking a group $G$ to a subgroup $H \times U(1)$.
Therefore let us turn on an adjoint VEV which depends on the complex
coordinate $z_2$ on the 7-brane, such that the gauge symmetry is
restored as $z_2\to 0$:
\be \Phi(z_1,z_2) = z_2 \, T_{U(1)}\, d\bar z_3 \wedge \ee
where $T_{U(1)}$ is the Cartan generator breaking $G$ to $H \times
U(1)$. In other words, we consider a vortex configuration for the
Higgs field $\Phi$. It is well known that there are solutions of the
Dirac equation localized on such a vortex, as we will now describe
more explicitly. (In the language of Higgs bundles, we will
construct a generator of the degree one hypercohomology of the Higgs
bundle, localized at $z_2 = 0$).

To find the massless fermions, we use separation of variables to
split the Dirac operator in a trivial six-dimensional part and a
two-dimensional part, and then solve the Dirac equation on the
$z_2$-plane with the $z_2$-dependent interaction term:
\be\label{unfoldDirac}
\begin{array}{ll}
   \delb_{\bar{z}_2}\, \psi_2^+ \psi_3^- \,
T^a\,+ \, z_2\,[T_{U(1)},T^a]\, d\bar z_3 \wedge \psi_2^-\psi_3^+ &= 0 \\[2mm]
  \delb_{\bar{z}_2}\, \psi_2^+\psi_3^+\,T^a \, -\, \bar{z}_2
\,[T_{U(1)},T^a]\,\imath_{\delb_{\bar z_3}} \psi_2^-\psi_3^- &= 0
\end{array}
\ee
as well as two more equations related by conjugation. By taking
suitable complex linear combinations, we can always take $T^a$ to be
an eigenvector of $T_{U(1)}$, and we will do so in the remainder.
The spinors $\psi_1^\pm,\chi^\pm$ are inert, and we only tensor with
them in the end to get complete wave functions. Then we get two
types of solutions. When the generator $T^a$ commutes with
$T_{U(1)}$, we get solutions which are holomorphic in $z_2$. They
transform in the adjoint of $H$ or $U(1)$ and their normalizability
depends on more global information of the cycle on which the
7-branes are wrapped. They will be further discussed in the next
subsection and we do not consider them here. Besides that, we may
also take $T^a$ to be a generator of the coset $G/(H\times U(1))$,
with
\be [T_{U(1)},T^a] =  q_a\, T^a. \ee
The precise normalization of the charge $q_a$ is not so important
here, only the sign is important. Then the second equation in
(\ref{unfoldDirac}) does not have normalizable solutions. The first
equation in (\ref{unfoldDirac}) on the other hand has solutions of
the form
\be\label{normalizablesolution}
\begin{array}{ll}
\exp(-q_a z_2\bar{z}_2)(d\bar z_3 - d\bar z_2)T^a \quad&
\exp(+q_{\bar a} z_2\bar{z}_2)(d\bar z_3 + d\bar
z_2)\overline{T^a}\eol[1.7mm] \exp(+q_a z_2\bar{z}_2)(d\bar z_3 +
d\bar z_2)T^a & \exp(-q_{\bar a} z_2\bar{z}_2)(d\bar z_3 - d\bar
z_2)\overline{T^a}
\end{array}
\ee
When $q_a=-q_{\bar a}$ is positive, the solutions in the first line
are normalizable but the solutions in the second line are not, hence
should be thrown away. Conversely when $q_a$ is negative, we throw
away the solutions in the first line.  In either case, we get zero
modes with the shape of a gaussian localized at the core of the
vortex, i.e. at the 7-brane intersection. Tensoring the normalizable
solutions in (\ref{normalizablesolution}) with constant modes of
$\chi^\pm$ and $\psi_1^\pm$, subject to the constraint that we
choose an even number of pluses in the superscripts, we therefore
find fermion zero modes of the $8d$ gaugino charged in the
off-diagonal representations appearing in (\ref{adjdec}), and
localized at $z_2=0$, i.e. localized on the 7-brane intersection.
These zero modes precisely fill out the fermionic content of a
six-dimensional hypermultiplet: multiplying by $\chi^-\psi_1^+$, we
get a chiral fermion transforming as $T^a$ and another transforming
as $\overline{T^a}$. Multiplying by $\chi^+\psi_1^-$, we get their
CPT conjugates.

Therefore to find the massless open string spectrum living on the
intersection of 7-branes, we simply have to know how the singularity
of the elliptic fibration gets enhanced over the intersection locus
of 7-branes, and decompose the corresponding adjoint representation.
This procedure gives a generalization of the usual rules for
intersecting 7-branes to the more general $F$-theory set-up
\cite{Katz:1996xe}.

Let us consider two examples, which are relevant for model building
purposes. Suppose that we have an $I_5$ singularity corresponding to
an $SU(5)$ gauge group, and we want to engineer matter by
intersecting it with a matter brane. The minimal version, which does
not introduce any extra gauge groups, is to add a locus of $I_1$
singularities. When the $I_1$ singularity intersects the $I_5$
singularity, it can get enhanced either to an $I_6$ singularity
corresponding to an $SU(6)$ gauge group, or an $I_1^*$ singularity
corresponding to an $SO(10)$ gauge group. The adjoint representation
of $SU(6)$ decomposes as
\be {\bf 35} = {\bf 24}_0 + {\bf 5}_{-1} + {{\bf
\overline{5\!}\,}}_{1} + {\bf 1}_0 \ee
Thus we get a six dimensional hypermultiplet in the fundamental of
$SU(5)$ on the intersection locus with enhanced $I_6$. For the
$I_1^*$ enhancement, we use the decomposition
\be {\bf 45} = {\bf 24}_0  + {\bf 10}_2 + {\bf
\overline{10\!}\,}_{-2} + {\bf 1}_0 \ee
Therefore on this intersection locus we get a six-dimensional
hypermultiplet in the ${\bf 10}$ of $SU(5)$.

For the second example, consider an $SO(10)$ singularity enhanced to
an $E_6$ singularity. Using the decomposition
\be {\bf 78} = {\bf 45}_0 + {\bf 1}_0 + {\bf 16}_{-3} + {{\bf
\overline{16\!}\,}}_3 \ee
we see that we get a hypermultiplet in the ${\bf 16}$ on the intersection.

In general the fermions localized along the intersection of 7-branes
further couple to the gauge bundles on the 7-branes. Again we can
use separation of variables to analyze this question. Let us turn on
a flux proportional to $T_{U(1)}$ (additional gauge fields could be
turned on, but this will suffice for our purposes). This will give
us a line bundle $E$ supported along $z_3=0$ and a bundle $F$
supported along $z_3 = \vev{\Phi(z_2)}$. Using (\ref{FWshift}), the
actual gauge fields are then associated to fake bundles $\tilde{E} =
E\otimes K_1^{-1/2}$ and $\tilde{F} = F\otimes K_2^{-1/2}$, where
$K_1$ and $K_2$ are the canonical bundles of $z_3=0$ and $z_3 =
\Phi(z_2)$ respectively.

Our solution is sharply localized in $z_2$ and $z_3$, so as long as the gauge field is smooth the equations
(\ref{unfoldDirac}) only receive a small perturbation.
The equations for $\psi_1^\pm T^a$ are
modified to
\be\label{IntersectionDolbeault} \delb_{\bar z_1} \psi_1^+ T^a +
A_{\bar z_1}[ T_{U(1)},T^a]\, d\bar z_1 \wedge \psi_1^+ \ = \ 0\ee
and its conjugate. Now we may take $A_{\bar z_1}$ to be
approximately independent of $z_2$, because the Higgs field still
creates a potential localizing the solution near $z_2=0$. Further although we can locally
pick $\psi_1^+ \sim 1$ and $\psi_1^- \sim d\bar z_1$, globally along the surface $\Sigma(z_1)$ parametrized by
$z_1$ we should consider them to live in $K_\Sigma^{1/2}$ and $K_\Sigma^{-1/2}$. We see that
$\psi_1^+$ and $\psi_1^-$ behave as sections of
\be \psi_1^+ \in S_\Sigma^+\otimes \tilde E^* \otimes \tilde F, \qquad \psi_1^-\in
S_\Sigma^- \otimes \tilde E \otimes\tilde F^*  \ee
Then by the usual argument,
the solutions to (\ref{IntersectionDolbeault}) and its conjugate are
precisely counted by the Dolbeault cohomology groups
\be\label{chiralzero} H^i(\Sigma,{\cal L}) \ee
where
\be {\cal L}= R_a(\tilde{E}) \otimes R'_a(\tilde{F})|_\Sigma \otimes
K_\Sigma^{1/2} \ee
Although ${\cal L}$ appears to contain various ill--defined bundles,
one can always combine them into something sensible. In
the case of mutually local branes, where $R'$ and $R$ correspond to the fundamental and
anti-fundamental representation respectively, we can write
\ba\label{goodbundle} {\cal L} &=& E^*\otimes F \otimes K_1^{1/2}
\otimes K_2^{-1/2} \otimes K_\Sigma^{1/2}|_\Sigma \eol
    &=& E^*\otimes F \otimes K_1|_\Sigma
\ea

Each solution of (\ref{IntersectionDolbeault}) must be dressed up
with a solution of (\ref{unfoldDirac}) in order to construct the
full wave function, which is a zero mode of the $8d$ gaugino
(\ref{8dgaugino}). Due to the constraints of normalizability and the
number of plus signs in (\ref{8dgaugino}) being even, there is in
fact a unique way to dress up each solution of
(\ref{IntersectionDolbeault}) to a zero mode of the $8d$ gaugino, so
the solutions are simply counted by (\ref{chiralzero}). This
concludes our derivation of the spectrum of fermion zero modes on
7-brane intersections.

Let us now examine the wave functions we have found in some more
detail. An important point is that the degree $i\,$ correlates with
the four-dimensional chirality. To see this, note that a generator
of the Dolbeault cohomology group $H^0(\Sigma,{\cal L})$ (i.e. a
zero mode of the form $\psi_{1}^+ \,T^a$) yields a zero mode of the
eight-dimensional Dirac operator of the form
\be\label{fullFermizero} \chi^-\psi_1^+(\psi_2^+\psi_3^- +
\psi_2^-\psi_3^+) T^a \ \sim\ \chi^-\Phi^{2,0}\, T^a + \chi^-
A^{0,1}\, T^a \ee
whereas a generator of $H^1(\Sigma,{\cal L})$  (i.e. a zero mode
$\psi_{1}^-\, T^a$) yields a zero mode of the form
\be \chi^+\psi_1^-(\psi_2^+\psi_3^- + \psi_2^-\psi_3^+)T^a \ \sim\
\chi^+\overline{ A^{0,1}}\,T^a+ \chi^+\overline{\Phi^{2,0}}\, T^a
\ee
Equivalently, it yields a chiral fermion transforming in the
conjugate representation $\overline{T^a}$. At any rate, from the
constraints of normalizability and the total number of plus signs
being even, we see that the degree of the cohomology group and the
$4d$ chirality are correlated, so that $i=0$ corresponds to a chiral
fermion and $i=1$ corresponds to an anti-chiral fermion.\footnote{
Note incidentally that if there had been normalizable solutions near
the intersection of the form $\psi_2^+\psi_3^+$ or
$\psi_2^-\psi_3^-$, we would find zero modes of the form
$\chi^+\psi_1^+\psi_2^+\psi_3^+T^a$ or
$\chi^-\psi_1^-\psi_2^-\psi_3^-T^a$ on the intersection. These
correspond to symmetries rather than deformations, and in the
present context would be interpreted as ghosts
\cite{Wijnholt:2005mp}. Fortunately we see that we cannot get them
for intersecting branes.} Further note that Serre duality maps a
zero mode with $i=0$ to a zero mode with opposite charges and $i=1$,
i.e. the opposite chirality for the four-dimensional chiral fermion.
Because we started with a single Weyl spinor in eight dimensions,
this means that generators related by Serre duality do not
correspond to independent four-dimensional fields, but to fields
related by CPT conjugation.

Now that we understand the chirality of our zero modes, we can check
when we can get a non-zero net chirality.  For this purpose we can
use the index theorem:
\ba\label{fluxgen} h^0(\Sigma, {\cal L}) - h^1(\Sigma, {\cal L}) &=&
\int_\Sigma \ch{{\cal L}}\wedge {\bf Todd}(T\Sigma)\eol
%&=& \int_\Sigma c_1({\cal G}) -\half c_1(K_\Sigma)
&=& \int_\Sigma c_1(\tilde F) - c_1(\tilde E) \ea
Thus the introduction of fluxes on the 7-branes is precisely what is
needed to make the hypermultiplets on the intersection chiral.
Further the net chirality is precisely given by the amount of flux
through the intersection. In the type IIb limit, this agrees
precisely with the known answer, which can be deduced eg. from
anomaly inflow arguments \cite{Minasian:1997mm,Cheung:1997az} or
from quantization of open strings stretching between $D$-branes
\cite{Polchinski:1998rr}. In fact our notation for the spinors was
chosen to emphasize the similarity with open string quantization.

Finally we would like to explain how this is related to integrals of
the $G$-flux. To compute the chiral spectrum we don't actually need
$E$ and $F$ separately.  All we actually need is the combination
$R(\tilde{E}) \otimes R'(\tilde{F})|_\Sigma$. To be concrete let's
discuss the case of $SU(5)$ gauge symmetry with matter in the $10$
and $\bar{5}$. Consider first the local geometry for an intersecting
$I_5$ and $I_1$ locus which gets enhanced to $I_6$. As we discussed,
there is a vanishing (anti-)holomorphic curve on top of $\Sigma$ for
each weight of the matter representation associated to it. Let's
assume that we have not turned on any holonomy for the $SU(5)$ gauge
field so that the group is unbroken (if not, the procedure we will
explain can be generalized by using all the vanishing curves instead
of just one). Then the $G$-flux close to the intersection is of the
form
\be\label{localGflux} {{\sf G}}\ \sim \ {\sf F}_1 \wedge \omega_1 +
{\sf F}_2 \wedge \omega_2 \ee
where $\omega_{1,2}$ are the two non-normalizable harmonic two-forms
associated to the overall $U(1)$'s for the $I_5$ locus and $I_1$
locus respectively. The $U(1)$'s may not appear in the low energy
theory, but a linear combination may correspond to a massive $U(1)$
and still appear in the $G$-flux as we will see in a later
subsection. Since the $U(1)$ charges of the BPS states associated to
the vanishing curves are given by $\pm 1$, we have
\be \int_C \omega_1 = +1, \qquad \int_C \omega_2 = -1 \ee
and we can integrate the $G$-flux over a vanishing curve to get
\be
\int_C {\sf G} = {\sf F}_1 - {\sf F}_2% \qquad \in \qquad {\rm Pic}(\Sigma) %H^{(1,1)}(\Sigma) \cap H^2(\Sigma,{\bf Z})
\ee
which we interpret as the curvature of $\tilde{E}^* \otimes
\tilde{F}$. By further integrating over $\Sigma$ and using
(\ref{fluxgen}), we get the net number of generations in the $\bfv$.

The other case is when the singularity is enhanced to $I_1^*$ along
the intersection of an $I_5$ and $I_1$ locus. This is not a
transversal intersection in $B_3$, however our arguments don't
depend on the form of this intersection\footnote{The local form of many collisions has been
worked out in \cite{Grassi:2000we}}, but rather on the intersection in the spectral cover picture.
Again let us assume unbroken
$SU(5)$ symmetry. Then we can pick one of the extra vanishing curves
$C$ over the intersection, and we get
\be \int_C {\sf G} = 2{\sf F}_1 - 2{\sf F}_2 \ee
The reason for the factor of two is because in the decomposition of
the $I_1^*$ singularity into individual $(p,q)$ 7-branes
\cite{DeWolfe:1998zf}, the BPS junction representing $C$ has two
ends on the $I_5$ locus and one end on each of the two extra
$I_1$-singularities, i.e. it has charge two under each of the two
$U(1)$'s. Further integrating over $\Sigma_\bt$ and using
(\ref{fluxgen}), we get the net number of generations in the $\bt$.

\newsubsection{Charged chiral matter from coincident branes}
\subseclabel{coincidentspectrum}

 There is a second way to get
charged chiral matter, by considering coincident 7-branes rather
than intersecting 7-branes. The reasoning is similar. We take a
7-brane with a non-abelian gauge symmetry wrapping a four-cycle
$B_2$. So far we assumed that only fluxes on the matter brane were
turned on, so as not to break any additional gauge symmetry on the
gauge brane. However we can also turn on generally non-abelian
holonomy on the worldvolume of the gauge brane. This corresponds on
the heterotic side to turning on a bundle on the trivial part of the
spectral cover. Eg. suppose we have an $E_6$ gauge symmetry along
$B_2$ and we turn on a $U(1)$ bundle $E$ so that the commutant in
$E_6$ is given by $SO(10) \times U(1)$. The $U(1)$ gauge field will
become massive by eating a closed string
axion.\footnote{Schematically this lifting arises as follows. From
the Chern-Simons couplings $\int C_{(4)}\wedge F \wedge F =-\int
dC_{(4)}\wedge \omega_3(A)$ on the 7-brane we deduce the existence
of a term $\int (*_8 dC_{(4)} - \omega_3(A))^2$. Then if we turn on
a line bundle on $B_2$ with first Chern class $c_1(E)$ and denote
the dual four-form as $\alpha_4$, the $U(1)$ gauge field will have a
four-dimensional coupling of the form $\int (A_\mu -\del_\mu a)^2$,
where $a$ is the RR axion obtained by expanding $C_{(4)}$ along
$\alpha_4$. This is completely analogous to a similar mechanism in
the heterotic string.} We decompose the adjoint representation of
$E_6$ under $ SO(10)\times U(1)$ as
\be
R_{\rm adj}(E_6)= {\bf 78} = \sum_a R'_{a}  \otimes R_{a} =
{\bf 45}_0 + {\bf 1}_0 + {\bf 16}_{-3} + {\bf {\overline {16\!}}\,}_3
\ee
Then chiral matter transforming in the $R_a'$ representation of
$SO(10)$ is given by zero modes of the eight-dimensional Dirac
equation. Using the spinor bundles in (\ref{S2spinors}), we get a
four-dimensional fermion for every generator of the cohomology
groups
\be\label{coincidentchiral} H^i(B_2,R_{a}(E) \otimes K_{B_2})\
\oplus\ H^i(B_2,R_a(E)) \ee
for $i=0,1,2$.\footnote{ More generally, if we also have a non-zero
VEV for $\Phi$, we should solve equations of type $\delb_A\,
\psi_1^+\psi_2^+\psi_3^- + [\Phi,a_1 \psi_1^+\psi_2^-\psi_3^+ + a_2
\psi_1^-\psi_2^+\psi_3^+]=0$. That is, we have a spectral sequence
with $E_2^{p,q} = H^p(B_2,R_a(E) \otimes K_{B_2}^{q})$, horizontal
differential $E_2^{p,q} \to E_2^{p+1,q}$ given by $\delb + A^{0,1}$,
and vertical differential $E_2^{p,0} \to E_2^{p,1}$ given by
$\Phi^{2,0}$. But when the $d_2$ differential $E_2^{0,1} \to
E_2^{2,0}$ of this spectral sequence is zero, then we have $E_2 =
E_\infty$ and we still get (\ref{coincidentchiral}). } As usual,
generators related by Serre duality are CPT conjugates, rather than
independent fields. In the above example, we would have
\ba N_\chi({\bf 16}) &=& h^0(B_2,L^{-3}\otimes K_{B_2}) +
h^1(B_2,L^{-3}) \eol N_{\chi}({\bf {\overline {16\!}}\,}) &=&
h^0(B_2,L^3\otimes K_{B_2})\ +h^1(B_2,L^{3}) \ea
where $L$ is the line bundle corresponding to the $U(1)$ gauge field
we turned on. These chiral fields clearly correspond to 7-brane
deformations and gauge field deformations respectively, and their
Serre duals are the corresponding anti-chiral fields. Generators of
$H^0(B_2,L^3)$ or $H^0(B_2,L^{-3})$ do not correspond to
deformations at all, but to symmetries. If these cohomology groups
are non-zero, the compactification has ghosts and is inconsistent.

In addition to this spectrum, we must find the massless matter
representations of $E_6$ originating from the intersection with
other 7-branes using the procedure we explained before, and add them
to the spectrum. An amusing feature is that this may effectively
increase the number of generations obtained from the intersection
with the matter brane. For instance, if we broke $E_6$ to a group
$G$ using an $SU(3)$ bundle, then from Higgsing it is clear that the
number of generations of $G$ obtained from the intersection is three
times the number of generations of $E_6$. A very similar mechanism
was used in \cite{Wijnholt:2007vn} to obtain the three generation
MSSM from a one generation model with an extended gauge group.

We can also ask about matter in real representations. We have
already seen that four-dimensional adjoint-valued chiral fields come
from $h^{0,1}(B_2)$ and $h^{2,0}(B_2)$. If we turn on a non-abelian
bundle $M$ on $\sigma_{B_2}$, we can ask for the number of bundle
moduli. This is given by the number of zero modes of the Dolbeault
operator acting on
\be {\rm Ad}(M) \otimes \Omega^{0,1}(K_{B_2}^{1/2}) \otimes
K_{B_2}^{-1/2} \ee
with the last factor accounting for the $R$-charge. Thus it is given
by the number of generators of $H^1(B_2,{\rm Ad}(M))$, in agreement
with the heterotic result \cite{Bershadsky:1997zs}.

Finally we have to give a prescription for relating line bundles on $B_2$
and $G$-flux. This is easy for coincident branes, the $G$-flux is simply
of the form ${\rm Tr}(F/2\pi) \wedge \omega$ where $\omega \in \Lambda$.

\newsubsection{Yukawa couplings}

The form of the SUSY Yukawa couplings can be deduced from the
reduction of the interaction terms in the ten-dimensional Yang-Mills
action (\ref{10Daction}). Schematically they are given by
\be \int \! d^2\theta\,d^4 x \, {\rm Tr}(\Phi_1\Phi_2\Phi_3)
\int_{S}\, {\rm Tr}(\varphi_1 \xi_2 \xi_3). \ee
where $\varphi_i, \xi_i$ denote bosonic and fermionic zero modes on
the 7-branes. Let us discuss the various special cases.

For coincident branes, the chiral fields came from generators of the
form $A^{0,1}$ or $\Phi^{2,0}$. We can compose two generators of
type $(0,1)$ and one of type $(2,0)$ to get a number:
\be\label{8DsuperW} \int_{}\, d_{abc}\, A^a \wedge A^b \wedge \Phi^c
\ee
This is just the cubic piece of the holomorphic Chern-Simons action
for 7-branes. In fact if we use the mode expansions for ${\sf C}_3$
(\ref{locallift}) and the complex structure moduli
(\ref{complexmodexp}), it can also be interpreted as a non-abelian
generalization of the
flux superpotential (\ref{fluxW}). A similar coupling for matter in
real representations was already discussed in
\cite{Bershadsky:1997zs}. We see that it holds more generally
provided the three-fold tensor product of the group indices contains
a singlet. (In the language of Higgs bundles, this corresponds to
the Yoneda pairing on the hypercohomology of the Higgs bundle).

Next let us consider intersecting 7-branes. As we discussed around
(\ref{unfoldDirac}), chiral fermions living on the intersection
$\Sigma$ must be interpreted as zero modes of the $8d$ gaugino by
dressing them with the normalizable wavefunctions for
$\psi_2^+\psi_3^-$ and $\psi_2^-\psi_3^+$. Furthermore as we also
discussed earlier, because of supersymmetry each such fermionic zero
mode of the $8d$ gaugino is paired with a bosonic zero mode of the
$8d$ fields $A^{0,1}$ and $\Phi^{2,0}$ with the same internal wave
function. Concretely we have the following dictionary:
\be\label{vertexop}
\begin{array}{rclrcl}
  \chi^-\psi_1^+\psi_2^+\psi_3^- & \to & \chi^- {\Phi^{2,0}} \qquad& \chi^-\psi_1^+\psi_2^-\psi_3^+ & \to & \chi^-
{A^{0,1}} \\[2mm]
  \chi^+\psi_1^-\psi_2^+\psi_3^- & \to & \chi^+ \overline{A^{0,1}} & \chi^+\psi_1^-\psi_2^-\psi_3^+ & \to & \chi^+\overline{{\Phi^{2,0}}}
\end{array}
\ee
Then clearly the Yukawa couplings are given by the same cubic
interaction inherited from the $8d$ gauge theory (\ref{8DsuperW}).
The same formula also applies for the overlap of zero modes which
are localized around the intersection with zero modes which are
spread over all of the 7-branes. Further, it is not hard to see that the Yukawa
couplings only depend on the Dolbeault cohomology classes of the zero
modes and not on the explicit representatives, so the Yukawa coupling corresponds to a
natural Yoneda product on the Dolbeault cohomology. By choosing
manifestly holomorphic representatives of the Dolbeault cohomology
classes of the zero modes, we can relate the integral
(\ref{8DsuperW}) to a purely holomorphic computation at
the intersection of the supports of the three
zero modes, without ever doing any integrals.
This is of course very analogous to computing a tree
level three-point function in the heterotic string, which also corresponds
to a Yoneda product on Dolbeault cohomology, and we will see
later that it is in fact dual to it.

This picture implies some intriguing results. Suppose for instance
we have an $E_6$ gauge group locally obtained from $E_8$ by turning
on two abelian adjoint fields. The locally we have two additional
unbroken $U(1)$'s, call them $U(1)_a \times U(1)_b$, which may be
broken in another patch. From the decomposition of the adjoint of
$E_8$ we get three copies of the ${\bf 27}$ with charges:
\be {\bf 27}_{(1,0)} + {\bf 27}_{(0,1)} + {\bf 27}_{(-1,-1)} \ee
Let's say that the adjoint field $\Phi_a$ vanishes for $z_1=0$, and
$\Phi_b$ vanishes for $z_2=0$, and we let $\psi_i$ be the spinor for
the $z_i$-plane as before. Then we could have chiral fields $A_{{\bf
27}_{(1,0)}} \sim \psi_1^-\psi_2^+\psi_3^+ \otimes {\bf 27}_{(1,0)}$
propagating on $z_1=0$, $A_{{\bf 27}_{(0,1)}} \sim
\psi_1^+\psi_2^-\psi_3^+ \otimes {\bf 27}_{(0,1)}$ propagating on
$z_2=0$, and $ \Phi_{{\bf 27}_{(-1,-1)}} \sim
\psi_1^+\psi_2^+\psi_3^- \otimes {\bf 27}_{(-1,-1)}$ propagating on
$z_1 = z_2$. Then from the intersection $z_1 = z_2 = 0$ we get a
coupling of the form
\be \int {\rm Tr}(A_{{\bf 27}_{(1,0)}} \wedge A_{{\bf 27}_{(0,1)}}
\wedge\Phi_{{\bf 27}_{(-1,-1)}} )\ee
Note that the indices of the forms and the gauge indices are
precisely right to allow for a non-zero contribution. More
precisely, the full wave function of a zero mode
(\ref{fullFermizero}) is a linear combination of $A^{0,1}\,T^a$ and
$\Phi^{2,0}\,T^a$, so we get a sum of terms of the above form with
the gauge indices permuted. Hence we see that we could engineer
certain interactions by playing with the matter curves, their
intersections and the localization of the chiral fields in the extra
dimensions. Similar ideas have played important roles in the
phenomenology literature on extra-dimensional models (see eg.
\cite{ArkaniHamed:1999dc} and references thereto).

As another example, suppose we have an $SO(10)$ gauge group on
$B_2$, and chiral matter on $\Sigma_{\bf 16}$ and $\Sigma_{\bf 10}$.
The Yukawa coupling for ${\bf 16} \times {\bf 16} \times {\bf 10}$
clearly gets localized on $\Sigma_{\bf 16} \cap \Sigma_{\bf 10}$.
Suppose that locally near such an intersection we can think of the
$SO(10)$ as arising from an $E_7$ gauge group which is broken by two
abelian adjoint fields $\Phi_a$ and $\Phi_b$. Then from decomposing
the adjoint of $E_7$ we get the following representations:
\be {\bf 16}_{(-2,1)} + {\bf 16}_{(0,-3)}+{\bf 10}_{(2,2)} +{\bf
1}_{(-2,4)} \ee
Again much like above we get a contribution to the Yukawa coupling
of the form
\be \int {\rm Tr}(\Phi_{{\bf 16}_{(-2,1)}} \wedge A_{{\bf
16}_{(0,3)}} \wedge A_{{\bf 10}_{(2,2)}}) \ee
from intersection points of $\Sigma_{\bf 16} \cap \Sigma_{\bf 10}$.
Again we may envisage geometric configurations that explain
hierarchies in the Yukawa couplings.

In the above examples we described the Yukawa couplings on
intersections of matter curves by configurations of two abelian
adjoint fields $\Phi_a$ and $\Phi_b$. This cannot always be done, it
is more generic actually that the Yukawa couplings are described by
a non-abelian adjoint field. Such more general configurations are
outside the scope of this paper, however in principle the resulting
Yukawa couplings still descend from the master formula
(\ref{8DsuperW}).

There are several other phenomenological scenarios that depend on
localization in the extra dimensions, and that can in principle be
implemented in $F$-theory. Localization can be helpful in
suppressing dangerous higher dimensions operators such as $\int
d^2\theta\, QQQL$ \cite{ArkaniHamed:1999dc}. It also provides
scenarios for mediation of supersymmetry breaking, such as gaugino
mediation \cite{Kaplan:1999ac,Chacko:1999mi}. For a review of some
of the possibilities of extra-dimensional models, see
\cite{Csaki:2004ay}.

Let us elaborate a bit on the remark above, that the Yukawa coupling
can be localized exactly at the intersection of matter curves. This
is a crucial difference between some of the old phenomenological
scenarios and the $F$-theory models. We have seen that when the $8d$
gauge theory approximation is valid, localized wave functions
approximately possess a gaussian tail, see eg. equation
(\ref{normalizablesolution}). In the phenomenology literature cited
above, if matter fields are localized at different positions, one
imagines getting small but non-vanishing Yukawa couplings from the
overlap of the tails, and this is an important ingredient for
explaining hierarchies in the couplings (see eg.
\cite{ArkaniHamed:1999dc}).

However as we already pointed out above, due to supersymmetry we do
not have to estimate the overlap of the wave functions, we can
calculate it exactly as a Yoneda pairing on Dolbeault cohomology.
More precisely, we can calculate superpotential couplings exactly up
to wave-function renormalization. The reason is familiar from type
IIb and heterotic models: the Yukawa couplings as well as other
$F$-term data depend only on the Dolbeault cohomology classes of the
wave functions, but not on the explicit representatives of the
Dolbeault cohomology classes. To compute the K\"ahler potential we
need the physical wave function, i.e. the harmonic representative.
But for the purpose of computing superpotential couplings we may
instead choose a different representative. In order to do the
computation exactly, it is more natural to choose a holomorphic representative,
or we could use a completely different model for the cohomology, like Cech cohomology.%
\footnote{It is helpful to consider de Rham cohomology by analogy. A
de Rham cohomology class is uniquely specified by its periods on
cycles (i.e. by the Poincar\'e duality pairing), so instead of
taking the harmonic representative we may take a delta-function
current localized on a Poincar\'e dual cycle. This representative is
clearly independent of the metric and makes the localization of the
triple product (i.e. the Yukawa couplings) manifest.}
Even better, if we are on an algebraic variety, as
is the case for all the examples we consider, then the computation
of the superpotential is equivalent to a purely algebraic
computation. In the holomorphic world, the computation is
reduced to a purely local calculation at the intersection. (In terms
of the spectral cover, we can easily see this localization using the
formulation in terms of Ext groups).

In any case, no overlap integrals need to ever be carried out to
compute superpotential terms. The gaussian tails of the harmonic
representatives have explicit anti-holomorphic dependence, and
therefore cannot play a role in the computation of superpotential
couplings. Indeed, the anti-holomorphic dependence in the tails can
simply be removed by a
complexified gauge transformation. If the algebraic supports of the
wave-functions do not intersect, then the overlap integral is simply
zero. Thus the localization of wave functions is a much sharper
effect in the presence of supersymmetry than in the old
phenomenological models with extra dimensions.

It follows that just as in type IIb or the heterotic string,
properties of the Yukawa matrices that are independent of
wave-function renormalization, such as the rank of the matrix and
texture zeroes, can be determined exactly by a purely holomorphic
(and often even algebraic)
computation. Furthermore, by the standard non-renormalization
argument, there are no perturbative corrections to the
superpotential. To see this, we define the complexified K\"ahler
moduli by
\be T_a\ =\ {m_{10}^4\over 2} \int_{D_a} J \wedge J - i \int_{D_a}
C_4 \ee
The volume of the four-cycle is measured in ten-dimensional Planck
units. The superpotential depends only holomorphically on the
K\"ahler moduli. Then because of the shift symmetry in the imaginary
part, $T_a$ can only appear in an exponential, so the superpotential
is only corrected non-perturbatively in the K\"ahler moduli. Such
non-perturbative corrections can be interpreted as $D3$-instantons.
We will give some additional discussion on such corrections in
section \ref{NPW}.

\newsubsection{D-terms}
\subseclabel{DTerms}

Up to now we have discussed purely holomorphic properties of
$F$-theory. We also have to say something about the $D$-term
constraints. In smooth compactifications, one requires that the
$G$-flux must be $J$-primitive:
\be J \wedge {\sf G} = 0 \ee
Decomposing ${\sf G} \sim {\sf F} \wedge \omega$, this condition is
reminiscent of
\be  i^*J \wedge {\sf F} =0 \ee
i.e. the standard $D$-term equation on a smooth abelian 7-brane. In
singular compactifications, or in the limit that the fiber is small
compared to the base, the non-abelian degrees of freedom are light
and we should use the non-abelian Hitchin's equations. Further we
should make sure that all the fermion zero modes that parametrize
symmetries correspond to gauginos. If not then the compactification
has ghosts and is inconsistent \cite{Wijnholt:2005mp}.

Let us specialize to the $K3$ fibrations over $B_2$ which are dual
to the heterotic string. Then the available K\"ahler forms are
\be J_{B_3} = t_1 \pi^* J_{B_2} + t_2 J_0 \ee
where $J_0$ is the Poincar\'e dual of the zero section
$\sigma_{B_2}$. Recall that $F$-theory is essentially
ten-dimensional supergravity coupled to eight-dimensional Yang-Mills
theory at certain singularities. Therefore it does not care about
$g_s$ and $\ell_s$ separately, but only about the combination given
by the ten-dimensional Planck length $l_{10}^4 = g_s \ell_s^4$ and
the Planck mass $m_{10} = 1/l_{10}$. In particular, in $F$-theory
volumes must be measured in ten-dimensional Planck units, because
these units are invariant under the $S$-duality transformations that
we must apply across branch cuts of the axio-dilaton on the IIb
space-time. For $F$-theory to be valid, both the volume of $B_2$ and
the ${\bf P}^1$-base of the $K3$ should be large in the
ten-dimensional Einstein frame.

On the other hand, the heterotic coupling is identified with the
volume of the ${\bf P}^1$. To see this, a $D3$-brane wrapped on the
base of the elliptically fibered $K3$ gets mapped to the fundamental
string of
the heterotic theory compactified on $T^2$. Its tension is %
\be T \sim l_8^{-2} (V_{{\bf P}^1})^{2/3}, \qquad T \sim l_8^{-2}
\lambda_8^{2/3}\ee
on the $F$-theory side and on the heterotic side respectively, where
$l_8$ is the effective eight-dimensional Planck length, $V_{{\bf
P}^1}$ is measured in ten-dimensional Planck units, and $\lambda_8$
is the eight-dimensional heterotic string coupling. Thus we find
that $\lambda_8 =V_{{\bf P}^1}$ up to numerical factors. As
expected, $F$-theory and the heterotic string have non-overlapping
regimes of validity. In particular it is possible that heterotic
constructions that were previously discarded correspond to valid
$F$-theory compactifications.

Further, using the conventions of \cite{Polchinski:1998rr}, in
general $F$-theory compactifications we have
\be\label{4DYMcoupling}  \alpha_{GUT} = {g^2_{GUT}\over 4\pi} =
{g_8^2 \over 4\pi V_{B_2}} = {l_{10}^4\over  V_{B_2}} \ee
\be G_N = {1\over 8\pi M_{Pl,4}^2}={2\kappa^2\over 16\pi V_{B_3}}
 = {\alpha_{GUT}^2 V_{B_2}^2\over 8
(2\pi)^2 V_{B_3}} \ee
where we defined $l_{10}^8 = g_s^2 \ell_s^8$ to be the
ten-dimensional Planck length. We also expect that
\be M_{GUT} \sim V_{B_2}^{-1/4} \ee
modulo threshold corrections. Note that in contrast to the heterotic
string, it is now in principle possible to take $M_{GUT}/M_{Pl,4}$
to zero while keeping $\alpha_{GUT}$ finite, since this only
requires the ratio $V_{B_2}^{3/2}/V_{B_3}$ to go to zero. This is
one of our main motivations for local models as we emphasized in the
introduction.

Now let's consider the available $G$-fluxes for $dP_9$ fibrations
over $B_2$. We could turn on fluxes for the Cartan generators of the
non-abelian gauge group localized on $\sigma_{B_2}$. As we discussed
in the context of coincident branes, this would partially break the
gauge symmetry. One may consider this as a mechanism for breaking
the GUT group to the Standard Model gauge group. However for testing
our formula for chiral matter in $F$-theory by comparing with the
heterotic string, we will also be interested in compactifications
where such fluxes are not turned on. Generically, the remainder of
the discriminant locus
\be \Delta'\ =\ \Delta - n[\sigma_{B_2}] \ee
is an $I_1$ locus and does not generate a massless four-dimensional
$U(1)$ vector multiplet, due to non-normalizability of the
associated local harmonic two-form\footnote{For duality with the
heterotic string, we also assume that real codimension two
singularities of the elliptic fibration are not localized on $B_2$,
as this would correspond to a non-perturbative gauge symmetry on the
heterotic side.}, so it may seem at first sight that there are no
other fluxes we could turn on. Said differently, a supersymmetric
$G$-flux must be of type $(2,2)$ and rational, and therefore it is
Poincar\'e  dual to (a rational linear combination of) algebraic
cycles in $Y_4$. At first sight, it seems that generically the only
available algebraic cycles must be either $(a)$ divisors in
$\sigma(B_3)$ or $(b)$ curves in $\sigma(B_3)$ with the elliptic
fibration on top. Both of these two types of fluxes are turned off
in the limit from $M$-theory to $F$-theory, hence do not exist in
$F$-theory, and therefore generically it would seem that there are
no supersymmetric fluxes available.

However the heterotic/$F$-theory duality map which we will discuss
in section \ref{Duality} allows us to map the question of $G$-fluxes
to an equivalent but more transparent problem. Namely, we get an
isomorphism between $G$-fluxes and certain fluxes arising in the
heterotic construction, with the explicit mapping being the
push-pull formula known as the `projected cylinder map' given in
appendix \ref{Nonprim}. On the heterotic side it will be clear that
even for generic values of the moduli, there is always an additional
rank one lattice of holomorphic quantized fluxes, generated by a
flux we will call $G_\gamma$. Moreover, for non-generic values of
the moduli there will be additional $G$-fluxes besides this generic
rank one lattice. (They were called the Noether-Lefschetz fluxes in
\cite{Donagi:2009ra}). The number of
allowed linearly independent fluxes is given by the rank of the
homology lattice $H^2(C,{\bf Z})$ of the spectral surface minus the rank of the
homology lattice of the base surface. By tuning the complex structure moduli,
these fluxes may also become of type $(1,1)$.

A rough counting of the rank
in a typical model, even requiring three generations, is obtained as follows.
We use Riemann-Hurwitz to relate the Euler
character of the spectral cover $C$ to the Euler character of the base $B_2$. By Lefschetz,
the odd cohomologies of $C$ tend to vanish, and it is evident
that the main contribution to $H^2(C,{\bf Z})$ comes from the Euler character of the branch locus of the
spectral cover. Imposing constraints of three generations and vanishing $c_1$ of the spectral line
bundle only
changes the number by order one. Taking some simple degree five spectral covers
over $B_2={\bf CP}^2$, and using the degree/genus formula to relate the degree of the
discriminant to the Euler character, we get numbers of order $10^3$ to $10^5$. So there
seems to be quite some flexibility in local models.

Following
\cite{Bousso:2000xa,Denef:2004ze}, the number of supersymmetric solutions is then estimated to be
given roughly by the volume of a ball in Euclidean space with
dimension $r$ given by the rank of this flux lattice, and a radius
$\sqrt{2}\Lambda^{1/2}$ that is set by global tadpole cancellation. It is estimated
(and to some extent verified in toy models) that for a generic such
flux, there is an $\cO(1)$ number of stable solutions. Thus the
number of solutions is roughly given by
\be \#\, {\rm solutions} \ \simeq\ {1\over (r/ 2)!}\,(2\pi  \Lambda)^{\rm r/2} \ \simeq \
\left( {4\pi e \Lambda\over r}\right)^{r/2}\ee
The tadpole cut-off $\Lambda$ is of order $\Lambda \sim \chi(Y_4)/24  \gtrsim r/24$.
It can be larger because in $r$ we only included the contribution to rank of the lattice
due to the visible sector.
Even with conservative values, this
can easily lead to the order of $10^{1000}$ possible solutions in our
local $F$-theory models. In heterotic language, this is a rough
estimate for the number of solutions to the hermitian Yang-Mills
equations on a rank five bundle on an elliptically fibered
Calabi-Yau with three generations. Clearly, the `universal' flux
lattice generated by ${\sf G}_\gamma$ is a rather special
sublattice.

At any rate, let us here focus on the special flux $G_\gamma$. By
construction it is guaranteed to be of Hodge type $(2,2)$ and
rational for general values of the moduli. However it is not a
priori clear that $G_\gamma$ is also $J$-primitive for some K\"ahler
class $J$ inside the K\"ahler cone. In appendix \ref{Nonprim} we
analyze conditions under which $G_\gamma$ is $J$-primitive.
From the analysis it follows that under certain assumptions,
$G_\gamma$ is actually $J$-primitive for all available K\"ahler
classes $J$:
\be \pi^*J_{B_2} \wedge G_\gamma = 0, \qquad J_0 \wedge G_\gamma =
0. \ee
The explanation for this is essentially that for generic moduli, the
light gauge fields sit in an unbroken non-abelian group and so there
are no Fayet-Iliopoulos parameters. Moreover, we will see in section
\ref{Duality} that the computation of the spectrum calculated on the
$F$-theory side using the gauge theory approach agrees exactly with
the heterotic calculation. In the case of $SU(5)_{GUT}$ models, the
net number of generations in the presence of the flux $\lambda\,
G_\gamma$ is given by
\be\label{numgen} N_{gen} \ = \ \lambda \int_{\Sigma_\bt} \gamma\ =\
-\lambda (6c_1-t)\cdot (c_1-t) \ee
where $\lambda \in {\bf Z} +\half$ due to quantization constraints.
We will use these fluxes to construct some simple $F$-theoretic GUT
models in section \ref{Examples}.

\newsubsection{Soliton quantization}
\subseclabel{SolitonQuantization}

Our derivations have mostly relied on the field theory approach. The
other idea we could have tried to use is to resolve the
singularities. Physically speaking, this corresponds to
compactifying to three dimensions. The $4d$ vector multiplet then
gains a pseudo-scalar corresponding to a Wilson line around the
extra $S^1$, and its expectation value parametrizes a Coulomb
branch. Going out on this $3d$ Coulomb branch corresponds
geometrically to the resolution mentioned above. When the resolved
cycles are large, the BPS $M2$-branes wrapped on them are heavy and
we can quantize them as solitons, as in \cite{Witten:1996qb}.
We can use this for
example to rederive the localized matter at loci of enhanced gauge
symmetry: assuming the background ${\sf C}_3$ is trivial, one tends
to get one resolved cycle for every pair of generators of $G/H
\times U(1)$, and quantizing a wrapped $M2$-brane on each such cycle
yields a half-hypermultiplet. In fact, we could even use this to rederive
our formulae for charged chiral matter. The basic idea is that in the
limit that the $M2$-branes are heavy, the dynamics is reduced to that of
charged particles moving in a magnetic field on the moduli space of the $M2$-brane,
here given by the matter curve.
Quantizing such $M2$-branes leads to the Dolbeault cohomology groups on the matter curve
discussed previously.
This approach is to some extent implicit in our discussion at the end of section
\ref{IBranes}.

The field theory (or Higgs bundle) approach and the soliton approach
have different ranges of validity. The soliton approach requires us
to go to $M$-theory and resolve the singularities, where we get a
quantum mechanics problem of massive BPS particles. However when the resolved cycles
are small, the Compton wave-length of the BPS particles is large
and we cannot use this approach. The field theory
approach instead smooths the singularities not by resolving, but by
incorporating the non-abelian degrees of freedom in the effective
action. Here we are writing equations on a different mathematical
object (a Higgs bundle), rather than on the $F$-theory Calabi-Yau
itself. But we can recover the singular Calabi-Yau using spectral
cover techniques, in particular the cylinder map. This approach is
justified when the Higgs field is slowly varying. In brane language,
this is the limit where the angles between intersecting branes are
small.

As long as K\"ahler and complex structure moduli are decoupled, both
approaches should give the same answer at the level of $F$-terms. The ability
to compare the different approaches gives very useful cross-checks on the results.
But for certain questions this decoupling may fail. One example is the
issue of poly-stability. Indeed, wall-crossing in the Hitchin system is
well-studied and it is known that there tend to be multiple chambers. By contrast, such chambers have not
been observed in the $M$-theory approach. The reason is that the flux is required to
satisfy $J \wedge G=0$ in the smooth $M$-theory regime, which is a closed condition.
Thus if we had a sequence of K\"ahler moduli $J_i$ accumulating to $J_\infty$,
such that $G \wedge J_i=0$ for each $i$,
then we also have $G \wedge J_\infty =0$, so there cannot be a wall in the regime of validity
of these equations. By contrast, the $D$-terms for an irreducible Higgs bundle
are equivalent to stability of the Higgs bundle. This is an open condition, which allows for walls.
For a more detailed analysis of what happens near a wall, see \cite{Donagi:2011jy}. This is not the only issue.
A more serious problem is that the blow-up modes may get lifted when $M2$-branes wrapped on
the corresponding vanishing cycles are condensed. This cannot be decided based on the geometry alone,
as it depends on the three-form configuration \cite{Donagi:2011jy}. In this case, we cannot compare with a smooth
supergravity solution.

Although it is useful to keep both pictures in mind, for our
purposes here the Higgs bundle approach has a number of advantages. It allows
for a much simpler analysis of the coupling to the background
three-form field ${\sf C}_3$. It allows us to derive interactions,
which we don't know how to do properly in the soliton approach. And the
regime of validity of the Higgs bundle approach can be attained in
$F$-theory, whereas the soliton approach always requires us to
extrapolate and go to a smooth $M$-theory compactification. This
may not be possible and when it is, information about
the $D$-terms derived from the soliton approach is not protected
under the extrapolation.

\newsubsection{Summary of local $F$-theory constructions}

To summarize, we can construct a class of local $F$-theory
compactifications for intersecting 7-branes with only the following
three ingredients:

\begin{enumerate}
  \item We take the four-fold to be a $dP_9$ fibration over a base $B_2$. For
  duals of heterotic spectral cover constructions $B_2$ can be an Enriques surface,
  a Del Pezzo surface, a Hirzebruch surface or blow-up thereof.
  \item The $dP_9$ fibration  is specified by a section $s: B_2 \to {\cal W}$ of a weighted projective
bundle ${\cal W} \to B_2$.
  This determines the ${\bf P}^1$ fibration
$B_3\to B_2$ and the discriminant locus $\Delta$, and hence the
positions
  and intersections of the 7-branes.
  \item In addition we can turn on a $G$-flux, where $[{\sf G}/2\pi]$ is of type $(2,2)$, (half-)integral,
  primitive.
  This specifies the magnetic fluxes on the 7-branes. For generic values of the moduli
  there is at least a rank one lattice of fluxes which do not further break the gauge
symmetry, and for special values of the moduli there may be
additional supersymmetric fluxes.
\end{enumerate}

\newpage

\newsection{Duality between $F$-theory and the heterotic string}
\seclabel{Duality}

\newsubsection{Spectral cover construction for heterotic bundles}

\newsubsubsection{Fourier-Mukai transform}

To specify an $N=1$ heterotic compactification in the supergravity
approximation we need a Calabi-Yau three-fold $Z$, and two bundles
$V_1,V_2$ on $Z$ with structure group in $E_8 \times E_8$ and
satisfying the Hermitian Yang-Mills equations\footnote{The first
equation is the $F$-term associated to the four-dimensional
superpotential $W = \int_{CY} \Omega^{3,0}\wedge \omega^3_{CS}(A)$,
and the second can be interpreted as a four-dimensional $D$-term $ F
\wedge J \wedge J=0$ \cite{Witten:1985bz}.}:
\be\label{HYM} F^{2,0} = F^{0,2} = 0 , \qquad  F_{i\bar{j}}\,
g^{i\bar{j}} = 0. \ee
We must further satisfy
\be\label{hettadpole} dH = {\alpha'\over 4}\,{\rm tr}(R\wedge R) -
{\alpha'\over 4}\,{\rm tr}_{E_8\times E_8}(F \wedge F) \ee
The topological obstruction to solving this equation is
\be\label{tadpolecancel} c_2(Z)\ =\ c_2(V_1) + c_2(V_2) \ee
However even if this topological condition is satisfied, clearly we
generally must turn on non-zero $H$. One may argue that a solution
may be constructed order by order in the $\alpha'$ expansion
starting with a Calabi-Yau metric and a solution of the Hermitian
Yang-Mills equations (\ref{HYM}) \cite{Witten:1985bz}. For some
special cases the existence of exact solutions may be inferred from
dualities or even proved mathematically \cite{Fu:2005sm}. In
addition one sometimes adds some five-branes wrapped on effective
curves in $Z$, even though this does not lead to a smooth
supergravity background. Such five-branes correspond to zero size
instantons and give further singular contributions to
(\ref{hettadpole}) and (\ref{tadpolecancel}).

In general constructing bundles satisfying (\ref{HYM}) is not an
easy matter. However if the three-fold admits an elliptic fibration
$\pi: Z \to B_2$ with a section $\sigma_{B_2}: B_2 \to Z$, then an
interesting class of bundles can be constructed using spectral
covers. The idea is very simple: suppose we have a stable
$SU(n)$-bundle $V$ over $Z$. First we restrict $V$ to the elliptic
fibers and learn how to describe bundles on each $T^2$, and then we
fiber this data over the base.

Restricting (\ref{HYM}) to a $T^2$-fiber suggest that the bundle
should be flat along fibers. This is actually not necessarily true
and even fails along at least a codimension one locus in the base in
all interesting examples, but let us assume we are in this situation
to build some intuition. Flat bundles on $T^2$ are classified by a
map $\pi_1(T^2) \to SU(n)$, that is by the Wilson lines around the
$T^2$. The fundamental group of $T^2$ is abelian, so these Wilson
loops commute, and by a gauge transformation the Wilson loops can be
taken to lie in the Cartan of $SU(n)$. Therefore, the restriction of
$V$ to the generic elliptic fibre splits as a sum of $n$ line
bundles of degree zero. Each line bundle is characterized by a point
on the dual $T^2$ (which parametrizes the holonomies), up to
residual symmetries which form
the Weyl group, therefore the moduli space is%
\be {\cal M}_{SU(n)} = [\Lambda_{SU(n)}^c \otimes
T^2]/W = {\bf WP}^n_{1,1,\ldots,1}\ee
where $\Lambda_{SU(n)}^c$ is the coroot lattice of $SU(n)$. The
restriction that the bundle be $SU(n)$ rather than $U(n)$ means that
the $n$ points on the dual torus are required to sum to zero under
the group law. Also, we may canonically identify the torus with its
dual. Similar results hold for bundles with other structure groups.

Fibering this data over the base, we see that an $SU(n)$ bundle can
be described by a set of $n$ points on the elliptic fibre summing to
zero, varying holomorphically over the base $B_2$, and thus sweeping
out a holomorphic surface $C$ which is an $n$-fold cover of $B_2$.
This is called the spectral cover. Intuitively this is familiar to
string theorists from T-duality of D-branes, which in this case maps
an $SU(n)$ ``9-brane'' to a ``7-brane'' by T-dualizing along the
elliptic fibre. Even though there are no physical branes in the
game, it is useful to keep this picture in mind. Moreover, as also
familiar from T-duality, to each of the $n$ points on the dual $T^2$
we can associate a line in the $n$-dimensional vector space
$H^0(V\otimes \cO(\sigma_{B_2})|_{T^2})$. These lines fit together
in a non-trivial holomorphic line bundle on $C$.

%%%%
 \begin{figure}[th]
\begin{center}
        %\resizebox{\textwidth}{!}{
            \scalebox{.3}{
               \includegraphics[width=\textwidth]{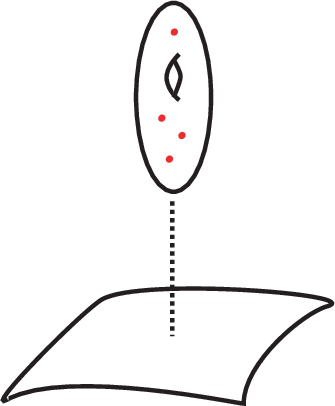}
               }
\end{center}
\vspace{-.5cm}
\begin{center}
\parbox{14cm}{\caption{ \it Part of the heterotic compactification
data consists of an elliptically
fibered Calabi-Yau, together with a set of points on each elliptic
fibre describing the Wilson lines of the ten-dimensional gauge
group.}\label{HetCY}}
\end{center}
 \end{figure}
In order to describe this more explicitly, we may proceed as follows
\cite{Friedman:1997yq}. We represent the three-fold $Z$ as a
Weierstrass equation:
\be
  y^2 = x^3 + f\, x v^4 + g v^6
\ee
Here $\{v,x,y\}$ are taken as sections of $\{\cO,
K_{B_2}^{-2},K_{B_2}^{-3}\}$ respectively, and $\{f,g\}$ are
sections of $\{K_{B_2}^{-4},K_{B_2}^{-6}\}$ respectively. Then the
data of $n$ points on each elliptic fiber summing to zero can be
encoded by writing an equation on the fiber which has exactly these
points as its solutions, and a pole of order $n$ at $v=0$ which we
identify with the intersection of the elliptic curve with the
section $\sigma_{B_2}$. Such an equation is a generic $n$th order
polynomial in $x$ and $y$:
\be\label{weightedsection} a_0\, v^n + a_2\, x v^{n-2} + a_3\, y
v^{n-3} + \ldots + a_n\, x^{n/2} = 0 \ee
(if $n$ is odd, the last term is $y x^{(n-3)/2}$). In order for this
equation to make sense globally on $B_2$, it follows that the $a_i$
must be sections of ${\cal N} \otimes K_{B_2}^i$ where ${\cal N}$ is
a line bundle on $B_2$. Since the $a_i$'s are defined only up to
multiplication on each fiber, they determine a section of the
weighted projective bundle over $B_2$
\be {\cal W}_{SU(n)} = {\bf P}(\cO \oplus K_{B_2}^{2} \oplus ...
\oplus K_{B_2}^{n}) \ee
with fiber ${\cal M}_{SU(n)}$. An analogous construction also works
for more general bundles. Thus the spectral cover $C$ is equivalent to
a section $s: B_2 \to {\cal W}_{SU(n)}$. This
description will provide an easy comparison of the analytic data
under $F$-theory/heterotic duality.

The relation between the spectral cover and the bundle $V$ on $Z$
can be put in a precise algebraic-geometric form which is known as
the Fourier-Mukai transform. The homology class of the spectral
cover $C$ can be expressed as
\be\label{coverhomology} [C] = n[B_2 ] + \pi^* [\eta] \qquad \in
\qquad H^{1,1}(Z,{\bf C}) \cap H^2(Z,{\bf Z}) ={\rm Pic}(Z) \ee
where $[\eta]$ is a class in $H_2(B_2,{\bf Z})$, and we used
Poincar\'e duality to identify the dual cohomology class. Comparing
with the description of $C$ using projective bundles over $B_2$, the
homology class of the zero set of a section agrees with
({\ref{coverhomology}) provided $c_1({\cal N}) = [\eta]$.  Further,
we need a line bundle $L$ on $C$. In order for the bundle $V$ to
have holonomy $SU(n)$ rather than $U(n)$, the line bundle $L$ is
required to satisfy
\be
c_1(V) = \pi_* c_1(L) + \half (c_1(C) - \pi^* c_1(B_2)) \equiv 0
\ee
Therefore, $c_1(L)$ is of the form
\be\label{ChernL} c_1(L) = -\half (c_1(C) - p_C^* c_1(B_2)) +
\lambda\,\gamma, \qquad \pi_*\gamma = 0 , \quad \gamma \in {\rm
Pic}(C)\ee
where $p_C$ is the natural projection $C\to B_2$, and $\lambda$ is a
(half-)integer. Generically the Picard group of $C$ is two
dimensional: one generator for the pull-back of the K\"ahler class
of $Z$, and the other generator given by $\Sigma = C \cap
\sigma_{B_2}$ which must be effective. Therefore when the complex
structure moduli take generic values, the only `traceless' classes
satisfying $\pi_*\gamma = 0$ that exist in general must be multiples
of the natural generator:
\be\label{hetgammadefinition} \gamma = n\, [\sigma_{B_2}\cdot C] -
p_C^*[ \eta - n\, c_1(B_2)] \ee
With this choice of $\gamma$ we can write $c_1(L)$ as
\be\label{Lquantization}
 c_1(L)= n(\lambda + \half) [\sigma_{B_2}\cdot C] + (\half - \lambda)p_C^*\eta
 +(n\lambda + \half)p_C^*c_1(B_2)
\ee
For $n$ odd (such as $n=5$), this is guaranteed to be an integral
class when $\lambda-\half$ is an integer.

For completeness let us briefly indicate how the bundle $V$ may be
reconstructed from this data in the case of $SU(n)$ holonomy. We
first introduce the space $\hat{Z} = Z \times_{B_2} Z$. There are
three natural divisors given by $\sigma_1 = \sigma \times_{B_2} Z$,
$\sigma_2 = Z \times_{B_2} \sigma$, and the diagonal divisor
$\Delta$ (not to be confused with the discriminant locus). We
further define $\hat{C} = C \times_{B_2} Z$ and the Poincar\'e line
bundle ${\cal P}$ on $\hat{C}$ as
\be\label{PoincareP} {\cal P} = \cO(\Delta - \sigma_1 - \sigma_2)
\otimes p_{B_2}^* K_{B_2}|_{\hat{C}} \ee
Then the bundle $V$ may be reconstructed by the Fourier-Mukai transform
\be
V = p_{Z*}(p_{\hat{C}}^*L \otimes {\cal P})
\ee
where $p_Z,p_{\hat{C}}$ denote the natural projections. With this
expression for $V$ one may compute the Chern classes of $V$
\cite{Friedman:1997yq,Curio:1998vu}. The result for $c_1(V)$ was
quoted in (\ref{ChernL}), and one finds $\pi_* c_2(V) = \eta$. For
the third Chern class one finds
\be c_3(V) = 2 \lambda\, \eta \cdot (\eta - n c_1(B_2)) \ee
The third Chern class is an important characteristic of the model as we will
review in a moment.

So far we have discussed solving the $F$-terms on $Z$, that is we
have discussed the construction of holomorphic bundles $V$ whose
curvature satisfies $F^{2,0} = F^{0,2} = 0$ and which admit a
connection which satisfies $F_{i\bar{j}}\,g^{i\bar{j}} = 0$ when
restricted to elliptic fibers. We must further show that it is
possible to solve the $D$-terms, $F_{i\bar{j}}\,g^{i\bar{j}} = 0$ on
all $Z$. As is well known, in an algebro-geometric setting one may
argue that there exists a unique solution provided the bundle $V$ is
(poly-)stable. Since Fourier-Mukai is an equivalence of categories,
the bundle $V$ is stable with respect to an appropriate K\"ahler
class when $L$ has rank $1$ and $C$ is irreducible. According to
\cite{Friedman:1997ih,Andreas:2003zb}, stability holds for
\be J = t_1\, \pi^* J_{B_2} + t_2\, J_0 \ee
where $J_0$ is the Poincar\'e dual of the section, and $t_1 >> t_2$.
That is, the base should be large compared to the $T^2$ fiber. Note
that both the fiber and base need to be large compared to the string
scale in order to keep $\alpha'$-corrections small.

Given a Calabi-Yau $Z$ and a bundle $V$ satisfying the Hermitian
Yang-Mills equations, we may deduce the low energy spectrum as
follows. We start with the ten-dimensional gaugino which transforms
in the adjoint of $E_8$, and we will concentrate on one $E_8$ factor
only. Then the four-dimensional fermions are zero modes of the Dirac
operator on $Z$ in the background with $SU(n)$ holonomy. Since $Z$
is a complex manifold, the zero modes of the Dirac operator are zero
modes of the Dolbeault operator coupled to the bundle $V$. Let us
denote the commutator of $H=SU(n)$ in $E_8$ as $G$, and decompose
the adjoint representation of $E_8$ as
\be {\bf 248} = \sum_a R_a(H) \otimes R'_a(G) \ee
Then the zero modes of the Dolbeault operator are given by the
generators of the cohomology groups
\be H^p(Z,R_a(V)) \otimes R'_a(G) \ee
Assuming $V$ stable, zero modes of grade $p=0,3$ occur only when
$R_a$ is the trivial representation. These are paired with
four-dimensional gauginos in the adjoint of $G$. Zero modes with
$p=1$ get paired with a left-handed four-dimensional chiral fermion
in the representation $R'_a(G)$, and zero modes with $p=2$ get
paired with a right-handed chiral fermion. Since supersymmetry was
preserved, we get a four-dimensional $N=1$ SUSY gauge theory with a
gauge group $G$ and matter in various representations $R'_a(G)$. The
net number of generations is given by
\be N_{\rm gen} = H^1(Z,V) - H^2(Z,V) = -\half c_3(V) \ee
assuming $H^p(Z,V)=0$ for $p=0,3$, which holds for stable bundles.
In addition, the reduction of the gravity multiplet on $Z$ gives
various other fields neutral under $G$.

 As would be expected from the brane-like interpretation for elliptically
fibered Calabi-Yaus $Z$, chiral matter is localized on the
intersection of the `$7$-branes.' This can easily be seen from the
Leray spectral sequence:
\be H^1(Z,V)\  \sim\  H^0(B_2,R^1)  \ee
where for each point $p$ on $B_2$
\be R^1_p\ =\  H^1(T^2_p,V|_{T^2_p}). \ee
Now recall that $V|_{T^2}$ splits as a sum of degree zero line
bundles $\sum_i L_i$, and $H^p(T^2,L_i)$ vanish unless $L_i$ is the
trivial line bundle. So the only contributions come from the locus
where one of the $L_i$ becomes a trivial line bundle, so that its
Wilson lines around the cycles of the $T^2$ vanish. This is
precisely the locus $\Sigma = C \cap \sigma_{B_2}$ where the
spectral cover intersects the section, and it is sometimes called
the `matter curve'. More precisely one can show that
\cite{Curio:1998vu,Diaconescu:1998kg}
\be
H^1(Z,V) = \Ext^1(i_*\cO_{B_2}, j_*L) = H^0(\Sigma,L \otimes N_{B_2}|_\Sigma).
\ee
Here $N_{B_2}$ is the normal bundle of $B_2$ in $Z$, and $N_C$ is
the normal bundle to $C$. Since $Z$ is Calabi-Yau, we have
$N_{B_2}=K_{B_2}, N_C = K_{C}$. For later comparison with
$F$-theory, it is useful to decompose the line bundle $L$ by
separating out the traceless piece:
\be L|_\Sigma = L_\gamma^\lambda \otimes N_{B_2}^{-1/2} \otimes
N_C^{1/2}|_\Sigma , \qquad \quad c_1(L_\gamma) = \gamma \ee
so that we can express the number of chiral fields as
\be\label{heteroticchiral} h^0(\Sigma,L_\gamma^\lambda|_\Sigma \otimes K_\Sigma^{1/2}). \ee

\newsubsubsection{Summary of the heterotic construction}

Suppose we are given a Calabi-Yau three-fold $Z$ with an elliptic
fibration $\pi: Z \to B_2$, and a section $\sigma_{B_2}: B_2 \to Z$.
Then an interesting class of $SU(n)$ bundles (and in fact also
bundles with more general structure groups) can be constructed with
only the following ingredients:
\begin{enumerate}
  \item An elliptically fibred threefold $\pi: Z \to B_2$, and a section $\sigma_{B_2}: B_2 \to Z$.
  \item An $n$-fold covering $p_{B_2}: C \to B_2$ with the homology
  class $[C] = n[\sigma_{B_2}] + [\pi^*\eta] \in H_4(Z,{\bf Z})$. Equivalently, we may specify
  a section of a weighted projective bundle $s: B_2 \to {\cal W}_{SU(n)}$. This
  involves specifying a line bundle ${\cal N}$ on $B_2$ with $c_1({\cal N}) = \eta$.
  The spectral cover describes the Wilson lines of the bundle $V$ along the $T^2$ fibres.
  \item A line bundle $L$ over $C$. Generically the only allowed line bundles on $C$
 have a first Chern class of the form
\be\label{ChernL} c_1(L) = -\half (c_1(C) - p^* c_1(B_2)) + \lambda
\, \gamma \ee
with $\gamma$ defined in (\ref{hetgammadefinition}). Thus for
generic complex structure moduli choosing the line bundle $L$
amounts to specifying $\lambda$, which must be chosen so that
$c_1(L)$ is integer quantized. In addition, one may turn on bundles
on $\sigma_{B_2}$ (the reducible part of the spectral cover), which
will further break the observed four-dimensional gauge symmetry.
\end{enumerate}

\newsubsection{Duality map in the stable degeneration limit}

\newsubsubsection{Matching the holomorphic data}

\subsubseclabel{MatchingHolomorphic}

The heterotic string compactified over $T^2$ is characterized by a
vector in an even self-dual lattice of signature $(18,2)$. However
we are only interested in a subset of this data, namely a bundle
with holonomy in a subgroup $H$ of $E_8$. This data may be isolated
from the other geometric data in the limit of large $T^2$. Recall
that the moduli space of stable $H$-bundles on $T^2$ is given by the
Looijenga weighted projective space
\be\label{E8moduli} {\cal M}_H = {\bf WP}^r_{s_0, \ldots, s_r} \ee
where $s_i$ are the Dynkin indices of the affine Dynkin diagram of
$H$, and $r$ is the rank of $H$. We can further fiber this data over
a base $B_2$, yielding a weighted projective bundle called ${\cal
W}_H$. An $H$ bundle over $Z$ (which is semi-stable on fibers)
determines a holomorphic section $s: B_2 \to {\cal W}_H$, or
equivalently a spectral cover $C$ which is identified with the zero
locus of the section. To reconstruct the bundle on $Z$, we also need
the twisting data. This is given by a line bundle on $C$. The line
bundle can be represented through its first Chern class. To make the
correspondence with $F$-theory clearer, the fiber of the covering $C
\to B_2$ is a discrete set of points which we denote by $f$. We can
use the Leray spectral sequence to identify
\be
H^2(C, {\bf Z}) \sim H^2(B_2, H^0(f))
\ee
This means that the flux can be represented as
\be F =  F_I \wedge \omega^I_0 \ee
where $F_I$ is a flux on $B_2$, and $\omega^I_0$ is a set of
generators of $H^0(f)$ which vary over $B_2$. It will be convenient
to take $\omega^0_0$ to be the diagonal generator which is the
pull-back of a zero-form on $B_2$, and let the remaining generators
satisfy $\pi_*\omega^I_0=0$. In particular, with
\be c_1(L) =  -\half (c_1(C) - p^* c_1(B_2))  + \lambda \gamma \ee
then the first two terms are proportional to $\omega^0_0$, and
$\gamma$ is built of the $\omega^I_0$ with $I\not = 0$.

%%%%
 \begin{figure}[th]
\begin{center}
        %\resizebox{\textwidth}{!}{
            \scalebox{.5}{
               \includegraphics[width=\textwidth]{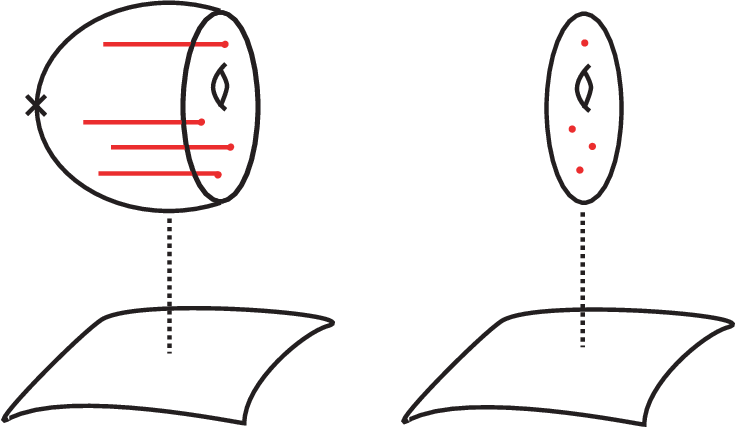}
               }
\end{center}
\vspace{-.5cm}
\begin{center}
\parbox{14cm}{\caption{ \it To every $dP_9$-surface we may associate
an elliptic curve with a set of points on it by intersecting a fixed
elliptic fiber of the $dP_9$ with the set of $-1$-curves. Conversely
by taking an elliptic curve with a set of points and thickening the
points to ${\bf P}^1$'s, we obtain a $dP_9$
surface.}\label{SpectralDP9}}
\end{center}
 \end{figure}
On the $F$-theory side we recovered the same ingredients, but with a
different interpretation. In the stable degeneration limit, the $K3$
fibration degenerates into two $dP_9$ fibrations $W_1,W_2$ over
$B_2$, glued along an elliptically fibered Calabi-Yau three-fold $Z$
which is identified with the heterotic three-fold. Concentrating on
$W_1$, we consider the unfolding of a $dP_9$ surface with an $E_8$
singularity, keeping a canonical divisor fixed. This can be
expressed by the degree six equation in ${\bf WP}_{1,1,2,3}$:
\be\label{DP9unfold}
 0= p_i(v,x,y)\, u^i
\ee
where $p_i$ is of degree $6-i$ and $p_0 = 0$ describes the
distinguished $T^2$-fiber. As we discussed in section \ref{ModelF},
requiring a section of singularities corresponding to an enhanced
gauge group $G$ implies certain restrictions on the $p_i, i>0$. The
coefficients in the $p_i$ are also determined by a choice of section
$s: B_2 \to {\cal W}_H$, up to a change of variables. In fact if $u$
appears only linearly, we can integrate out the variable $u$ without
loosing any information about the complex structure moduli
\cite{Katz:1997eq,Berglund:1998ej}. That is, the same information is
contained in the pair of equations
\be\label{criticalpoints} p_0(v,x,y)=0  ,\qquad p_1(v,x,y) = 0 \ee
This yields a collection of points on the $T^2$ at $u=0$, which we
interpret as the spectral cover. Conversely the $dP_9$ surface may
be obtained as follows: we take the elliptic curve $p_0=0$ with a
collection of points in it determined by the heterotic bundle and
encoded as an equation $p_1=0$. Then we thicken each of these points
to lines by adding the variable $u$, with each line intersecting the
$T^2$ at $u=0$ in a point\footnote{This construction generalizes for
spectral covers for groups other than $SU(n)$, and is called the
cylinder map \cite{Curio:1998bv}.}. This yields $p_0 + u p_1=0$.
Thus we have a completely explicit dictionary.

The twisting data is interpreted as turning on a ${\sf C}_3$ field
with non-zero $G$-flux. We have seen very explicitly above that
there is a canonical map which associates to each point in the fiber
$f$ of the spectral cover $C \to B_2$ a ${\bf P}^1 \subset DP_9$,
and dually with each zero-form in $H^0(f)$ a two-form in
$H^{1,1}(DP_9)$. Thus we have a natural map
\be
\begin{array}{ccc}
H^{i,j}(C) & \longrightarrow & H^{i+1,j+1}(Y_4) \\[3mm]
\updownarrow & & \uparrow \\[3mm]
H^{i,j}(B_2,H^0(f)) \quad& \longrightarrow & \quad
H^{i,j}(B_2,H^{1,1}(dP_9))
\end{array}
%H^{i,j}(C) \sim H^{i,j}(B_2,H^0(f)) \quad \to \quad H^{i,j}(B_2,H^{1,1}(DP_9)) \subset H^{i+1,j+1}(Y_4)
\ee
The map is actually somewhat ambiguous for $\omega^0$, because
$dP_9$ has two two-forms (dual to the base and the fiber) that it
could get mapped to. But as we discussed in section \ref{ModelF},
the corresponding $G$-fluxes do not exist in $F$-theory anyways even
off-shell, so requiring that we get a sensible flux eliminates the
ambiguity. In particular the `traceless' piece of the magnetic flux
on the spectral cover gets mapped unambiguously to a non-zero
$G$-flux on the $dP_9$ fibration. For more details of the mapping
between the spectral line bundle and the $G$-flux, see
\cite{Curio:1998bv} and appendix \ref{Nonprim}. In a similar vein, the continuous moduli of the
spectral line bundle, which live in $h^{0,1}(C)$, and deformations
of the spectral cover, which live in $h^{2,0}(C)$, get mapped in
$F$-theory to continuous moduli of the 7-brane gauge fields and deformations of the
7-branes, which live in $h^{1,2}(Y_4)$ and $h^{3,1}(Y_4)$
respectively as was summarized in table \ref{moduli}.

\newsubsubsection{Matching the spectrum and Yukawa couplings}

Now we would like to argue that the computation of the spectrum
agrees with heterotic computations for $F$-theory duals of spectral
cover constructions, in the stable degeneration limit. (The
arguments in this section can be understood more concisely as saying
that the spectrum and superpotential computed from the Higgs bundle
are the same as the spectrum and superpotential computed from the
spectral cover).

In $F$-theory we have a $dP_9$ fibration over a base $B_2$, with a
certain section of singularities leading to a four dimensional gauge
group $G$, but of generic type $I_1$ elsewhere. Suppose we want to
compute the number of chiral fields in the representation $R(G)$. As
we have discussed in section \ref{ModelF}, these are localized along
a curve $\Sigma$ where the singularity gets enhanced. This means
that the 7-branes wrapping $B_2$ (which we called the gauge branes)
intersect another 7-brane (which we called the matter brane) over a
curve $\Sigma \subset B_2$. On the heterotic side we must get the
corresponding gauge symmetry enhancement over the same curve $\Sigma
\subset B_2$. Thus it coincides with one of the matter curves on the
heterotic side, the locus where one of the spectral covers $C$
(analogous to our matter brane) intersects the section
$\sigma_{B_2}$ (analogous to the gauge 7-branes).

Now we need the magnetic fluxes on the 7-branes, restricted to
$\Sigma$. We consider first the matter curves where the ${\bf
10\!}\,$ of $SU(5)$, the ${\bf 16}$ of $SO(10)$, the ${\bf 27}$ of
$E_6$ and the ${\bf 56}$ of $E_7$ are localized. On the heterotic
side this corresponds to the intersection of $\sigma_{B_2}$ with the
spectral cover for the fundamental representation of the $SU(n)$
holonomy group, where $n=5,4,3,2$ respectively. The $F$-theory
fluxes were described on the heterotic side by a line bundle $L$ on
the spectral cover, with first Chern class
\be
c_1(L) =  -\half (c_1(C) - p^* c_1(B_2)) + \lambda \gamma
\ee
According to the discussion in the previous subsection, using the identification
$H^{1,1}(C) \sim H^{1,1}(B_2, H^0(f))$, the flux $\gamma$ gets mapped to
\be \gamma =  F_I \wedge \omega_0^I \quad \to\quad   G_\gamma = F_I
\wedge \omega_2^I \quad \in\quad  H^{2,2}(Y_4) \ee
where $F_I$ is a flux on $B_2$, and the index $I$ labels the
generators of $H^0(f)$. Further, the remaining piece of $c_1(L)$
gets mapped to zero. Thus it is evident that the magnetic flux for
$R_a(\tilde{E}) \otimes R'_a(\tilde{F})|_\Sigma$ extracted from the
$G$-flux, using the rules described in section \ref{ModelF}, is
exactly given by $\lambda \gamma|_\Sigma = -\lambda \eta \cdot
\Sigma$. As for the heterotic string, we denote the line bundle on
$\Sigma$ whose first Chern class is $\gamma$ as $L_\gamma$. Now
plugging into our formula for the number of zero modes
(\ref{chiralzero}), we get
\be h^i(\Sigma, L_\gamma^{\, \lambda}  \otimes K_{\Sigma}^{1/2}|_\Sigma) \ee
This is exactly the same as the answer we obtained on the heterotic
side (\ref{heteroticchiral}).

In the $SU(5)$ case it is also interesting to consider the spectral cover
$C_{\bf 10}$ for the
anti-symmetric representation of $SU(5)$. The intersection $\Sigma'
= C_{\bf 10} \cap \sigma_{B_2}$ is the locus in $B_2$ where the
gauge symmetry gets enhanced from $SU(5)$ to $SU(6)$, so this
corresponds on the $F$-theory side to the locus where $I_5$ and
$I_1$ collide transversally to create an $I_6$ singularity.

The heterotic prediction for the amount of chiral matter in the
${\bf 5}$ or $\overline{\bf 5\!}\,$ of $SU(5)$ is
\be H^p(Z,\Lambda^2V) = H^{p-1}(\Sigma',M \otimes
K_{B_2}|_{\Sigma'}) \ee
where $M$ is a rank one sheaf on $C_{\bf 10}$ obtained by
Fourier-Mukai transform from $\Lambda^2V$. The spectral cover
$C_{\bf 10}$ is singular along a codimension one locus and $M$ may
fail to be a line bundle there. This singular locus intersects
$\Sigma'$ in a finite number of points so $M|_{\Sigma'}$ may also
fail to be a line bundle. Nevertheless because the holonomy group
is $SU(5)$ rather than $U(5)$, the anti-symmetric sits in $SU(10)$ rather than
$U(10)$, and we may again decompose
\be c_1(M) =  -\half (c_1(C_{\bf 10}) - p_{C_{\bf 10}}^* c_1(B_2)) +
\lambda'\, \kappa \ee
where $\kappa$ is a class in $H^{1,1}(C_{\bf 10})$ with $p_{C_{\bf
10}\,*} \kappa =0$, and $\lambda'$ is a (half-)integer. Since $M$ is
not a line bundle, its first Chern class is somewhat ambiguous, but
with the appropriate definition this formula should be satisfied.
The $G$-flux constructed from $\lambda'\, \kappa$ should be the same
as the $G$-flux constructed from the class $\lambda \, \gamma$ on
the spectral cover associated to the fundamental representation.
Thus the difference between the 7-brane fluxes on the $F$-theory
side should be given by $\lambda' \,\kappa|_{\Sigma'}$. Following
our previous arguments then, the cohomology groups on both sides of
the duality simplify to
\be H^{p-1}(\Sigma',L_\kappa^{\lambda'}\otimes K^{1/2}_{\Sigma'})
\ee
for $p=1,2$, where $L_\kappa$ satisfies $c_1(L_\kappa) =
\kappa|_{\Sigma'}$.

 We can also check that the chiral spectrum
from coincident 7-branes agrees with the chiral spectrum computed on
the heterotic side. The Freed-Witten shift can be ignored in this
case because the branes are wrapped on the same four-cycle. On the
heterotic side we have a reducible spectral cover consisting of
multiple copies of $\sigma_{B_2}$, together with the bundle $R_a(E)$
on it. The sheaf $ {\sigma_{B_2}}_*R_a(E)$ is the Fourier-Mukai
transform\footnote{This differs slightly from some of the literature
because we included a factor of $K_{B_2}$ in our Poincar\'e sheaf
${\cal P}$ (\ref{PoincareP}).}
of the bundle $V=\pi^*R_a(E)$ on $Z$,
so the heterotic answer in this case is
\ba
 H^p(Z,\pi^*R_a(E)) &=& \Ext^p(\cO_Z, \pi^*R_a(E)) \eol[2mm]
 &=& \Ext^p({\sigma_{B_2}}_* \cO_{B_2}, {\sigma_{B_2}}_*R_a(E))
 \eol[2mm]
 &\sim&
H^p(B_2,R_a(E)) \oplus H^{p-1}(B_2,R_a(E) \otimes K_{B_2}) \ea
Here we used the fact that the Fourier-Mukai transform preserves the
Ext groups. Again this agrees with what we obtained in $F$-theory.

Finally, me may check that the Yukawa
couplings computed on both sides must agree. After Fourier-Mukai transform,
the Yukawa couplings on the heterotic side take the same form as (\ref{8DsuperW}):
\be
\int_{B_2} d_{abc}\, A^a \wedge A^b \wedge \Phi^c
\ee
Here $\Phi$ takes values in $K_{B_2}$ on both sides of the duality.
Further, the procedure we have given for computing the wave
functions of $A^{0,1}$ and $\Phi^{2,0}$ on $B_2$ only used $B_2$
itself, the data of where on $B_2$ gauge symmetry gets enhanced
(i.e. the matter curves), and the fluxes on the matter curves. Thus
we manifestly end up with the same wave functions on $B_2$, and the
Yukawa couplings must agree as well.

\newsubsection{Duality of the classical superpotential}

\subseclabel{SuperpotentialDuality}

We would like to briefly discuss the behaviour of the flux
superpotential under $F$-theory/heterotic duality. Recall that on
the $F$-theory side we had
\be
W_{\rm flux} = {1\over 2\pi} \int \Omega^{4,0} \wedge {\sf G}
\ee
and further, on large and smooth four-folds we had a set of
$D$-terms
\be
 J \wedge {\sf G} = 0
 \ee
Generically this stabilizes all but one of the moduli.  The
solutions of the $F$-terms equations are given by integer quantized
$(2,2)$-classes. These are somewhat rare, and one typically has to
tune all the complex structure moduli to find them (i.e. solve a
Noether-Lefschetz problem). All but one of the K\"ahler moduli can
in principle be stabilized with the $D$-term potential. Note however
that this does not provide a potential for the volume modulus,
because if $J \wedge {\sf G} = 0$, then $x J \wedge {\sf G}=0$ for
all $x$. However these equations will receive corrections, which we
may try to use to stabilize also this last modulus.

Now consider $F$-theory on $K3$. First note that it is not possible
to turn on any internal fluxes, since $K3$ has only even dimensional
harmonic forms and flux proportional the volume form of $K3$ is
forbidden. Moreover a $G$-flux that lives purely in eight-dimensions
does not exist in $F$-theory. So all $G$-fluxes can be interpreted
as gauge field fluxes for the $18+2$ gauge fields in eight
dimensions\footnote{This includes the possibility of fluxes for the
NS and RR three-forms.}. Similarly, the $\Omega^{(4,0)}$ form must
be decomposed into the internal $(2,0)$ form of the $K3$ and a
$(2,0)$ form in eight dimensions. The flux superpotential reduces to
the natural pairing of this $(2,0)$ form and the $18+2$ abelian
fluxes. This data can be further fibered over a base $B_2$.

Analogously, in the heterotic string in ten dimensions we have the
superpotential:
\be
W = \int \Omega^{3,0} \wedge (H + i\,dJ) + \int \Omega^{3,0} \wedge \omega_3(A),
\ee
where $dJ\not = 0$ allows for the possibility of torsion
\cite{Becker:2003gq,Lopes Cardoso:2003af}, and $\omega_3(A)$ is the
holomorphic Chern-Simons form
\be
\omega_3(A) = \epsilon^{\bar{i}\bar{j}\bar{k}}
(A_{\bar{i}} \del_{\bar{j}} A_{\bar{k}} + {2\over 3} A_{\bar{i}} A_{\bar{j}} A_{\bar{k}})
\ee
It simply reduces to $\epsilon^{\bar{i}\bar{j}\bar{k}}A_{\bar{i}}
F_{\bar{j}\bar{k}}$ for abelian gauge fields. After compactification
on $T^2$, we can turn on certain fluxes corresponding to
$\del_\lambda g_{\mu\nu}$ or $\del_\lambda B_{\mu\nu}$ with two
indices on the $T^2$, or $F_{\mu\nu}$ with one index on the $T^2$.
However this corresponds to varying the Narain moduli over the
eight-dimensional space-time and the same data exists on the
$F$-theory side also as we have discussed, but is not interpreted as
$G$-flux. Flux of type $F_{\mu\nu}$ with two indices on the $T^2$
might be allowed a priori, but is of type $(1,1)$ and the
superpotential doesn't depend on it. The remaining fluxes can be
interpreted as fluxes for the $18+2$ gauge fields in eight
dimensions coming from the Cartan of $E_8 \times E_8$ and from modes
of the metric and $B$-field on $T^2$. The $(3,0)$ form reduces to a
$(2,0)$ form in eight dimensions, and from reduction of the
ten-dimensional superpotential we get a pairing between this $(2,0)$
form and the $18+2$ fluxes. The data can be further fibered over
$B_2$. Thus the flux superpotentials match naturally under duality.

This brings up the following puzzle: in $F$-theory, we can stabilize
all the complex structure moduli at tree level. On the heterotic
side, stabilizing the complex structure and vector bundle moduli has
been problematic, and one usually invokes worldsheet instanton
effects. Clearly, if the $F$-theory arguments are correct, we must
be able to stabilize these moduli at tree level also, because as we
just saw the superpotentials on both sides are isomorphic. The fact
that $H$ or $dJ \not = 0$ stabilizes some moduli has already
received some attention. Here we would like to focus on the
holomorphic Chern-Simons superpotential.

The moduli of the heterotic bundle $V$ translate into continuous
moduli of the line bundle on the spectral cover, which are absent in
generic models, and deformations of the spectral cover. In
constructions in the heterotic literature, the spectral cover moduli
are flat directions for the holomorphic Chern-Simons superpotential,
and they are not stabilized perturbatively. Instead one invokes
worldsheet instanton effects, as their one-loop pre-factor depends
on vector bundle moduli (as well as complex structure moduli), but
they have the disadvantage of being exponentially suppressed at
large volume and hard to compute. Clearly we would prefer to
stabilize these moduli perturbatively, and duality with $F$-theory
predicts that this must be possible. So why exactly are the spectral
moduli unobstructed? Let us examine this issue in $F$-theory/7-brane
language.

Varying the flux superpotential with respect to vector bundle
moduli, we find that
\be 0=DW={1\over 2\pi}\int_{S_2}  \Phi^{2,0} \wedge  {\sf F} +
\ldots \ee
where we used $\delta \Omega \sim \Phi \wedge \omega$ and ${\sf G}
\sim {\sf F} \wedge \omega$. Generally, $ {\sf F}$ has both $(0,2)$
components and $(1,1)$ components. However, the critical critical
locus of the superpotential is the locus in moduli space where $
{\sf F}$ is purely of type $(1,1)$. Therefore, moduli of the Higgs
field (which correspond to vector bundle moduli in the heterotic
language) in fact {\it can} be stabilized if we choose fluxes that
have a $(0,2)$-component for generic values of the moduli. The
number of equations is the same as the number of variables, so the
generic solution is a completely rigid bundle. Actually, if we
change the complex structure of the Calabi-Yau then we generically
destroy this solution, so we should really think of it as
stabilizing combinations of complex structure and vector bundle
moduli. (The same is true for the one-loop pre-factor of an
instanton contribution to the superpotential. The pre-factor depends
on both complex structure and vector bundle moduli, and therefore
generates a potential for combinations of them).

Indeed, upon reflection it is obvious that the spectral line bundles
that are normally used in heterotic constructions are very special:
they are actually obtained by restricting line bundles on the
Calabi-Yau three-fold to the spectral cover. They are therefore of
type $(1,1)$ for any values of the spectral cover moduli, and so do
not induce a potential for these moduli. This explains why it is
frequently stated in the heterotic literature that one needs
worldsheet instantons to stabilize such moduli.

However it is now clear that by choosing more general spectral line
bundles which are not generically holomorphic, we can induce a
potential for these moduli at tree level. Moreover, we see that we
get the same type of exponential growth in the number of solutions
as on the $F$-theory side, simply by compactifying the $10d$ $E_8$
Yang-Mills theory. This should not be surprising, because the second
betti number $b_2(C)$ for the two $E_8$ spectral covers tends to
provide the main contribution to $b_4(Y_4)$ under
heterotic/$F$-theory duality, but it does not appear to have been
previously appreciated. It implies that the $F$-theory landscape has
so far been missed on the heterotic side, and moreover the
exponential growth in the number of solutions can be seen directly
in the visible sector, predicting an exponentially large number of
solutions with the spectrum of the MSSM. We gave a rough estimate
for the number of solutions in section \ref{DTerms}. For more
discussion of these new heterotic vacua, including some toy
constructions, see \cite{Donagi:2009ra}.

The heterotic string also has a set of $D$-terms $F \wedge J \wedge
J=0$ in ten dimensions. In eight dimensions we get a term $J\wedge F
=0$ where $F$ are $E_8$ fluxes. Compatibility with $T$-duality
suggests there should be such a term for all $18+2$ fluxes.

\newsubsection{Non-perturbative corrections to the superpotential}
\subseclabel{NPW}

As we reviewed earlier, the classical superpotential in $F$-theory
does not receive corrections to any order in large volume
perturbation theory, however it may receive non-perturbative
corrections due to $D$-instantons. Let us discuss the possibilities
and their heterotic analogues. Our discussion is similar to
\cite{Witten:1996bn}.

The easiest way to get the correspondence is to follow BPS states
across a chain of dualities in seven dimensions:
\begin{center}
\begin{tabular}{ccc}
\label{HetFDualitysequence} {\it F}-theory/$K3 \times S^1_R$ \quad &
$=$ \quad {\it M}-theory/$K3$ \quad &= \quad { Heterotic}/$T^3$
\end{tabular}
\end{center}
which we will further compactify to four dimensions by fibering over
a base $B_2$ and taking $R\to \infty$. Let us first consider the
equivalence on the left. We could get non-trivial instanton effects
from $M2$-branes wrapping a three-cycle which includes one of the
circles of the $T^2$. These would correspond to instantons made of
$(p,q)$ strings on the $F$-theory side. Such three-cycles are rare
however. If the three-cycle lives in $B_3$ completely then the
instanton will have infinite action as $R\to \infty$. Therefore we
can concentrate on the $M5$-branes. An $M5$-brane wrapped on $K3$
gets mapped to a $D3$-brane wrapping the ${\bf P}^1$ base of the
$K3$, and we can wrap it on an additional curve $\alpha_2 \subset
B_2$ to get an instanton. The other option is to wrap the $M5$-brane
on the $T^2$ fiber of the $K3$ and some additional four-cycle
$\alpha_4$ which does not contain the ${\bf P}^1$-base of the $K3$.
This gets mapped to a $D3$-brane instanton which wraps the
four-cycle $\alpha_4$. If this coincides with the location of gauge
7-branes and if the bundles on the four-cycle also agree, such
instantons may be interpreted as gauge theory instantons.

Now we consider the equivalence on the right. The $M2$-brane
instantons, if they exist, get mapped to instanton versions of
Dabholkar-Harvey states, whose worldline wraps a (possibly trivial)
one-cycle in $B_2$. This includes ordinary worldsheet instanton
effects obtained from wrapping a string worldsheet on a geometric
curve. An $M5$-brane wrapped on $K3$ gets mapped to the heterotic
fundamental string. An $M5$-brane wrapping any other cycle of the
$K3$ gets mapped to an $NS5$-brane wrapping some cycle in $T^3$.
Therefore, the $D3$-instantons wrapping $\alpha_2 \times {\bf P}^1$
get mapped to worldsheet instantons wrapping $\alpha_2$ in the
heterotic string, and the $D3$-instanton wrapping $\alpha_4$ get
mapped to space-time instanton effects in the heterotic string.

On the heterotic side, we could also consider worldsheet instantons
wrapping the $T^2$ fiber. The volume of this $T^2$ is a K\"ahler
modulus which gets mapped to a `transcendental' complex structure
parameter on the $F$-theory side, and the limit of infinite volume
for $T^2$ corresponds to the stable degeneration limit on the
$F$-theory side where the $K3$ degenerates into two $dP_9$ surfaces
(and hence becomes algebraic). The associated worldsheet instanton
corrections on the heterotic side correspond to classical
contributions to the superpotential on the $F$-theory side which
vanish in the stable degeneration limit.

The rules for $D$-instanton contributions to the superpotential were
originally given in
\cite{Becker:1995kb,Harvey:1999as,Witten:1999eg}. The prescription
is simply to calculate the partition function of the instanton
obtained by integrating over all the collective coordinates. In type
IIa or type IIb backgrounds, where the instanton typically
intersects wrapped branes, this includes localized degrees of
freedom on the intersection of the instanton with the background
branes. The collective coordinates, including the localized ones,
can be described using the concept of `Ganor strings'
\cite{Ganor:1996pe}. This calculus has recently been clarified in
\cite{Florea:2006si,Ibanez:2007rs,Blumenhagen:2006xt}.

For finite string coupling, the presence of branch-cuts of the
axio-dilaton seems to pose problems for the $D$-instanton approach.
The gauge field on the $D3$-instanton is not invariant
under $Sl(2,{\bf Z})$ monodromies, so the sum over electric fluxes
does not make sense. Further, there are no weakly coupled
Ganor strings that we could quantize, so the partition function of $37$
strings ceases to make sense for finite string coupling as well.

These issues have been resolved recently in
\cite{Blumenhagen:2010ja,BDW}. Let us briefly summarize some of the results.
The instanton contribution
consists of an exponential part and a pre-factor:
\be\label{NPsuperW}
\Delta W \ = \ f(m)\, e^{-T}
\ee
where $T$ denotes K\"ahler moduli and $m$ denotes complex structure moduli.
The crucial point is that the pre-factor does not depend on K\"ahler moduli, due to shift symmetries.
Therefore by varying K\"ahler moduli we can extrapolate to a regime where we
get a well-defined computation without changing the pre-factor. There are currently two such limits known.
Namely we could try to extrapolate the calculation either to $11d$ supergravity, or to
a heterotic computation (even in certain cases when there is no heterotic dual).
One may also contemplate extrapolating to perturbative IIb, but by contrast to
the above limits, the IIb limit involves changing the complex structure. This changes the
pre-factor, so the perturbative IIb results will be modified.

In the $11d$ supergravity approach we get an $M5$-instanton, and assuming the instanton is
smooth then by the general prescription of
\cite{Becker:1995kb,Harvey:1999as,Witten:1999eg} the instanton contribution to the superpotential
is given by the partition function. The $M5$ worldvolume theory contains scalars $\phi$, fermions $\psi$ and a chiral
two-form $B^+$, so the partition function is of the form
\be
Z_{M5} \ = \ Z_{\phi}^M\, Z_{\psi}^M\, Z_{B^+}^M
\ee
and the contribution to the superpotential is obtained by factoring out four universal bosonic and two universal
fermionic zero modes. (Here we ignored the fact that $Z_{\psi}^M$ and $Z_{B^+}^M$ are partition vectors, so it would be more
accurate to write $Z_{M5}  =  Z_{\phi}^M\, \vev{Z_{\psi}^M, Z_{B^+}^M}$ for a suitable inner product).
By contrast, the $D3$ partition function is of the form
\be
Z_{D3} \ = \ Z_\phi\, Z_\psi\, Z_F\, Z_{\lambda_{37}}
\ee
where $\phi,\psi,F$ denote the $D3$ worldvolume fields, and $\lambda_{37}$ denotes fermions obtained from
quantizing the 37 Ganor strings. Although this seems to look different, the expressions can nevertheless be matched.
Roughly speaking, one finds
\be
Z_{\phi}^M \ \to \ Z_\phi^{(5)}, \qquad Z_{\psi}^M\ \to \ Z_\psi, \qquad Z_{B^+}^M \ \to \ Z_\phi^{(1)}\,Z_F \, Z_{\lambda_{37}}
\ee
Here we split the six scalars on the $D3$ as $1+5$, and correspondingly $Z_\phi$ as $Z_\phi^{(1)} Z_\phi^{(5)}$.
In other words, it was found that in the IIb limit, the chiral two-form
incorporates the $D3$ gauge field $F$, its magnetic dual $\tilde F$ as well as the localized fermionic degrees of
freedom $\lambda_{37}$ due to $37$-strings seen in the IIb weak coupling limit in a single, globally
well-defined entity. Furthermore, the relation between $ Z_{\lambda_{37}}$ and $Z_{B^+}^M$ is given by a cylinder mapping. Thus by replacing
$Z_{F,\,\tilde F}\,  Z_{\lambda_{37}}$ with $Z_{B^+}$ for finite string coupling, it is possible to address
both of the afore-mentioned problems and
make sense of $D3$-instantons in a global way.

The partition function $Z_\phi^M$ consists of a classical piece $\exp(-{\rm vol}(M5))$ and a one-loop determinant, given by the determinant of the Laplacian acting on the scalars. Similarly, $Z_\psi^M$ is given by the determinant of
${\not \!\! D\,} + {\not \!\! G}$, the Dirac operator extended by a term involving the $G$-flux. When the $G$-flux vanishes, this is related to Ray-Singer torsion.
The partition function $Z_{B^+}$ is more subtle. It was first treated
in \cite{Witten:1996hc}, where it was essentially identified with the theta function $\Theta(\tau, z)$ on the intermediate
Jacobian of the $M5$-brane (pulled back to the moduli space of the compactification), and further aspects
relevant for phenomenology were discussed in \cite{BDW}. In particular, in perturbative IIb, and also in the heterotic string, there is an interesting
relation between anomalous $U(1)$ gauge symmetries, selection rules for the
superpotential, and charged fermionic zero modes. It was found to be beautifully transported
to the presence or absence of $G$-flux induced tadpoles for $Z_{B^+}$ in $M$-theory. (Note there are no charged fermionic zero modes in this picture,
as $B^+$ is the only field that transforms under gauge transformations).
A downside of the $M5$ approach however is that in practice the
partition function is still
hard to calculate explicitly, as one needs to know the complex structure $\tau$ of the intermediate Jacobian and the periods
$z$ of the $F$-theory three-form field as a function of the moduli, or calculate the zero locus of $Z_{B^+}$ on the moduli space
(i.e. the pull-back of the theta divisor).

Alternatively, we could use heterotic/$F$-theory duality \cite{BDW}.
Then the $D3$-instanton gets mapped to a worldsheet instanton, whose partition function is of the form
\be
Z_{WS} \ = \ Z_\phi\, Z_\psi\, Z_\lambda
\ee
where $\phi,\psi$ denote world-volume scalars and right-moving fermions, and $\lambda$ denotes the left-moving fermions.
Again, this expression can be matched with $Z_{M5}$. The heterotic partition function $Z_\phi$ should be
factorized into a contribution $Z_\phi^{(3)}$ from three of the scalars and a contribution $Z_\phi^{(5)}$ from the remaining
scalars. Then we have
\be
Z_{\phi}^M \ \to \  Z_\phi^{(5)}, \qquad Z_{\psi}^M \ \to \ Z_\psi, \qquad Z_{B^+}^M \to Z_\phi^{(3)}\, Z_\lambda
\ee
In the heterotic picture, the most interesting aspects
are reproduced by the partition function $Z_\lambda$ of the left-movers.
This approach allows one to make contact with explicit computations of $Z_\lambda$
as a function of some of the moduli \cite{Buchbinder:2002ic}, and agreement
with the $M5$ picture is a consequence $2d$ bosonization and a cylinder mapping, this time applied to the instanton worldvolume.
It also resolve some tensions with vanishing results from perturbative type IIb. The resolution is that these vanishing
results are caused by extra light anomalous
$U(1)$ gauge bosons that appear in the IIb limit but are absent (or rather very massive) in $F$-theory \cite{BDW}.
This is quite similar to the way that certain Yukawa couplings are forced to vanish in IIb, but are non-vanishing in
$F$-theory, and further confirms that perturbative IIb results should be interpreted with care.

We may try to use non-perturbative effects to stabilize the
last modulus (denoted by $S$) that we could not stabilize in section
\ref{SuperpotentialDuality}. This last modulus is by far the most
interesting one, because its VEV also serves as our expansion
parameter, and it remains somewhat controversial whether the
associated conceptual problems have really been solved.

Including the leading non-perturbative correction, the
superpotential is of the form
\be W \ = \ W_0\ +\ e^{-S} \ee
As is by now well-known on the $F$-theory side, this yields an AdS
vacuum with all moduli stabilized, with $\vev{S}$ large provided
$W_0$ is extremely small \cite{Kachru:2003aw}. On the heterotic
side, this is exactly the old gaugino condensation story
\cite{Dine:1985rz}, but here is is often stated that $W_0$ is not
small enough to trust the solution. This is mainly a limitation of
the explicit models considered. From the viewpoint of
heterotic/$F$-theory duality, we can make $W_0$ (or really $e^{{\cal
K}/2}|W_0|$) equally small provided we include the full landscape of
constructions on the heterotic side. $F$-terms and $D$-term supersymmetry breaking
and uplifting to de
Sitter space can also be studied on both sides the duality. Whether
these effects are really small enough would need to be investigated in more
detail, and this is still controversial even on the $F$-theory side.

\newpage

\newsection{Examples with GUT groups}
\seclabel{Examples}

In this section we consider some explicit three generation $SU(5)$
and $SO(10)$ GUT models in $F$-theory. One may easily come up with
some models by lifting from the heterotic literature and translating
into $F$-theory language. These examples are not yet realistic for a
number of reasons. The primary reason is that we still need to
specify a mechanism to break the GUT group to the Standard Model
gauge group. In addition, generically these models will have extra
non-chiral matter, $R$-parity violating couplings, and so on.
Nevertheless we think these examples are useful to illustrate the
ideas, and leave a more detailed analysis of the phenomenology for
the future.

\newsubsection{Examples with $SU(5)$ gauge group}
\subseclabel{SU5Examples}

 We take a $dP_9$ fibration over
a base $B_2$, and denote by $N_{B_2}$ the normal bundle for $B_2$ in
$B_3$, with Chern class $c_1(N_{B_2}) = -t$. The slightly odd
notation is essentially for `historical' reasons, as it matches with
the notation of \cite{Friedman:1997yq}. We use $s$ to denote a
coordinate on the normal bundle. In section \ref{ModelF} the
singularity was located at $v=0$ and $v/u$ can be taken the
coordinate on the normal bundle. Comparing the line bundles, we see
that $N_{B_2} = K_{B_2}^6 \otimes {\cal N}|_{\sigma_{B_2}}$ and
hence the relation between $t$ and $\eta$ is given by
\be t = 6 c_1(B_2) - \eta. \ee

In order to write down a model, we need to do two things: first, we
need to specify a suitable Weierstrass equation, and second we need
to specify a flux. The Weierstrass equation for $dP_9$ is of the
form
\be
y^2 = x^3 + f\, x + g
\ee
where $f$ and $g$ are sections of $K_{B_3}^{-4}$ and $K_{B_3}^{-6}$
respectively. Near $\sigma(B_2)$ we have $K_{B_3} \sim K_{B_2}
\otimes N_{B_2}^{-1}$ and we can expand the Weierstrass equation
\be
y^2 = x^3 + x \sum_{i=0}^4 f_{4c_1+(i-4)t}\, s^i + \sum_{j=0}^6 g_{6c_1+(j-6)t}\, s^j
\ee
The $f_{4c_1-nt}$ are sections of a line bundle over $B_2$ with
Chern class $4\, c_1(B_2) - n\, t$, and the $g_{6c_1-nt}$ are
sections of line bundles with Chern class $6\, c_1(B_2) - n\, t$. In
order to specify an $SU(5)$ singularity along $B_2$, we need a
section of the projective space bundle ${\cal W}_{SU(5)} \to B_2$
with fibers
\be {\cal M}_{SU(5)}  = {\bf WP}^4_{(1,1,1,1,1)} \ee
That is, we need to specify five sections of line bundles with
appropriate Chern classes as discussed in section \ref{ModelF}. In
\cite{Bershadsky:1996nh} these sections are denoted as
\be\label{SU5sections}
h_{c_1-t},\quad H_{2c_1-t},\quad q_{3c_1-t}, \quad f_{4c_1-t}, \quad g_{6c_1-t} %\ \in
%\ H^0(B_2, {\bf P}(\cO \oplus \cL^{-2} \oplus \cL^{-3} \oplus \cL^{-4} \oplus \cL^{-5}))
\ee
In section 2 we instead denoted them as
\be a_5, \quad a_4, \quad a_3, \quad a_2, \quad a_0 \ee
The $f$'s and $g$'s are expressed in terms of these five sections as
\cite{Bershadsky:1996nh}
\be
g_{6c_1-6t} \sim h_{c_1-t}^6, \quad f_{4c_1-4t} \sim h_{c_1-t}^4, \quad \ldots
\ee
Near $\sigma_{B_2}$, i.e. to leading order in $s$, the discriminant
locus can be expressed as \cite{Bershadsky:1996nh}
\be \Delta \sim s^5 \, a_5^4\,  P_{8c_1-3t} + \cO(s^6), \qquad f
\sim a_5^4, \qquad g \sim a_5^6 \ee
where $P_{8c_1-3t}$ is a section of a line bundle with $c_1 = 8 c_1
(B_2) - 3 t$. Explicitly, we have \cite{Friedman:1997yq}
\be P_{8c_1-3t} = a_0 a_5^2 - a_2 a_3 a_5 + a_3^2 a_4\ee
Note that this section is quadratic in $a_3$ and $a_5$, so it is
certainly not a generic representative of $8c_1 - 3 t$.

From the discriminant, we can easily read off the equations of the
matter curves. Using the Kodaira classification, the zero locus of
$a_5$ corresponds to an enhancement from $SU(5) \to SO(10)$, so
anti-symmetric matter is localized here. The curve $\{a_5=0\}$ is
denoted by $\Sigma_\bt$. The zero locus of $P_{8c_1-3t}$ corresponds
to the enhancement $SU(5) \to SU(6)$, so this is where fundamental
matter is localized. The curve $\{P_{8c_1-3t}=0\}$ is denoted by
$\Sigma_\bfv$.

As an example \cite{Diaconescu:2005pc}, let us take $B_2$ to be a
$dP_8$ surface, and $\eta = 6 c_1(B_2)$. Then there exist
holomorphic sections (\ref{SU5sections}) with the required Chern
classes, so the spectral cover exists, and $[\Sigma] = \eta - 5
c_1(B_2) = c_1(B_2)$ is effective, in fact it is just the canonical
class (which is an elliptic curve). From equation (\ref{numgen}),
the net number of generations is given by
\be
N_{\rm gen} = -\lambda \eta\cdot (\eta - 5 c_1(B_2)) = -6\lambda \ee
so taking $\lambda=-\half$ we get three generations.

As a second example, let us take $B_2$ to be an Enriques surface. We
refer to \cite{BPVdV} for facts about Enriques surfaces. A generic
Enriques surface always contains two effective divisors $D_1,\, D_2$
with intersection numbers
\be D_1^2 = D_2^2 = 0,\qquad D_1 \cdot D_2 = 1,\qquad c_1 \cdot D_1
= c_1 \cdot D_2 = 0.\ee
To construct an $SU(5)$ GUT model, we may take eg. $\eta = D_1 +
3D_2 + 5 c_1\sim D_1 + 3D_2 + c_1$. Then there exist sufficient
holomorphic sections (\ref{SU5sections}) with the required Chern
classes. If the spectral cover is smooth, then we may apply formula
(\ref{numgen}):
\be N_{\rm gen} = -\lambda \eta\cdot (\eta - 5 c_1(B_2)) = -6\lambda
\ee
For $\lambda = -1/2$ we get exactly three chiral generations.

After the first version of this paper appeared, where this model was
included in a footnote, we have performed a more detailed analysis
which shows that the spectral cover in this homology class may
generically be taken to be smooth. The argument is given below.

There are several ways to represent Enriques surfaces. We will
represent an Enriques surface as an elliptically fibered surface
with a rational base. Let us discuss some of its properties. First
we recall the notion of a multiple fiber. Given an elliptic
fibration $\pi:S \to {\bf P}^1$ whose generic fibers are smooth, a
multiple fiber $\pi^{-1}(p)$ is a special fiber such that a multiple
$m \cdot \pi^{-1}(p)$ is linearly equivalent to the generic fiber,
with $m \geq 2$. An Enriques surface admits an elliptic fibration
over ${\bf P}^1$, and such a fibration has exactly two multiple
fibers $2F$ and $2F'$, with
\be K_S = \cO(F-F') \ee
The divisors $F$ and $F'$ are often called a half-pencil.

Furthermore, one may show that the elliptic fibration admits a
2-section, that is an irreducible curve $G$ with $G\cdot f=2 $ for
every fiber $f$. Now there are two possibilities for $G$. Either
$G^2 = -2$, which is non-generic because the existence of
$-2$-curves requires tuning the complex structure moduli (and
moreover means $S$ is nodal, i.e. singular); or $G^2=0$. In the
latter case, the linear system $|2G|$ has dimension one and the
generic element is an elliptic curve, hence defines another elliptic
fibration for which $G$ is one of the half-pencils. Moreover $G
\cdot F=1$ where $F$ was a half-pencil of the first elliptic
fibration. Therefore a generic Enriques has two effective divisors
$D_1$ and $D_2$, with $D_1^2 = D_2^2=0$, $D_1 \cdot D_2=1$, where
$D_1$ and $D_2$ can be explicitly thought of as half-pencils of two
distinct elliptic fibrations $\pi_1$ and $\pi_2$. We will use the
explicit description of these divisors given above for constructing
models.

Above we gave an explicit description of a certain sublattice of the
Picard lattice of a generic Enriques surface. Let $D_1, D_1'$ be the
two half pencils corresponding to a pencil $E_1$. Then
\be D_1'-D_1= c_1, \quad E_1=2\,D_1. \ee
Similarly let $D_2, D_2'$ correspond to a second pencil $E_2$, with
\be D_1\cdot D_2=1. \ee
In the following it will be useful to have explicit names for the
intersection points. Hence we will define the following four
distinct points on $S$:
\be D_1 \cdot D_2' = \{p\}, \qquad D_1'\cdot D_2 = \{p'\}, \qquad
D_1 \cdot D_2 = \{ q\}, \qquad D_1'\cdot D_2' = \{ q'\} \ee

Our explicit model was given by
\be \eta = D_1 + 3D_2 , \qquad \lambda = 1/2\ee
This is closely related to the model mentioned in the first version
of this paper, which had $\eta = D_1 + D_2 , \lambda = 3/2$. Our
analysis below goes through for this model just as well, but it has
fewer complex structure moduli. (The linear system $D_1+D_2$ is a
pencil (i.e. $h^0=2$), and $p,p'$ are its base points. Indeed, a
basis for the pencil consists of $D_1+D_2$ and $D_1'+D_2'$, and
their intersection is $p$ and $p'$. Moreover, these are the only
points in the base locus.)

Our spectral surface is defined by the (projectivized) equation:
\be b_0 u^5 + b_2 u^3 v^2 + b_3 u^2 v^3 + b_4 u v^4 + b_5 v^5 = 0
\ee
The $b_{even}$ are elements of the linear system $D_1+3D_2$, which
has $h^0=4$. An explicit basis consists of:
\be b_{even}:\qquad D_1+3D_2,\quad  D_1'+D_2'+2D_2,\quad
D_1+D_2+2D_2',\quad D_1'+3D_2'. \ee
The base locus is non empty: it consists of the points $p$ and $p'$
with multiplicity one. Similarly, the $b_{odd}$ are elements of the
linear system $D_1 + 3 D_2 + c_1= D_1' + 3D_2$. An explicit basis
consists of
\be b_{odd}:\qquad D_1' +3D_2,\quad  D_1'+ 2D_2'+D_2,\quad
D_1+D_2'+2D_2,\quad D_1+3D_2'. \ee
Again the base locus is non-empty. It is not hard to see that it
consists of the points $q,q'$ with multiplicity one. Thus even
though we will choose the $b_i$ to be distinct linear combinations
of the generators of these linear systems, there are two special
points (namely $p$,$p'$) where the $b_{even}$ all vanish, and two
more special points (namely $q$,$q'$) where the $b_{odd}$ all
vanish.

Since the $b_i$ above have some special properties, one might worry
that the linear system $|C|$ has base points. This is indeed the
case, it is not hard to see that the base points are given by
\be (x,u,v) \in (S,{\bf P}(\cO \oplus K_S))\ |\ x=p,p' {\rm \ and\ }
v=0, x=q,q' {\rm \ and\ } u=0 \ee
Therefore we cannot apply Bertini to conclude that the generic
spectral surface is smooth. Since the formula for the net generation
number (\ref{numgen}) was computed under the assumption that the
spectral surface is smooth, it seems that we may have to do more
work to write down suitable fluxes and check the generation number.
But fortunately we can show that the spectral surface is in fact
smooth generically despite the presence of base points in the linear
system, and so we can automatically still use the universal flux and
apply the formula (\ref{numgen}) for the net number of generations.

To see this, consider an element in the linear system of $C$:
\be b_0 u^5 + b_2 u^3 v^2 + b_3 u^2 v^3 + b_4 u v^4 + b_5 v^5 = 0
\ee
The generic such surface is smooth away from the base locus, so the
only special loci we need to worry about correspond to the base
points of $|C|$ found above. When $x=p$ or $p'$, we get
\be {\del\over \del v} (b_0(p) u^5 + .. + b_5(p) v^5) =2 u^3 v
b_3(p) + 5 b_5(p) v^4 \ee
Since $b_3(p)$ and $b_5(p)$ are generically non-zero, the only
dangerous points also have $v=0$. However
\be {\del\over \del x}((b_0(x) u^5 + .. + b_5(x) v^5)|_{x=p,v=0} =
{\del\over \del x}((b_0(x) u^5)|_{x=p} \ee
which is generically non-zero, even at $x=p$ or $p'$ because the
base locus of $D_1 + 3 D_2$ only has multiplicity one. Similarly one
checks that $x=q,q'$ and $u=0$ are smooth points generically.

Thus we have verified that the generic spectral cover in this linear
system is smooth, and hence the application of (\ref{numgen}) for
the number of generations in our Enriques models with `universal'
fluxes is justified.

\newsubsection{Examples with $SO(10)$ gauge group}

We can repeat much of the discussion for $SU(5)$ with few changes.
Again we consider the Weierstrass equation
\be
y^2 = x^3 + x \sum_{i=0}^4 f_{4c_1+(i-4)t}\, s^i + \sum_{j=0}^6 g_{6c_1+(j-6)t}\, s^j
\ee
In order to get an enhanced $SO(10)$ symmetry for $s=0$, we need to
specify a section of the weighted projective bundle ${\cal W} \to
B_2$ with fiber
\be
{\cal M}_{SU(4)} = {\bf WP}^3_{(1,1,1,1)}
\ee
That is we need to specify four sections
\be\label{SO10sections} h_{2c_1-t}, \quad q_{3 c_1-t}, \quad
f_{4c_1-t}, \quad g_{6 c_1-t} \ee
The $f$'s and $g$'s are recovered as
\be
f_{4c_1-2t} \sim h_{2c_1-t}^2, \quad g_{6c_1-3t} \sim h_{2c_1-t}^3, \quad
g_{6c_1-2t} =q_{3 c_1-t}^2-f_{4c_1-t}h_{2c_1-t}
\ee
The leading terms in $s$ are
\be
\Delta = s^7 h_{2c_1-t}^3q_{3 c_1 -t}^2 + \cO(s^8), \quad f \sim s^2 h_{2c_1-t}^2, \quad
g \sim s^3 h_{2c_1-t}^3
\ee
The ${\bf 16}$'s are localized at $h_{2c_1-t}=0$ where the symmetry
is enhanced to $E_6$, and the ${\bf 10}$'s are localized at $q_{3
c_1 -t}=0$ where the symmetry is enhanced to $SO(12)$.

As an example (not present in the literature as far as we know), let
us take the base to be a del Pezzo surface with a $-1$-curve,
denoted by $E$, and take $\eta = 7 c_1(B_2) -2 E$. Then the
quantization condition (\ref{Lquantization}) is satisfied if
$\lambda$ is integral, and there exist sections (\ref{SO10sections})
with the required Chern classes. From the analogue of (\ref{numgen})
for $SO(10)$ we have
\be N_{\rm gen} = -\lambda \eta \cdot (\eta - 4 c_1(B_2)) = -(21
c_1^2 -24)\lambda \ee
Therefore by taking $S = dP_8$ and $\lambda = 1$ we get an explicit
three-generation model.

\newpage

\newsection{Breaking the GUT group to the SM}
\seclabel{GUTbreaking}

So far we have discussed how to engineer GUT groups. To get a
realistic model however we need some way to break the GUT group to
the SM gauge group. As is well-known, it is typically hard in string
theory to obtain representations that are large enough to achieve
this. For instance in the heterotic string let's suppose we would
like to get four-dimensional fields in the adjoint representation of
the GUT group. These would originate from Wilson lines on the
Calabi-Yau.  But on manifolds of $SU(3)$ holonomy there are no
harmonic one-forms, so in this setting we cannot get any
four-dimensional fields in the adjoint of the GUT group.

On the $F$-theory side, we could get adjoint matter in four
dimensions from zero modes of the gauge field or of the adjoint
field of the eight-dimensional gauge theory. In duals of the
heterotic string, the gauge 7-brane is wrapped on a base $B_2$ which
has $h^{0,1}=h^{2,0}=0$, hence we get no such zero modes. In order
to get adjoint fields we must wrap our gauge brane on a surface of
general type or a $K3$ surface. For instance the $K3$ surface can be
realized as a quartic in ${\bf P}^3$; since the canonical bundle is
trivial and the normal bundle has many sections, it is not hard to
see that we can easily get an elliptic fibration with $I_5$ singular
fibers along such a surface. However just as in conventional
four-dimensional models we would then have to face the
doublet-triplet splitting problem. Hence we prefer to look for an
alternative mechanism.

Another idea, which was already considered in the early days of
heterotic model building (see eg. \cite{Witten:1985bz}), is to turn
on certain $U(1)$ fluxes. We have essentially already seen this in
the context of coincident branes. For instance in the case of an
$SU(5)$ model, we could turn on an internal flux on $\sigma_{B_2}$
for the gauge field that corresponds to hypercharge. The commutant
of this $U(1)$ in $SU(5)$ is clearly $SU(3) \times SU(2) \times
U(1)$. However turning on such a flux will typically spoil gauge
coupling unification. As we discussed earlier, the $U(1)$ generator
whose flux is turned on will swallow an RR axion and become massive.
This can be avoided by turning on a $U(1)$ flux in the same
cohomology class in the hidden sector. The axion then couples to the
sum of these $U(1)$'s, and the difference will remain massless. As
discussed in \cite{Witten:1985bz}, because hypercharge is now a
linear combination of the `original' hypercharge generator and a
$U(1)$ in the hidden sector, the model is not truly unified and this
mechanism would typically change the relation of the $U(1)$ coupling
to the $SU(2)$ and $SU(3)$ couplings at the GUT scale\footnote{On
the other hand, such a coupling to the hidden sector provides an
interesting possibility for mediation of SUSY breaking
\cite{Verlinde:2007qk}.}.

A third approach for breaking the GUT group, which does not have the
usual baggage of four-dimensional GUTs and has the cleanest
phenomenological features, is to use discrete Wilson lines. Namely
if $B_2$ admits a non-trivial fundamental group, then we could turn
on a discrete $G$-flux, or perhaps we could fiber the $dP_9$ over
$B_2$ in such a way that the GUT group is globally broken to the
Standard Model group. If we restrict to the usual models with
heterotic duals, then the only allowed $B_2$ which has non-trivial
fundamental group is the Enriques surface, and it does not lead to
consistent models due to lack of stability in the hidden sector.
However locally it is not hard to construct such models. As an
example, consider the three generation $SU(5)$ model from section
\ref{SU5Examples} based on the Enriques surface. It allows a $Z_2$
Wilson line to break the $SU(5)$ GUT group to $SU(3)\times
SU(2)\times U(1)$. Presumably this local model has global embeddings
which are not dual to the heterotic string.

However, we may give a stronger argument against such models: any
smooth model in $F$-theory with discrete Wilson lines has light
lepto-quarks. To see this, suppose more generally that we break the
GUT group through a line bundle $L$. Then the spectrum of
lepto-quarks descending from the eight-dimensional gauge and adjoint
fields is determined by the formulae in section
\ref{coincidentspectrum}. In particular, the Euler character
\be \chi(B_2,L) = \half c_1(L)^2 - \half c_1(L) c_1(K) + {1\over
12}(c_2(S) + c_1(K)^2) \ee
must vanish. Now let us assume that $c_1(L)$ vanishes as for
discrete Wilson lines. Then the above is just equal to
$\chi(B_2,\cO)=h^{0,0}(B_2) - h^{0,1}(B_2) + h^{0,2}(B_2)$. But we
assume that $h^{0,1}(B_2) = 0$ (and actually also $h^{0,2}(B_2)=0$)
in order to avoid massless adjoint fields. It follows that vanishing
$c_1(L)$ implies that $\chi(B_2,L)$ cannot vanish, so models with
discrete Wilson lines also necessarily come with light exotic
matter, whatever the surface that we wrap the 7-branes on.

The fact that models based on rational surfaces do not seem to allow
for discrete Wilson line breaking raises a puzzle though, because
they arise in heterotic/$F$-theory duality. On the heterotic side
one may certainly construct elliptically fibered three-folds with a
rational base and with a finite fundamental group. These three-folds
do not have a section, only a multi-section. However they are
quotients by an automorphism of elliptically fibered Calabi-Yaus
with a section, so we can construct the $F$-theory dual of the
cover. What does the automorphism get mapped to?

Consider a freely acting involution $\tau$ from the elliptically fibered
three-fold to itself. Then $\tau$ can be decomposed as
\be \tau = t_\xi \circ \alpha
\ee
where $\alpha$ maps the zero section of the elliptic fibration to
itself and $t_\xi$ is translation by a section $\xi$ different from
$\sigma_{B_2}$. The automorphism $\alpha$ induces an involution
$\alpha_{B_2}$ on the base $B_2$ which necessarily has fixed points.
Now $t_\xi$ acts trivially on the Wilson lines on each $T^2$ fiber,
so it does not appear to induce any action on the dual $T^2$ or the
$dP_9$ surface constructed from the dual $T^2$ and the Wilson lines
of the $E_8$ bundle. Therefore the action of $\tau$ on the heterotic
side seems to induce only the action of $\alpha_{B_2}$ on the
$F$-theory side, which has fixed points, and we would have to
understand how to deal with the fixed points. Unfortunately,
$F$-theory is currently only understood as a large volume expansion.
When the $F$-theory base $B_3$ has singularities, there is no small
parameter available and no clear way to understand the physics.

\vspace{1cm}
\noindent
{\it Acknowledgements:}

R.D. is partially supported by NSF grant DMS 0612992, NSF Focused
Research Grant DMS 0139799 `The Geometry of Superstrings,' and NSF
Research and Training Grant DMS 0636606. MW is supported by a Marie
Curie Fellowship of the European Union. Some of this work took place
at the August 2007 Simons workshop at SUNYSB and the March 2007
workshop at the Galileo Galilei Institute in the beautiful city of
Florence. MW would further like to thank the Ecole Polytechnique,
Harvard University and the University of Pennsylvania for
hospitality while this work was in progress and the opportunity to
present some of these results. It is a pleasure to
thank the members of these groups
for useful discussions. We would also like to thank S. Katz for comments
on the manuscript.

\vspace{2cm}

\newpage

\appendix

\renewcommand{\newsection}[1]{
\addtocounter{section}{1} \setcounter{equation}{0}
\setcounter{subsection}{0} \addcontentsline{toc}{section}{\protect
\numberline{\Alph{section}}{{\rm #1}}} \vglue .6cm \pagebreak[3]
\noindent{\bf Appendix {\Alph{section}}:
#1}\nopagebreak[4]\par\vskip .3cm}

\newsection{Spinors and complex geometry}
\seclabel{Cspinors}

In this appendix we would like to review some properties of spinors
on complex manifolds. We will not be very rigorous; instead we will
use the fastest route available. See \cite{LMSpin} for a more
thorough treatment.

Suppose we are given a $m$-dimensional Riemannian manifold $M$ with a spin structure,
i.e. a lifting of the structure group $SO(m) \to Spin(m)$. On such a manifold,
we may construct the spin bundle $S$ associated to the spinor representation of $Spin(m)$.
Given a local ortho-normal frame $e_a$ and a set of $\Gamma$-matrices satisfying
$\{\Gamma^\mu,\Gamma^\nu\} = 2 g^{\mu\nu}$, the Dirac operator is defined to be the first
order differential operator given by
\be
 \not \!D = \Gamma^a \nabla_a
\ee
where $\nabla_a$ is the lift of the Levi-Civita connection.

When $M$ is $2n$-dimensional, the spinor representation of $Spin(2n)$
is reducible, and $S$ decomposes as $S = S^+ \oplus S^-$. The Dirac operator
interchanges these representations, i.e. $\not \! D : S^\pm \to S^\mp$. If
$M$ is a K\"ahler manifold, i.e. if the holonomy can be further reduced to
$U(n) \subset SO(2n)$, then we can relate the Dirac operator to certain standard
operators appearing in holomorphic geometry.

Let us first consider one-dimensional K\"ahler spaces. We can then define a
spinor to be an object which gets mapped to minus itself under a
$2\pi$ rotation on every holomorphic tangent plane. This identifies
it as a section of the bundle $S = T^{-1/2} \oplus T^{1/2}$ where
$T$ is the holomorphic tangent bundle. Under a rotation by $\pi$
(i.e. a reflection $z \to -z$) spinors transform by $\pm i$. The
sign is called its chirality. The bundle $T^{-1}$ is also known as
$K$, the canonical bundle. Moreover if we have a K\"ahler metric
then we can identify $T$ with the bundle of $(0,1)$ forms
$\Omega^{(0,1)}$  by mapping sections as $f^z \,\del_z \to f^z g_{z
\bar{z}}\, d\bar{z}$. Therefore we can also write
\be S = K^{1/2} \oplus \Omega^{(0,1)}(K^{1/2}). \ee
Up to normalization, the Dirac operator $\not \! D$ therefore corresponds to
$\delb_A + \delb_A^\dagger$, where $\delb_A$ is the Dolbeault
operator coupled to $K^{1/2}$. More explicitly we can write this as
\be \not \!D =
\left(
  \begin{array}{cc}
    0 &  -\del_z - A_z\\
    \delb_{\bar{z}} + A_{\bar{z}} & 0 \\
  \end{array}
\right)
\ee
We can also couple the spinors to various bundles, by adding further gauge fields.

We can generalize this to higher dimensions by using a splitting
principle. That is we decompose the holomorphic tangent bundle for a
complex $n$-fold formally into a sum of $n$ line bundles and tensor
the corresponding spinor bundles together. For instance on a complex
three-fold we would decompose $T = T_1\oplus T_2 \oplus T_3$ and tensor the
$T_i^{-1/2} \oplus T_i^{1/2}$ together. The result, after
reconstructing representations of the full $U(n)$ holonomy, is
\be
S^+ = \sum_{p\ even} \Omega^{(0,p)}(K^{1/2}) \qquad  S^- = \sum_{p\ odd} \Omega^{(0,p)}(K^{1/2})
\ee
The Dirac operator is then formally thought of as the sum of the
Dirac operators associated to each $T_i$.

\newsection{Branes and twisted Yang-Mills-Higgs theory}
\seclabel{TYMH}

In this appendix we briefly review the Yang-Mills theories living on
branes in string theory, with an emphasis on curved embeddings of
the brane in space-time.

The collective coordinates of $Dp$-branes are are given by the
field content of maximally supersymmetric Yang-Mills theory in $p+1$
dimensions. They may all be obtained by starting with $N=1$
Yang-Mills theory in ten dimensions and reducing it to $p+1$
dimensions. For applications to $F$-theory we would like to
understand how to reduce ten-dimensional Yang-Mills theory to a
complex submanifold denoted $B$. The ten-dimensional action is of
the form
\be\label{10Daction} \int d^{10}x\,  -{i\over 2g^2} {\rm Tr}(\bar{\psi}\not\!\! D\, \psi)
-{1\over 4g^2} {\rm Tr} (F_{\mu\nu}F^{\mu\nu}) \ee
Now let us do the reduction, assuming the case of a 7-brane wrapped
on a surface $S$ in a Calabi-Yau three-fold.  Following Hitchin
\cite{Hitchin:1986vp} we write
\be\label{HYMreduce} A^{0,1} = A_{\bar z_1}(z_1,z_2) + A_{\bar
z_2}(z_1,z_2) + \Phi_{\bar 3}(z_1,z_2) \ee
Here we used the splitting principle to express the tangent bundle
as $T = T_1 \oplus T_2 \oplus K_S \oplus T{\bf R}^{1,3}$. Thus our
data consists of a connection $A$ on a bundle $E$ on $S$, together
with the complex `Higgs field' $\Phi$, which we may view as a map
$\Phi: E \to E \otimes K_S$. Similarly spinors now becomes sections
of
\be (T_1^{-\half} \oplus T_1^{\half})\otimes (T_2^{-\half} \oplus
T_2^{\half})\otimes (K_S^{-\half} \oplus K_S^{\half}) = \sum_p
\Omega^{(0,p)}(K_B^{1/2}) \otimes (K_S^{-\half} \oplus K_S^{\half})
\ee
tensored with four-dimensional spinors.

The ten-dimensional gaugino variation is of the form
\be \delta \psi \simeq F_{\mu\nu}\Gamma^{\mu\nu}\epsilon, \ee
Thus requiring a BPS solution means that we have to solve
$F_{\mu\nu}\Gamma^{\mu\nu}\epsilon=0$, or equivalently
\be F^{0,2}=0, \qquad g^{i\bar j} F_{i\bar j} = 0 \ee
These are the hermitian Yang-Mills equations. To reduce this to
eight dimensions, we now substitute (\ref{HYMreduce}):
\be \delta\psi \simeq  \left( F_{\mu\nu}\Gamma^{\mu\nu} + 2\, D_\mu
\Phi_a \Gamma^{\mu a} + [\Phi_a,\Phi_b] \Gamma^{ab}\right)\epsilon=0
\ee
where we use $\mu,\nu$ for real indices tangent to the brane, and
$a,b$ for real indices normal to the brane. The $F$-terms and
$D$-terms of the effective four-dimensional gauge theory can be read
off from the right-hand side. In particular the $F$-terms come from
the lack of integrability of $\bar{D} = \delb + A^{0,1} + \Phi$.
Preservation of supersymmetry thus requires $\bar{D}^2 =0$. By
decomposing we get the following equations:
\be\label{HiggsYM} F^{0,2}=0, \qquad \delb \Phi_{\bar 3} +
[A^{0,1},\Phi_{\bar 3}] = 0, \qquad
 \ee
Similarly the $D$-terms can be written as\footnote{We would like
to thank J.~Heckman for pointing out the term involving $\Phi$,
which is crucial for a correct understanding of the $D$-terms but
which we had initially ignored.}
\be g^{i\bar j} F_{i\bar j} +  [\Phi_{\bar 3}^\dagger,\Phi_{\bar
3}]\ =\ 0 \ee

Let us take a closer look at these equations. The first equation in
(\ref{HiggsYM}) says that the gauge field is a connection on a
holomorphic bundle (i.e. the transition functions may all be chosen
holomorphic). As is well-known, we may then apply a complexified
gauge transformation so that the anti-holomorphic component
$A^{0,1}$ vanishes in holomorphic frames. The $(0,1)$ part of the
gauge covariant derivative then reduces to the Dolbeault operator
$\delb$, and the second equation in (\ref{HiggsYM}) says that $\Phi$
must be a holomorphic section.

The $D$-terms are not invariant under the complexified gauge
transformations, and require us to choose a hermitian metric, or
equivalently a reduction of the complexified structure group to a
compact subgroup. Recall that given a hermitian metric $h$, we
can pick a canonical connection by
requiring the covariant derivative to be compatible with the
hermitian metric and with the complex structure. This determines
$A^{1,0} \sim -(\del h) h^{-1}$. Thus assuming we have fixed the
$F$-term data, we see that the $D$-terms may be viewed as an
equation for the hermitian metric $h$ on $E$. This equation is a
highly non-linear PDE, which is virtually impossible to solve
explicitly. Nevertheless, existence and uniqueness of a solution can
be reformulated as an algebro-geometric criterion.

A subbundle $F \subset E$ is said to be a Higgs subbundle if
$\Phi(F) \subset F \otimes K$. A Higgs bundle is said to be
$J$-stable if
\be \mu(F) < \mu(E) \ee
for every Higgs subbundle, where the slope is defined as usual, $\mu
= J$-degree/rank. A Higgs bundle is poly-stable if it is a direct
sum of stable Higgs bundles with the same slope. If the Higgs bundle is polystable, then
the $D$-terms should have a unique solution. (For abelian bundles,
this requires adding an explicit Fayet-Iliopoulos-like term to the
equation). The corresponding hermitian metric is sometimes called
the hermitian-Einstein metric.

For smooth $F$-theory compactifications, we encountered the primitiveness
condition $J \wedge G = 0$. It is clearly reminiscent of $i^*J \wedge F=0$ on the spectral
cover, in the gauge where $A$ and $\Phi$ are diagonalizable. This
might seem to suggest that $D$-flatness corresponds not to stability, but to the
statement that the hermitian connection on the spectral line bundle
is $J$-primitive. However this is not correct. It is actually
well-known that the push-down of the hermitian
connection on the spectral line bundle should not be identified with
the solution of the $D$-terms of the Higgs bundle. To see this,
note that if the spectral cover is locally defined by $y^2 -z=0$
then
\be F_{z\bar z}\ \sim\ F_{y\bar y}/(z\bar z)^{1/2} \ee
which diverges at the branch locus, here given by $z=0$. Thus the
push-down of the hermitian connection on the
spectral line bundle is singular at the branch locus. However,
Donaldson-Uhlenbeck-Yau provides a smooth solution to the $D$-term
equations on the Higgs bundle. In particular
we have $[\Phi^\dagger,\Phi]\not =0$ for the actual hermitian metric
solving the $D$-terms.\footnote{A second issue in the comparison is
that $J$ is not quite a pull-back of a class on the base.} We essentially
already discussed the source of this apparent discrepancy in section \ref{SolitonQuantization}.
It arises because the supergravity derivation of $J \wedge G=0$
requires a large and smooth four-fold. This is different from the regime
where we get parametrically light $M2$-branes, which is where the
$8d$ gauge theory description can be trusted.

In \cite{Donagi:2011jy} we therefore proposed that when there are parametrically light $M2$-branes,
the correct criterion is existence
of the hermitian-Einstein metric in the Higgs bundle picture. More generally,
we expect that $D$-flatness in $F$-theory compactifications
is described by a stability condition. This is something we can
study in the four-fold picture. Stability can be phrased in terms of
Fayet-Iliopoulos terms. Thus a question that we can ask is how the
Fayet-Iliopoulos terms are related in the different pictures, and if
there are certain classes of interesting compactifications where
stability should be automatic. We turn to this in the next section.

For further reading on the structures discussed
here, see \cite{Hitchin:1986vp,Simpson,DSpectral,Katz:2002gh,Donagi:2003hh,Donagi:2011jy}.

\newsection{Definition of $G_\gamma$}
\seclabel{Nonprim}

In this appendix we give a detailed definition of the class
$G_\gamma$ on the $F$-theory space $Y_4$, obtained via the cylinder
map from a $(1,1)$-class $\gamma \in H^{1,1}(C,\bf{Z})$ on the
heterotic spectral cover $C$. We also consider Fayet-Iliopoulos
terms, which are an important ingredient for defining stability
conditions. We will see that the classical expression for the Fayet-Iliopoulos
terms in $F$-theory can be matched at least qualitatively with the tree level
and one-loop expressions on the heterotic side.

Let us briefly recall the general set-up. On the heterotic side, the
$E_8$-bundle restricted on each elliptic fiber is determined by its
Wilson line, an element
\be   f \in \Hom(\Lambda_{E_8} , T^2) \ee
taken modulo ${\cal W}_{E_8}$. Here $\Lambda_{E_8}$ represents the
root lattice of $E_8$, the dual of the coroot lattice. This lattice
is actually self-dual, so we don't need to be too careful on this
point. Now let us fiber over $B_2$, and let us consider not modding
out by ${\cal W}_{E_8}$. Then the fibers $\Hom(\Lambda_{E_8} , T^2)$
fit together in a flat vector bundle of rank eight, with structure
group ${\cal W}_{E_8}$. In order to represent this data by a
spectral cover of finite degree, we have to choose a suitable
representation. The physically relevant representation is the
adjoint representation of $E_8$. This is a
 minuscule representation: its weights are the roots, and these lie in a single Weyl group
orbit  $\mbox{Roots}  \subset  \Lambda_{E_8} $. The the image
\be f(\mbox{Roots}) \subset T^2   \ee
gives a collection of $240$ points on the $T^2$, one for each root
of the adjoint representation of $E_8$. The Weyl group ${\cal
W}_{E_8}$ acts as the monodromy group on these 240 points. Fibering
over $B_2$ yields the degree 240 spectral cover of an $E_8$ bundle.
For a generic $E_8$ bundle, this is the smallest non-trivial
permutation representation available, i.e. it is the smallest
representation we can use for the monodromy action on a spectral
cover. If the monodromy is smaller than $E_8$, there can be other
available permutation representations and thus there will be
spectral covers of smaller degree which capture the same
information.

On the $F$-theory side, the same data determines a $dP_8$ surface.
The anti-canonical divisor is identified with the $T^2$ above, and
the $240$ lines of the $dP_8$ intersect the $T^2$ in the 240 points
above. Blowing up, we get a $dP_9$ with a distinguished zero
section. Generically the $9$th point is not on any of the lines, and
the $240$ lines of the $dP_8$ lift to $240$ sections of the $dP_9$,
disjoint from the zero section. When the $9$th point does lie on a
line, the curve on the $dP_9$ corresponding to this line is its
total transform. This is a numerical section, i.e. an effective
curve with self-intersection $-1$ and intersection number $1$ with
the elliptic fiber. Its intersection number with the zero section
still vanishes. But this is now a reducible curve, consisting of two
components: the zero section plus the proper transform of the
original line. The Weyl group ${\cal W}_{E_8}$ acts as a monodromy
group on these $240$ lines. Fibering over $B_2$ yields the cylinder.
In \cite{Curio:1998bv} this was the variety $R'$ whose fibers over
each point in $B_2$ are the $240$ lines in the $dP_8$'s. Since we
work here with $Y_4$, we consider instead the total transform $R$ of
$R'$. This is the subvariety of $Y_4$ whose fibers over each point
in $B_2$ are the $240$ numerical sections in the $dP_9$'s. By the
above observation, the intersection of $R$ with the zero section
$\sigma_{B_3}$ is the union  of lines (in either $R$ or
$\sigma_{B_3}$) over the intersection curve $\Sigma = C \cdot
\sigma(B_2)$. Explicitly:
\be\label{linesinR} R \cdot \sigma_{B_3} = p_R^*[\Sigma]_C =
\rho^*[\Sigma]_{B_2} \ee

Physically it is probably more natural to identify the roots with
$-2$ classes in the $dP_8$. However they are equivalent to the lines
up to a shift by the canonical class of the $dP_8$, which does not
affect the arguments below due to the tracelessness condition of the
fluxes. The lines are effective classes and are easier to keep track
of.

Let us fix the notation for this appendix:
\ba \pi_Y:  Y_4 \to B_3   & &  {\rm elliptic\ fibration\ } \eol
\sigma: B_3 \to Y_4 & &  {\rm the\ section\ } \eol
 \rho: B_3 \to B_2   & &  P^1 {\rm \ fibration} \eol
 \sigma_{B_2}  & &  {\rm its\ section\ , embedded\ in\  either\ } B_3 {rm or} Y \eol
Z \subset Y_4 & & \pi_Y^{-1} {\rm \ of\ a\ section\ of\ }\rho. \eol
i_Z: Z \hookrightarrow Y & & {\rm the\ natural\ inclusion} \eol p:
Y_4 \to B_2 & & dP_9 {\rm \ fibration.} \eol \pi_C: C \to B_2 & &
{\rm the\ heterotic\  spectral\  cover} \eol \pi_Z: Z \to B_2 & &
{\rm the\ restriction\  of} p   \eol p_R: R \to C & & {\rm the\
``cylinder",\ or\ union\ of\ lines\ in\ the\ } dP_8{\rm 's} \eol
        & & {\rm (i.e.\ numerical\ sections\ of\ } dP_9{\rm 's, \ disjoint\ from\ } \sigma {\rm )\ parametrized}\eol
        & & {\rm  by\ points\ of\ } C. \eol
\Sigma = C \cdot \sigma_{B_2} & & {\rm the\ matter\  curve\ } \eol
j: (C = R \cap Z) \subset R & &  {\rm  the\ inclusion\ ``at\
infinity"} \eol i_R: R \hookrightarrow Y & & {\rm the\ natural\
inclusion.} \eol \ea
The spectral line bundle is mapped to $G$-flux. Given a class
\be \gamma \in H^2(C, {\bf Z}), \qquad \pi_{C*}\gamma = 0 \ee
in \cite{Curio:1998bv} the dual $G$-flux was defined as
\be\label{Ggammadefinition} G\ \mathop{=}^{?} \ i_{R*} p_R^* \gamma
\ \in\  H^4(Y_4, {\bf Z})
 \ee
Actually, there is also a shift by $r/2$ in the quantization law on
the heterotic side, and by $c_2/2$ on the $F$-theory side, but one
can easily correct for this and we will not mention it any further.
In \cite{Curio:1998bv}, $Y_4$ was considered to be a
$dP_8$-fibration over $B_2$. As we discussed in section
\ref{MatchingHolomorphic}, in the case of $dP_9$-fibrations there is
an ambiguity in the mapping from $H^2(C) \to H^4(Y_4)$. On the other
hand, not all generators of $H^4(Y, {\bf Z})$ can be realized
off-shell as $G$-fluxes in $F$-theory. By requiring that $\gamma$
gets mapped to an allowed $G$-flux, we can fix the ambiguity.

A more conceptual way to arrive at the same conclusion is as
follows. As we vary the $dP_9$ surface over $B_2$, the 240 lines on
the $dP_8$ (which becomes sections of $dP_9$ after blow-up) are
exchanged by monodromies which take value in the $E_8$ Weyl group.
Although the Weyl group does not act on the spectral cover or its
cohomology, we can decompose the cohomologies $R^0{\pi_C}_*{\bf R}$
and $R^2{p}_*{\bf R}$ into isotypic pieces. On the heterotic side,
the 240 dimensional permutation representation of ${\cal W}_{E_8}$
on each fiber of the $240$-fold spectral cover can be decomposed in
the following irreducible representations
\be {\bf 240} = {\bf 1} + {\bf 8} + {\bf 35} + {\bf 84} + {\bf 112}
\ee
The 1-dimensional piece corresponds to the sum over the Weyl group
orbit. The 8-dimensional piece corresponds to the action of $W$ on
the root lattice of $E_8$.

Similarly, on the $F$-theory side, the second cohomology of the
$dP_9$ decomposes in the following irreducible representations
\be {\bf 10} = {\bf 1_b} + {\bf 1_f} + {\bf 8}. \ee
Now the 1-dimensional pieces come from the  base and fiber,
respectively, while the 8-dimensional piece is the part of the
second cohomology of the $dP_9$ orthogonal to the base and fiber.
This was denoted by $H^2_\Lambda$ in \cite{Curio:1998bv}.

Now the cylinder map in (\ref{Ggammadefinition}) does not exactly
identify the 8-dimensional representations on both sides. To get the
map that does this, we need to subtract a singlet on the $F$-theory
side. This is the `projected cylinder map' of \cite{Curio:1998bv}.

An allowed $G$-flux has three indices along $B_3$ and one index
along the elliptic fiber. More precisely, allowed $G$-fluxes must be
orthogonal to
\begin{description}
   \item[(i)]  classes in $ \sigma_{B_3 *}H_4(B_3,{\bf Z})$;
   \item[(ii)] classes in $ \pi_Y^*H^4(B_3,{\bf Z})$.
 \end{description}
However, with the above definition of $G$ we find that
\be\label{NotAFlux} G \cdot_Y \sigma_{B_2} = \gamma \cdot_C \Sigma
\not = 0 \ee
where $\Sigma = C\cdot \sigma_{B_2}$. Therefore the flux we defined
above is not allowed in $F$-theory, and the map from $\gamma$ to
$G$-flux has to be modified by projecting on $\sigma_{B_3
*}H_4(B_3)^\perp$. We claim that the correct definition of the
$G$-flux dual to $\gamma$ is\footnote{We are indebted to Taizan
Watari for pointing out that the expression (\ref{Ggammadefinition})
for the $G$-flux in a previous version of this paper needed to be
projected as in \cite{Curio:1998bv} and (\ref{NewGgammadefinition})
in order to produce an allowed flux.}
\be\label{NewGgammadefinition} G_\gamma\ \equiv\ i_{R*} p_R^* \gamma
- n_\gamma [dP_9], \qquad n_\gamma = \gamma \cdot_C \Sigma \ee
where $[dP_9]= p^{-1}(pt)$ denotes the class of a $dP_9$ fiber of
$Y_4$.

Our first task is to show that this is an allowed $G$-flux that is
orthogonal to all classes of type {\it (i)} and {\it (ii)}. Let us
make a list of these classes. Since $B_3$ is a ${\bf
P}^1$-fibration, $H_4(B_3,{\bf Z})$ is spanned by
\be H_4(B_3,{\bf Z})= {\rm span} \{ \sigma_{B_2}, \rho^{*}(w)\}\
\quad {\rm where} \ w \in H_2(B_2,{\bf Z}) \ee
Similarly,
\be H_2(B_3,{\bf Z})= {\rm span} \{ \sigma_{B_2 *}w, \rho^{*}(pt)\}\
\quad {\rm where} \ w \in H_2(B_2,{\bf Z}) \ee
and therefore,
\be \pi_Y^*H^4(B_3,{\bf Z}) = {\rm span}\{i_{Z*}\pi_Z^*w,
p^{*}(pt)\} \ \quad {\rm where} \ w \in H_2(B_2,{\bf Z}) \ee
Now we simply proceed by computing the intersections with all these
classes. We have
\ba i_{R*}p_R^*\gamma \cdot i_{Z*}\pi_Z^*w &=& i_{C*}\gamma \cdot_Z
\pi_Z^*w \eol &=& \gamma \cdot_C \pi_C^*w \eol &=& 0 \ea
Similarly
\be i_{R*}p_R^*\gamma \cdot p^*(pt) = p_R^*\gamma \cdot_R
p_R^*\pi_C^*(pt) =0 \ee
It is also easy to show that
\be [dP_9]\cdot i_{Z*}\pi_Z^*w = 0, \qquad [dP_9]\cdot p^{*}(pt) =
0\ee
since we may choose the support of these classes to be disjoint.
Therefore $G_\gamma$ is indeed orthogonal to classes of type {\it
(ii)}.

As for orthogonality against classes of type {\it (i)}, we use
(\ref{linesinR}) and compute:
\be i_{R*}p_R^*\gamma \cdot \sigma_{B_3*}\rho^*w =
\rho^*(\gamma\cdot \Sigma) \cdot \rho^*w = 0 \ee
We have already pointed out in \ref{NotAFlux} that
$i_{R*}p_R^*\gamma \cdot \sigma_{B_2} = \gamma \cdot \Sigma$, and it
is not hard to check that
\be [dP_9] \cdot \sigma_{B_3*}\rho^*w =0, \qquad [dP_9]\cdot
\sigma_{B_2}=1. \ee
Therefore $G_\gamma$ is also orthogonal to classes of type {\it
(i)}. It is therefore an allowed flux in $F$-theory, as claimed.

Next we discuss the wedge product $G_\gamma\wedge J$, where $J$ is
the K\"ahler class. Any K\"ahler class $J$ on $B_3$ can be
decomposed as
\be J \ =\ p^*J_{B_2} + a\, \pi^*_Y J_0, \ee
where $J_{B_2} \in H^2(B_2, {\bf R})$, $a$ is a real number, and
$J_0$ is any one class in $H^2(B_3)$ not in the image of $\rho^*:
H^2(B_2) \to H^2(B_3)$. We will take $J_0$ to be the class  in
$H^2(B_3)$ of the divisor $\sigma_{B_2}$. We could also take the
section at infinity $J_\infty$, which is related to $J_0$ by
$J_0 - J_\infty = \rho^*c_1(N_{B_2/B_3})$. Note that $\pi_Y^*J_\infty$ is the
class in $H^2(Y)$ of the divisor $Z$. We will see that $G_\gamma
\wedge J$ is closely related to certain moment maps associated to
$U(1)$ gauge symmetries, known as Fayet-Iliopoulos parameters.

Since we are working with harmonic forms, we can work modulo
torsion, and calculating $G_\gamma \wedge J$ is equivalent to
finding all the intersection products
\be \int_{Y_4} G_\gamma \wedge J \wedge D \ee
for any $D \in H^2(Y_4)$. Let us make a list of such classes. To
identify all the divisors in $Y_4$, we use the Leray spectral
sequence associated to the fibration $Y_4 \to B_2$, following
\cite{Curio:1998bv}. The divisors in $Y_4$ are spanned by
\be H^2(Y_4) \sim H^2(B_2,R^0p_*) \oplus H^0(B_2,R^2p_*) \ee
Some of these generators may be lifted by higher order differentials
in the spectral sequence, but we are interested in the span and this
level of analysis suffices. Generators of $H^2(B_2,R^0p_*)$
correspond to pull-backs of divisors in $B_2$ by $p^*$. The second
cohomology of the $dP_9$ can be split up into three pieces:
\be R^2p_*\ =\ [{\bf P}^1]_{base} + [T^2]_{fibre} + \Lambda_{E_8}
\ee
This decomposition makes sense globally over $B_2$ because $Y^4 \to
B_3$ has a section. The rank eight lattice $\Lambda_{E_8}$ is
denoted by $H^2_\Lambda$ in \cite{Curio:1998bv}. Using this
decomposition, we see that the divisors in $Y_4$ are spanned by
\be\label{Y4DivSpan} H^2(Y_4,{\bf Z})\ =\ {\rm span}\{ \sigma_{B_3},
\, \pi_Y^*H^2(B_3,{\bf Z}),\, \Lambda \} \ee
where $\Lambda = H^0(B_2,\Lambda_{E_8})$ is the coroot lattice
defined in (\ref{Lambdadef}).

We claim that $G_\gamma \wedge J$ is automatically orthogonal to
divisors of the first two types in (\ref{Y4DivSpan}). This is easy
to see: $J$ itself is a class in $\pi_Y^*H^2(B_3,{\bf Z})$, and its
intersection with a divisor in the span of $\sigma_{B_3}$ and
$\pi_Y^*H^2(B_3,{\bf Z})$ gives a four-cycle of type {\it (i)} or
{\it (ii)}. But $G_\gamma$ is always orthogonal to such four-cycles,
and hence the claim follows. The remaining intersections we are of
the form
\be \int_Y G_\gamma \wedge J \wedge \omega, \qquad \omega \in
\Lambda \ee

Although for a generic $E_8$ bundle the spectral cover will be
irreducible (since the 240 sheets form a single orbit of ${\cal
W}_{E_8}$), physically we are typically interested in a situation
where the monodromy group is smaller and the cover decomposes into
several pieces. For instance suppose that we restrict the
monodromies to lie in the Weyl group of some $SU(5)_H$ subgroup of
$E_8$. The adjoint representation of $E_8$ decomposes as
\be {\bf 248}\  =\ ({\bf 24},{\bf 1}) + ({\bf 1},{\bf 24}) +
(\bfv,\bt) + (\bfb,\btb) + (\bt,\bfb) + (\btb,\bfv) \ee
under $SU(5)_H \times SU(5)_{GUT}$. Thus in this example, the degree
240 spectral cover splits up into pieces of degree 20, 1, 5 and 10,
with various multiplicities (determined by the dimension of the
corresponding $SU(5)_{GUT}$ representation). If we additionally
restrict the monodromy so as to get extra abelian gauge symmetries,
the spectral cover would split up further.

%let us decompose the spectral cover and the cylinder into its
%irreducible pieces $C_k$ and $R_k$, and denote the restriction of
%$\gamma$ to each component by $\gamma_k$. Note that
%$\pi_{C_k*}\gamma_k$ does not necessarily vanish on each irreducible
%component. If it does, we say that $\gamma$ is strongly primitive.

In order to proceed, we further subdivide the divisors in $\Lambda$.
The lattice $\Lambda_{E_8}$ may be split into two pieces (up to
torsion, which is irrelevant for us): a varying piece, also known as
the Mordell-Weil lattice; and a fixed piece which is locally
constant, also known as the vertical component:
\be \Lambda_{E_8} = \Lambda_{MW} \oplus \Lambda_{vert} \ee
The vertical piece is generated by the exceptional divisors of the
ADE singularity of the generic $dP_9$. Although $\Lambda_{vert}$ is
locally constant, as we go from patch to patch there may be some
non-trivial monodromies. Whether such monodromies are present should
correspond to the `split' versus `non-split' distinction in Tate's
algorithm.

%We further introduce the notion of a Weyl group of the fibration. By
%this we will mean the subgroup of ${\cal W}_{E_8}$ which normalizes
%$\Lambda_{vert}$. The monodromy group of the fibration, or at least
%the part that acts non-trivially on $\Lambda_{MW}$ but trivially on
%$\Lambda_{vert}$, is a subgroup of the Weyl group.

Let us denote divisors in $H^0(B_2,\Lambda_{vert})$ by $E$. By
definition, the remaining generators in $H^0(B_2,\Lambda_{MW})$ can
be identified with certain linear combinations of the non-vanishing
lines in $dP_9$ varying over $B_2$, fitting together into a divisor
of $Y_4$. That
is, they correspond to certain linear combinations of the
irreducible components of the cylinder. They are interpreted as
`extra' $U(1)$'s, i.e. generators of ${\bf g}/[{\bf g},{\bf g}]$ where
${\bf g}$ is the Lie algebra of the $4d$
gauge group. Let us denote generators of $H^0(B_2,\Lambda_{MW})$ by
$\omega_X$.

Now recall that we are supposed to compute (\ref{FIterms}) for any
generator of $\Lambda = H^0(B_2,\Lambda_{E_8})$. Since $\Lambda$ is
spanned by $E$ and $\omega_X$, we will compute these intersections.
We first note that
\be\label{note1} i_*p_R^*\gamma \cdot p^*J_{B_2} \ = \ \sum_k
i_*p_R^*(\gamma_k \cdot \pi_{C_k}^*J_{B_2}) \ee
Geometrically, this represents a collection of lines sitting over a
finite number of points in $B_2$. Let us concentrate on one such
$dP_9$ fiber. Let us also assume that $\Lambda_{vert}$ is actually
constant, so there are no monodromies that can further break the
gauge group (the simply laced cases), which is the case usually
relevant for phenomenological applications. Now all the lines that
lie in a single orbit of the Weyl group have the same intersection
number with a given exceptional cycle of the $dP_9$. Furthermore,
the intersection pairing is preserved by the monodromies. Therefore
we get
\be i_*p_R^*\gamma \cdot p^*J_{B_2}\cdot E \ = \ \sum_k n_k
(\gamma_k \cdot \pi_{C_k}^*J_{B_2}) \ee
where $n_k$ is the intersection number of a line in $R_k$ with $E$.
For our purpose, the only thing that matters is that this number is
the same for all the sheets in a single Weyl orbit. But then by
tracelessness of $\gamma$, the sum over all $k$ belonging to a
single Weyl orbit vanishes. Hence
\ba i_*p_R^*\gamma \cdot p^*J_{B_2} \cdot E &=&  0 \ea

The other intersections are easier to check. We have
\ba [dP_9] \cdot p^*J_{B_2} \cdot E &=& 0 \ea
since we can take the support of the $dP_9$ fiber and $p^*J_{B_2}$
to be disjoint. Further
\ba G_\gamma \cdot \pi_Y^*J_{0} \cdot E &=& 0 \ea
follows because because $\pi_Y^*J_{\infty}$ and $E$ have disjoint support, and
the difference between $\pi_Y^*J_0$ and $\pi_Y^*J_\infty$ has zero intersection by
the calculation above.
(Recall we took the singularities to be along the zero section of
$B_3$, and we defined $J_\infty$ to be the section at infinity, which is
disjoint from it). Therefore $G_\gamma \wedge J$ is automatically
orthogonal to $E$.

The only remaining intersections are $G \wedge J \wedge \omega^X$.
In particular, if there are no $\omega^X \in H^0(B_2, \Lambda_{MW})$
then the $G$-flux should be $J$-primitive. These remaining
intersections correspond to the Fayet-Iliopoulos terms in
$F$-theory:
\be\label{FIterms} \xi^X_F\ \simeq\ m_{10}^4\int_Y G_\gamma \wedge J
\wedge \omega^X, \qquad \omega^X \in H^0(B_2,\Lambda_{MW}) \ee
For a derivation, see section 5.2 of \cite{Donagi:2008kj}. They must vanish in the
$11d$ supergravity regime, but they may be non-vanishing in $F$-theory.
In order to compare
with the heterotic string, it is better to use $4d$ quantities, so we rewrite
the Fayet-Iliopoulos term as
\be\label{FIterms2} \xi^X_F\ \simeq\ M_{Pl}^2 {1\over {\cal V}_F} \int_Y G_\gamma \wedge J_F
\wedge \omega^X \ee
Here ${\cal V}_F$ is the volume of $B_3$ in $10d$ Planck units, and we also absorbed a factor
of $m_{10}^2$ in $J_F$, so that the K\"ahler moduli correspond to volumes measured in Planck units.
By supersymmetry, this expression is also related to the matrices
$\Pi_M^X$ in \cite{Donagi:2008kj} which describe couplings between
$U(1)_X$ gauge fields and RR axions. As discussed in \cite{Donagi:2011jy},
we expect that the Fayet-Iliopoulos parameters can be non-vanishing away from the $11d$ supergravity limit,
and that they can be used
to define stability conditions on an $F$-theory compactification.

On
the heterotic side, the Fayet-Iliopoulos terms are given by
\be \xi^X_{Het}\ \sim\  M_{Pl}^2{1\over {\cal V}_h} \int_Z  c_1(L^X)
\wedge J_h \wedge J_h \ee
where ${\cal V}_h ={\rm vol}(Z)$, and all volumes are measured in string units.
Here $L^X$ is a fractional power of the determinant of a sub-bundle
$V' \subset V$ against which we are testing stability. The heterotic
Green-Schwarz terms also give a loop correction to this expression
of the form \cite{Dine:1987xk,Blumenhagen:2005ga}
\be
\delta \xi^X_{het} \ \sim \ M_{Pl}^2 {1\over S} \int c_1(L^X) \wedge (c_2(V) - \half c_2(T))
\ee
which describes one-loop corrections related to $U(1)$ anomalies. For spectral cover bundles,
we can write this suggestively as
\be
\delta \xi^X_{het} \ \sim \ \int_{B_2} c_1(L)\cdot (\eta - 6 c_1) \ =
\ \int_{B_2}c_1(L) \wedge c_1(N_{B_2/B_3})
\ee
where $L^X = \pi_{Z}^*L$, and we used $\pi_{Z*}(c_2(V) - \half c_2(T)) = \eta - 6 c_1$.

To match the expressions qualitatively, we assume for simplicity that $\omega^X$ is completely localized
on some singularity, and we take $B_3 = B_2 \times {\bf P}^1$. For simplicity we also take the size of
the elliptic fiber on the heterotic side of order one.  The heterotic and $F$-theory
quantities are related as
\be
{\cal V}_F\ \sim\ {\rm vol}(B_2) \times {\rm vol}({\bf P}^1)\ \sim\ {\cal V}_h \, \lambda_h^{-1}
\ee
and also
\be
\lambda_h J_{B_2,F}\ \sim\ J_{B_2,h}, \qquad \lambda_h^2\ \sim\ {\cal V}_h/S
\ee
Then we find
\be
M_{Pl}^2 {1\over {\cal V}_F} \int_Y G_\gamma \wedge J_{B_2,F}
\wedge \omega^X  \ \sim\  M_{Pl}^2 {1\over {\cal V}_h} \int_{B_2} c_1(L) \wedge J_{B_2,h}
\ee
and similarly
\be
M_{Pl}^2 {1\over {\cal V}_F} \int_Y G_\gamma \wedge J_{0}
\wedge \omega^X  \ \sim\ M_{Pl}^2 {1\over S} \int_{B_2} c_1(L) \wedge (\eta - 6\, c_1(S))
\ee
In other words, the two pieces of the classical $F$-theoretic expression match qualitatively
with the tree level and the one-loop contribution in the heterotic string respectively.
If the $U(1)$ anomalies vanish, as in the MSSM, then the one-loop contribution will vanish. In
addition, there could be non-perturbative corrections on both sides

We can relate both heterotic and $F$-theoretic expressions to spectral cover data.
The calculation is slightly intricate. We didn't match the expressions from $F$-theory and the
heterotic string precisely, but they should likely match. It would be interesting
to check this more precisely.

It is interesting to note that since relative size of the $4d$ string coupling $S$ and the K\"ahler moduli $T$
is varied as we extrapolate from the heterotic string to $F$-theory, the loop-corrected slope also varies and
we could write down models which are stable in the heterotic regime and unstable in the $F$-theory regime, or vice versa. This appears to be one of the few settings where we can study stability issues as a function of the string coupling.

\end{document}